\documentclass[runningheads,twoside]{llncs}

\usepackage{eqnarray} 
\usepackage{tikz}
\usetikzlibrary{matrix,arrows,decorations.pathmorphing,calc,shadings,decorations.pathreplacing,patterns}
\usepackage{bbm}
\usepackage{amssymb,multirow,commath,graphicx,setspace,makecell}
\usepackage{amsfonts,bbding,amsmath}
\usepackage{latexsym,url}
\usepackage{shuffle}
\usepackage{microtype}
\usepackage{chngcntr}
\usepackage{enumerate}
\usepackage{thmtools}
\usepackage{enumitem}
\usepackage[font=scriptsize]{caption}

\usetikzlibrary{matrix,arrows,decorations.pathmorphing}
\def\natnum{{\mathbb N}}

\def\email#1{Email: {\tt #1}}
\def\proof{\noindent{\bf Proof.}\enspace}

\def\qed{~~\hbox{\hskip 1pt \vrule width 4pt height 8pt depth 1.5pt\hskip 1pt}}
\spnewtheorem{thm}{Theorem}{\bf }{\it }
\spnewtheorem{prop}[thm]{Proposition}{\bf }{\it }
\spnewtheorem{prob}[thm]{Open Problem}{\bf }{\it }
\spnewtheorem{cor}[thm]{Corollary}{\bf }{\it }
\spnewtheorem{lem}[thm]{Lemma}{\bf }{\it }
\spnewtheorem{defn}[thm]{Definition}{\bf }{\rm }
\spnewtheorem{rem}[thm]{Remark}{\bf }{\rm }
\spnewtheorem{exmp}[thm]{Example}{\bf }{\rm }
\spnewtheorem{clm}[thm]{Claim}{\bf }{\it }
\spnewtheorem{nota}[thm]{Notation}{\bf }{\rm }

\newcommand{\PBT}{\mbox{PBT}}
\newcommand{\PBTD}{\mbox{PBTD}}
\newcommand{\TD}{\mbox{TD}}

\newcommand\spn[1]{{\left\langle#1\right\rangle}}
\newcommand{\wh}{\widehat}
\newcommand{\pnats}{\mathbb{N}}
\newcommand{\nats}{\mathbb{N}_0}
\newcommand{\ve}{\varepsilon}
\newcommand{\sm}{\setminus}

\newcommand{\ra}{\rightarrow}

\newcommand{\eset}{\emptyset}

\newcommand{\cF}{{\mathcal F}}
\newcommand{\cI}{{\mathcal I}}
\newcommand{\cJ}{{\mathcal J}}
\newcommand{\cC}{{\mathcal C}}
\newcommand{\Var}{\mbox{Var}}

\newcommand{\NC}{\mbox{NC}}
\newcommand{\N}{\mbox{N}}
\newcommand{\R}{\mbox{R}}
\newcommand{\ga}{\gamma}

\newcommand{\Const}{\mbox{Const}}
\newcommand{\SR}{\mbox{SR$\Pi$}}
\newcommand{\QR}{\mbox{QR$\Pi$}}
\newcommand{\AG}{\mbox{AG}}

\begin{document}

\title{The Teaching Complexity of Erasing Pattern Languages With Bounded Variable Frequency} 

\titlerunning{Erasing Pattern Languages With Bounded Variable Frequency}

\author{
Ziyuan Gao} 

\authorrunning{
Z.~Gao} 

\institute{ 
Department of Mathematics, National University of Singapore \\
10 Lower Kent Ridge Road, Singapore 119076, Republic of Singapore \\
\email{matgaoz@nus.edu.sg}}

\maketitle

\begin{abstract}
Patterns provide a concise, syntactic way of describing a set of strings, but their 
expressive power comes at a price: a number of fundamental decision problems concerning 
(erasing) pattern languages, such as the membership problem and inclusion problem, are known to be 
NP-complete or even undecidable, while the decidability of the equivalence problem is still open; in 
learning theory, the class of pattern languages
is unlearnable in models such as the distribution-free (PAC)
framework (if $\mathcal{P}/poly \neq \mathcal{NP}/poly$). 
Much work on the algorithmic learning of pattern languages has thus focussed 
on interesting subclasses of patterns for which positive learnability results may be achieved. 
A natural restriction on a pattern is a bound on its variable frequency -- the maximum number $m$ such
that some variable occurs exactly $m$ times in the pattern.  This paper examines the effect of 
limiting the variable frequency of all patterns belonging to a class $\Pi$ 
on the worst-case minimum number of labelled examples needed to uniquely identify any pattern 
of $\Pi$ in cooperative teaching-learning models. 
Two such models, the teaching dimension model as well as the preference-based teaching model,
will be considered. 
\end{abstract}

\section{Introduction}

In the context of this paper, a \emph{pattern} is a string made up of symbols from two disjoint sets, 
a countable set $X$ of \emph{variables} and an alphabet $\Sigma$ of \emph{constants}.  The \emph{non-erasing
pattern language} generated by a pattern $\pi$ is the set of all words obtained by substituting
nonempty words over $\Sigma$ for all the variables in $\pi$, under the condition that for any
variable, all of its occurrences in $\pi$ must be replaced with the same word; the \emph{erasing pattern 
language} generated by $\pi$ is defined analogously, the only difference being that the variables in 
$\pi$ may be replaced with the empty string. 
Unless stated otherwise, all pattern languages in the present paper refer to erasing pattern languages.
In computational learning theory, the non-erasing pattern languages were introduced by Angluin \cite{Angluin80} 
as a motivating example for her work on the identification of uniformly decidable families of languages in the 
limit. 
Shinohara \cite{Shinohara82b} later introduced the class of erasing pattern languages, proving that
the class of all such languages generated by \emph{regular} patterns (patterns in which every variable occurs at most once)
is polynomial-time learnable in the limit. 
Patterns and allied notions - such as that of an extended regular expression \cite{Aho90,Campeanu03,Schmid16,FS17}, 
which has more expressive power than a pattern -- have also been studied in other fields, including word
combinatorics and pattern matching.  For example, the membership problem for pattern languages is closely 
related to the problem of matching `patterns' with variables (based on various definitions 
of `pattern') in the pattern matching community \cite{Baker96,Amir07,FernauMMS15,DayFMN17,DayFMNS18}. 

The present paper considers the problem of uniquely identifying pattern languages from
labelled examples -- where a labelled example for a pattern
language $L$ is a pair $(w,*)$ such that $*$ is ``$+$'' if $w$ belongs to $L$ and
``$-$'' otherwise -- based on formal teaching-learning models. 
We shall study two such 
models in the computational learning theory literature: the well-known teaching dimension (\TD) 
model \cite{GoldmanK95,ShinoharaM91} 
and the preference-based teaching (\PBT) 
model \cite{GRSZ2016} (c.f.\ Section \ref{sec:tdpbtd}).  Given a model $\mathcal{T}$ and any class 
$\Pi$ of patterns to be learnt, the maximum size of a sample (possibly $\infty$) needed for a learner to 
successfully identify any pattern in $\Pi$ based on the teaching-learning algorithm of $\mathcal{T}$ is 
known as the teaching complexity of $\Pi$ (according to $\mathcal{T}$).      
The broad question 
we try to partly address is: what properties 
of the patterns in a given class $\Pi$ of patterns influence the teaching complexity of $\Pi$
according to the \TD\ 
and \PBT\ models? 
More specifically, let $\Pi_m$ be a class of patterns $\pi$ such that the maximum number
of times any single variable occurs in $\pi$ (known here as the \emph{variable frequency}
of $\pi$) is at most $m$; how does the teaching complexity of $\Pi_m$ vary with
$m$?  The variable frequency of a pattern is quite a natural parameter that has been 
investigated in other problems concerning pattern languages.  For example, 
Matsumoto and Shinohara \cite{Matsumoto97} established an upper bound on the query complexity of 
learning (non-erasing) pattern languages in terms of 
the variable frequency of the
pattern and other parameters; 
Fernau and Schmid \cite{FernauS15} 
proved that the membership problem for patterns remains 
NP-complete even when the variable frequency is restricted to $2$ (along with other
parameter restrictions). 
 
In this paper, one motivation for concentrating on the variable frequency of a pattern  
rather than, say, the number of distinct variables occurring in the pattern, comes
from examining the teaching complexity of some basic patterns.  Take the constant
pattern $0$, where $0$ is a letter in the alphabet $\Sigma$ of constants.  The language generated 
by this pattern cannot be finitely distinguished (i.e., distinguished 
using a finite set of labelled examples) 
from every other pattern 
language, 
even only those generated by a pattern with at most one variable.  
Indeed, any finite set $\{(0,+),(w_1,-),\ldots,(w_k,-)\}$ of labelled examples for the pattern $0$ is also 
consistent with the pattern $0x^m$ where $m=\max_{1\le i\le k}|w_i|$.
The latter 
observation depends crucially on the 
fact that a variable may occur any number of times in a pattern, and less so on the number of distinct 
variables occurring in a pattern.  A similar 
remark applies to 
the pattern languages generated by
patterns with a constant part of length at least $2$ 
\cite[Theorem 3]{bayeh17}.  
On the other hand, if one were to teach the singleton
language $\{0\}$ w.r.t.\ all languages generated by patterns with variable frequency at most $k$ for
some fixed $k$, then a finite distinguishing set for $\{0\}$ could consist of $(0,+)$ plus all negative
examples $(0^n,-)$ with $2 \leq n \leq k+1$.  This seems to suggest that the maximum variable 
frequency of the patterns in a class of patterns may play a crucial role in determining
whether or not the languages generated by members of this class are finitely distinguishable. 

The first section of this work studies the teaching complexity of \emph{simple block-regular} 
patterns, which are equivalent to patterns of the shape $x_1 a_1 x_2 a_2 \ldots a_{n-1}$ $x_n$, 
where $x_1,\ldots,x_n$ are distinct variables and $a_1,\ldots,a_{n-1}$ are constants.
They make up one of the simplest, non-trivial classes of patterns that have a restriction on
the variable frequency.  Bayeh et al. \cite{bayeh17} showed that over alphabets of size
at least $4$, the languages generated by such patterns are precisely those
that are finitely distinguishable; we refine this result by determining, over any alphabet, the 
\TD\ and \PBT\ dimensions of the class of simple block-regular patterns.  Further, we calculate
the \TD\ of these patterns w.r.t.\ the class of regular patterns and provide 
an asymptotic lower bound for the \TD\ of any given simple block-regular pattern w.r.t.\ 
the whole class of patterns. 
In the subsequent section, we proceed to the more general problem of 
determining, for
various natural classes $\Pi$ of patterns that have a uniformly bounded variable frequency, 
those members of $\Pi$ that are finitely distinguishable.
It will be proven that all $m$-quasi-regular patterns (i.e.\ every variable of the pattern
occurs exactly $m$ times) and $m$-regular (i.e.\ every variable 
occurs at most $m$ times) non-cross patterns are finitely distinguishable w.r.t.\ the class of
$m$-quasi-regular and $m$-regular non-cross patterns respectively; moreover, 
the \TD\ of the class of $m$-regular non-cross patterns is even finite and in fact sublinear
in $m$. 
Next, we present partial results on the problem of determining the subclass of
$m$-regular 
patterns that have a finite \TD. 
Over any infinite alphabet, \emph{every} $m$-regular pattern is finitely distinguishable --
contrasting quite sharply with the previously mentioned theorem that over alphabets with at
least $4$ letters, the only patterns with a finite \TD\ are the simple block-regular
ones.
Over binary alphabets, on the other hand, there are patterns that are not finitely distinguishable even when 
the variable frequency is restricted to $4$.  

Due to space constraints, most proofs have been deferred to the appendix.  

\section{Preliminaries}

$\natnum_0$ denotes the set of natural numbers $\{0,1,2,\ldots\}$ and
$\natnum=\natnum_0\setminus\{0\}$. 
Let $X = \{x_1,x_2,x_3,\ldots\}$ be an infinite set of variable symbols. An alphabet is a finite or countably infinite set of symbols, disjoint from $X$. Fix an alphabet $\Sigma$.  A \emph{pattern} is a nonempty finite string over 
$X \cup \Sigma$.  The class of patterns over any alphabet $\Sigma$ with $z = |\Sigma|$
is denoted by $\Pi^z$; this notation reflects the fact that all the properties of patterns
and classes of patterns considered in the present work depend only on the size of the alphabet 
and not on the actual letters of the alphabet.  
The 
\emph{erasing pattern language} $L(\pi)$ generated by a pattern $\pi$ over $\Sigma$ 
consists of all strings generated from $\pi$ when replacing variables in $\pi$ with any string over $\Sigma$, 
where all occurrences of a single variable must be replaced by the same string \cite{Shinohara82b}. 
Patterns $\pi$ and $\tau$ over $\Sigma$ are said to be \emph{equivalent}
iff $L(\pi) = L(\tau)$;
they are \emph{similar} iff $\pi = \alpha_1 u_1 \alpha_2 u_2 \ldots u_n \alpha_n$
and $\tau = \beta_1 u_1 \beta_2 u_2 \ldots u_n \beta_n$ for some $u_1,u_2,\ldots,u_n \in \Sigma^+$
and $\alpha_1,\ldots,\alpha_n,\beta_1,\ldots,\beta_n \in X^*$. 
Unless specified otherwise, we identify any pattern $\pi$ belonging
to a class $\Pi$ of patterns with every other $\pi' \in \Pi$ such that $L(\pi) = L(\pi')$. 
$\Var(\pi)$ (resp.~$\Const(\pi)$) denotes the
set of all distinct variables (resp.~constant symbols) occurring in $\pi$. 

For any symbol $a$ and $n \in \natnum_0$, 
$a^n$ denotes the string equal to $n$ concatenated copies of $a$. 
For any alphabets $A$ and $B$, a \emph{morphism}
is a function $h:A^* \ra B^*$ with $h(uv) = h(u)h(v)$
for all $u,v \in A^*$.  A \emph{substitution} 
is a morphism $h:(\Sigma \cup X)^* \ra \Sigma^*$
with $h(a) = a$ for all $a \in \Sigma$. 
By abuse of notation, we will often use the same symbol
$h$ to represent the morphism $(X \cup \Sigma)^* \mapsto \Sigma^*$
that coincides with the substitution $h$ on individual variables
and with the identity function on letters from $\Sigma$.
$\cI_{h,\pi}$ denotes the mapping of closed intervals of positions of 
$\pi$ to closed intervals of positions of $h(\pi)$
induced by $h$;
$\pi(\ve)$ denotes the word obtained from $\pi$ by substituting 
$\ve$ for every variable in $\pi$.
Let $\sqsubseteq$ denote the \emph{subsequence} relation on $\Sigma^*$: 
$u \sqsubseteq v$ holds iff there are numbers $i_1 < i_2 < \ldots < i_{|u|}$
such that $v_{i_j} = u_j$ for all $j \in \{1,\ldots,|u|\}$.
Given any $u,v \in \Sigma^*$, the \emph{shuffle product} of $u$ and $v$,
denoted by $u \shuffle v$, is the set $\{u_1v_1u_2v_2\ldots u_kv_k: u_i,v_i \in \Sigma^*
\wedge u_1u_2\ldots u_k = u \wedge v_1v_2\ldots v_k = v\}$.
Given any $A, B \subseteq \Sigma^*$, the \emph{shuffle product} of $A$
and $B$, denoted by $A \shuffle B$, is the set $\bigcup_{u \in A\wedge v \in B}u \shuffle v$.
If $A = \{u\}$ 
, we will often write $A \shuffle B$
as $u \shuffle B$. 

\section{Teaching Dimension and Preference-based Teaching Dimension}\label{sec:tdpbtd}

\emph{Machine teaching} 
focusses on the problem of designing, for any given learning algorithm, an optimal training 
set 
for every concept belonging to a class of concepts to be learnt \cite{ZSZR18}.  
Such a training set is sometimes known as a \emph{teaching set}.
In this work, an ``optimal'' teaching set for a pattern $\pi$ is one that has the minimum number of
examples labelled consistently with $\pi$ needed for the algorithm to successfully identify $\pi$ (up to equivalence).
We study the design of optimal teaching sets for various classes of pattern languages w.r.t.\
(i) the classical \emph{teaching dimension} model~\cite{GoldmanK95,
ShinoharaM91}, 
where it is only assumed that the learner's hypotheses are always consistent with the given 
teaching set; (ii) the \emph{preference-based teaching} model \cite{GRSZ2016}, where the learner has, for
any given concept class, a particular ``preference relation'' on the class, and the learner's
hypotheses are always not only consistent with the given teaching set, but also not less 
preferred to any other concept in the class w.r.t.\ the preference relation.

Fix an alphabet $\Sigma$.
Let $\Pi$ be any class of patterns, and suppose $\pi \in \Pi$.   
A \emph{teaching set for $\pi$ w.r.t.\ $\Pi$} is a set $T \subseteq \Sigma \times \{+,-\}$ that is consistent with 
$\pi$ but with no other pattern in $\Pi$
(up to equivalence), that is, $w \in L(\pi)$ for all $(w,+) \in T$ and $w \notin L(\pi)$ for all $(w,-) \in T$.
The \emph{teaching dimension of $\pi$ w.r.t.\ $\Pi$}, denoted by $\TD(\pi,\Pi)$ is defined as
$
\TD(\pi,\Pi)=\inf\{|T|: T\mbox{ is a teaching set for }\pi\mbox{ w.r.t.\ }\Pi\}.
$
Furthermore, if $\Pi' \subseteq \Pi$, then the \emph{teaching dimension of $\Pi'$ w.r.t.\ $\Pi$}, denoted by $\TD(\Pi',\Pi)$, is defined as 
$
\TD(\Pi',\Pi)=\sup\{\TD(\pi,\Pi): \pi \in \Pi'\}.
$
The \emph{teaching dimension of $\Pi$}, denoted by $\TD(\Pi)$, is defined as $\TD(\Pi,\Pi)$.

In real-world learning scenarios, even the smallest possible teaching set for a given
concept relative to some concept class may be impractically large.  
Learning algorithms often make predictions based on a set of assumptions
known as the \emph{inductive bias}, which may allow the algorithm
to infer a target concept from a small set of data even when there is more than
one concept in the class that is consistent with the data.  
Certain types of bias impose an a priori preference ordering on the learner's hypothesis 
space; for example, an algorithm that adheres to the Minimum Description Length (MDL)
principle favours hypotheses that have shorter descriptions based on some given
description language.  The preference-based teaching model, to be defined shortly, considers learning
algorithms with an inductive bias that specifies a preference ordering of the learner's 
hypotheses. 
 
Let $\prec$ be a {\em strict partial order} on $\Pi$,
i.e., $\prec$ is asymmetric and transitive. 
The partial order that makes every pair $\pi,\pi' \in \Pi$ (where $L(\pi) \neq L(\pi')$)
incomparable 
is denoted by $\prec_\eset$. For every $\pi \in\Pi$, let 
$\Pi_{\prec \pi} = \{\pi'\in\Pi: \pi' \prec \pi\}$
be the set of patterns over which $\pi$ is strictly preferred (as mentioned
earlier, equivalent patterns are identified with each other).
A {\em teaching set for $\pi$ w.r.t.~$(\Pi,\prec)$} is defined as 
a teaching set for $\pi$ w.r.t.~$\Pi\sm\Pi_{\prec \pi}$. Furthermore define 
$\PBTD(\pi,\Pi,\prec) = \inf\{|T| : T\mbox{ is a teaching set for $\pi$ w.r.t.~$(\Pi,\prec$})\}
\in \natnum_0\cup\{\infty\}$.
The number $\PBTD(\Pi,\prec) = \sup_{\pi \in \Pi}\PBTD(\pi,\Pi,\prec) \in \natnum_0\cup\{\infty\}$ 
is called the {\em teaching dimension of $(\Pi,\prec)$}. 
The {\em preference-based teaching dimension of $\Pi$} is given by
$
\PBTD(\Pi) = \inf\{\PBTD(\Pi,\prec) : \mbox{$\prec$ is a strict partial order on $\Pi$}\}. 
$
For all pattern classes $\Pi$ and $\Pi'$ with $\Pi' \subseteq \Pi$,
$K(\Pi') \leq K(\Pi)$ for $K \in \{\TD,\PBTD\}$ (i.e.\ the \TD\ and \PBTD\ are monotonic) and 
$\PBTD(\Pi) \leq \TD(\Pi)$ \cite{GRSZ2016}. 

\section{Simple Block-Regular Patterns} 


Fix an alphabet $\Sigma$ of size $z \leq \infty$.
A pattern $\pi \in \Pi^z$ is said to be \emph{simple block-regular} if
it is of the shape $X_1 a_1 X_2 a_2 \ldots a_{n-1} X_n$, where $X_1,\ldots,X_n \in X^+$,
$a_1,\ldots,a_{n-1} \in \Sigma$, and for all $i \in \{1,\ldots,n\}$,
$X_i$ contains a variable that does not occur in any other variable block $X_j$ with $j \neq i$.
Every simple block-regular
pattern is equivalent to a pattern $\pi'$ of the shape
$y_1a_1y_2a_2\ldots$ $a_ky_{k+1}$, where
$k \geq 0$, $a_1,a_2,\ldots,a_k \in \Sigma$ and $y_1,y_2,\ldots,y_{k+1}$
are $k+1$ distinct variables \cite[Theorem 6(b)]{Jain10}.  $\SR^z$ denotes
the class of all simple block-regular patterns in $\Pi^z$.  $\SR^z$
is a subclass of the family of \emph{regular} patterns (denoted by $\R\Pi^z$),
which are patterns in which every variable occurs at most once.

As mentioned in the introduction, the simple block-regular patterns constitute precisely
the subclass of finitely distinguishable patterns over any alphabet of size at least $4$ 
\cite[Theorem 3]{bayeh17}.  The language generated by a simple block-regular pattern
is known as a \emph{principal shuffle ideal} in word combinatorics \cite[\S 6.1]{Lothaire97}, 
and the family of all such languages is an important object of study in the PAC learning model \cite{AAEK13}.   

The goal of this section is to determine the teaching complexity of the class of simple 
block-regular patterns over any alphabet $\Sigma$ w.r.t.\ three classes: $\SR^{|\Sigma|}$ itself, 
$\R\Pi^{|\Sigma|}$ and $\Pi^{|\Sigma|}$.
It will be shown that $\TD(\SR^{|\Sigma|}) < \TD(\SR^{|\Sigma|},\R\Pi^{|\Sigma|})$ $< \TD(\SR^{|\Sigma|},\Pi^{|\Sigma|})$.
To this end, we introduce a uniform construction of a certain negative example
for any given pattern $\pi$; as will be seen shortly, this example is powerful enough to 
distinguish $\pi$ from every simple block-regular pattern whose constant
part is 
a proper subsequence (not necessarily contiguous) of the constant part of $\pi$.    

\begin{nota}\label{nota:hatw}
For any word $w = \delta_1^{m_1}\delta_2^{m_2}\ldots
\delta_k^{m_k}$, where $\delta_1,\ldots,\delta_k \in \Sigma$
and $\delta_i \neq \delta_{i+1}$ whenever $1 \leq i < k$, 
$m_1,\ldots,m_k \geq 1$ and $k \geq 1$, define
\begin{equation}\label{eqn:simplebrnegativehatw}
\wh{w} := \underbrace{\delta_1^{m_1-1}\delta_2^{m_2}\delta_1}_{} \underbrace{\delta_2^{m_2-1}\delta_3^{m_3}\delta_2}_{}
\ldots \underbrace{\delta_i^{m_i-1}\delta_{i+1}^{m_{i+1}}\delta_i}_{} \ldots 
\underbrace{\delta_{k-1}^{m_{k-1}-1}\delta_k^{m_k}\delta_{k-1}}\underbrace{\delta_k^{m_k-1}}.
\end{equation}
\end{nota}

\noindent (In particular, if $m \geq 1$, then $\wh{\delta_1^m} = \delta_1^{m-1}$.)

\begin{lem}\label{lem:simplebrnegative}
Fix any $z \in \natnum \cup \{\infty\}$ and any $\pi,\tau \in \SR^z$ with
$\pi(\ve) \neq \ve$.
Then $\wh{\pi(\ve)} \notin L(\pi)$.  
Furthermore, if $\tau(\ve) \sqsubset \pi(\ve)$, then 
$\wh{\pi(\ve)} \in L(\tau)$.  
\end{lem}

\proof
Suppose $\pi(\ve) = \delta_1^{m_1}\delta_2^{m_2}\ldots\delta_k^{m_k}$,
where $\delta_1,\ldots,\delta_k \in \Sigma$ and $\delta_i \neq \delta_{i+1}$ 
whenever $1 \leq i < k$, $m_1,\ldots,$ $m_k \geq 1$ and $k \geq 1$. 
That $\wh{\pi(\ve)} \notin L(\pi)$ may be argued as follows: if
$k = 1$, then $\wh{\pi(\ve)} = \delta_1^{m_1-1} \sqsubset \pi(\ve)$ is immediate;
if $k \geq 2$, then one shows by induction that for $i = 1,\ldots,k-1$, $\delta_1^{m_1}
\delta_2^{m_2}\ldots\delta_i^{m_i}\delta_{i+1}\not\sqsubseteq \underbrace{\delta_1^{m_1-1}
\delta_2^{m_2}\delta_1}_{}\underbrace{\delta_2^{m_2-1}\delta_3^{m_3}\delta_2}_{}
\ldots$ $\underbrace{\delta_i^{m_i-1}\delta_{i+1}^{m_{i+1}}\delta_i}_{}$.
For the second part of the lemma, suppose $\tau(\ve) = \delta_1^{n_1}
\delta_2^{n_2}\ldots\delta_k^{n_k}$,
where $0 \leq n_i \leq m_i$ for all $i \in \{1,\ldots,k\}$ and
$n_{i_0} \leq m_{i_0}-1$ for some least number $i_0$.  
Taking $w = \pi(\ve)$ in Equation (\ref{eqn:simplebrnegativehatw}), 
observe that $\delta_i^{n_i} \sqsubseteq
\delta_i^{m_i-1}\delta_{i+1}^{m_{i+1}}\delta_i$ for all $i < i_0$,
$\delta_{i_0}^{n_{i_0}} \sqsubseteq \delta_{i_0}^{m_{i_0}-1}$,
and $\delta_j^{n_j} \sqsubseteq \delta_j^{m_j}\delta_{j-1}\delta_j^{m_j-1}$
for all $j > i_0$.  Thus, since $\tau$ is simple block-regular, one has that 
$\wh{\pi(\ve)} \in L(\tau)$.~\qed

\medskip
\noindent
Lemma \ref{lem:simplebrnegative} now provides a tool for establishing the 
\TD\ of $\SR^z$.

\begin{thm}\label{thm:tdsimpleblockregular}
For any $z \in \natnum \cup \{\infty\}$, $\TD(\SR^z) = 2$ and $\PBTD(\SR^z) = 1$. 
\end{thm}

\proof
Fix any $0 \in \Sigma$.  
The pattern $\pi := x_1 0 x_2$ needs to be taught with
at least one negative example in order to distinguish it from $x_1$.
Suppose a teaching set for $\pi$ contains $(w_1w_2\ldots w_k,-)$, where $w_1,\ldots,w_k 
\in \Sigma$.  For any $m \geq 3$, $w_1w_2\ldots w_k \notin L(\pi')$, where $\pi' := x_1w_1x_2w_2x_3\ldots x_k w_k x_{k+1} 0 x_{k+2} 
0$ $\ldots 0 x_{k+m}$.  Since $\pi'$ is simple block-regular and $L(\pi') \neq L(\pi)$, 
at least one additional example is required to distinguish $\pi$ from $\pi'$. 
Hence $\TD(\SR^z) \geq 2$.  

Let $\pi$ be any simple block-regular pattern.  Since $x_1$ can be taught 
with the single example $(\ve,+)$, we will suppose that
$\pi(\ve) \neq \ve$.
A teaching set for $\pi$ consists of the two examples $(\pi(\ve),+)$
and $(\wh{\pi(\ve)},-)$.  By Lemma \ref{lem:simplebrnegative}, $(\wh{\pi(\ve)},-)$
is 
consistent with $\pi$ and $(\wh{\pi(\ve)},-)$ distinguishes $\pi$ 
from all patterns $\pi'$ such that $\pi'(\ve) \sqsubset \pi(\ve)$,
while $(\pi(\ve),+)$ distinguishes $\pi$ from all patterns
$\pi''$ such that $\pi''(\ve) \not\sqsubseteq \pi(\ve)$.       

Let $\prec$ be a preference relation on
$\SR^z$ such that for any $\pi,\tau \in \SR^z$ with $L(\pi) \neq L(\tau)$,
$\pi \prec \tau$ iff $\left|\pi(\ve)\right| < \left|\tau(\ve)\right|$.
Every $\pi \in \SR^z$ can be taught w.r.t.\ $(\SR^z,\prec)$ 
using the example $(\pi(\ve),+)$: for every $\tau \in \SR^z$ such that
$L(\tau) \neq L(\pi)$ and $\pi(\ve) \in L(\tau)$, $\tau(\ve) \sqsubset \pi(\ve)$;
thus $|\tau(\ve)| < |\pi(\ve)|$ and so $\pi \succ \tau$. 
~\qed

\medskip
\noindent
Not surprisingly, the \TD\ 
of a simple block-regular
pattern is in general larger w.r.t.\ the whole class
of regular patterns than w.r.t.\ the restricted class of 
simple block-regular patterns. 
It might be worth noting that a smallest teaching set for a 
simple block-regular pattern $\pi$ need not necessarily contain 
$\pi(\ve)$ as a positive example, 
as the proof of the following result (c.f.\ Appendices \ref{appen:prooflemtdsimplebregpatbinary}
and \ref{appen:tdsimplebregpatatleast3}) shows.

\begin{thm}\label{thm:tdsimplebrregpat}
$\TD(\SR^z,\R\Pi^z) = 3$.
\end{thm}

\noindent
To prove the lower bound in Theorem \ref{thm:tdsimplebrregpat}, it suffices to observe that any teaching set
(w.r.t.\ the whole class of regular patterns) for a non-constant 
regular pattern not equivalent to $x_1$ 
must contain at least two
positive examples and one negative example; for a very
similar proof, see \cite[Theorem 12.1]{bayeh17}. 
We 
prove the upper bound.
If $z = 1$, then $\R\Pi^z$ is the union of $\SR^z$ and all constant patterns (up to 
equivalence).  By the proof of Theorem \ref{thm:tdsimpleblockregular}, any
$\pi \in \SR^z$ can be distinguished from every non-equivalent $\tau \in \SR^z$
with one positive example or one positive and one negative example; to distinguish $\pi$
from any constant pattern, at most one additional positive example is needed.
Suppose $z \geq 2$.
The proof will be split 
into the cases (i) $|\Sigma| = 2$
and (ii) $|\Sigma| \geq 3$. 

\begin{lem}\label{lem:tdsimplebregpatbinary}
If $\pi \in \SR^2$, then $\TD(\pi,\R\Pi^2) \leq 3$. 
\end{lem}

\def\prooflemtdsimplebregpatbinary{
\proof
Suppose $\pi = x_1\delta_1x_2\delta_2\ldots \delta_{n-1}x_n$, where $\delta_1,\delta_2,\ldots,
\delta_{n-1} \in \Sigma$. 
We build a teaching set $T$ for $\pi$ w.r.t.\ $R\Pi^2$.  Let
$\tau$ denote any regular pattern that is consistent with $T$.
Let $w_1$ be the word obtained from $\pi$ as follows: first, substitute $\overline{\delta}_1$
for $x_1$ and substitute $\overline{\delta}_{n-1}$ for $x_n$; second, for every
substring of $\pi$ of the shape $\delta x_i \delta$, where $\delta \in \Sigma$,
replace $x_i$ with $\overline{\delta}$; all other variables are replaced with $\ve$. 
Next, let $w_2$ be the word obtained from $\pi$ such that for every substring
of $\pi$ of the shape $\delta x_i \overline{\delta}$, where $\delta \in \Sigma$,
$x_i$ is replaced with $\delta$; all other variables are replaced with $\ve$.
Let $\varphi_1$ (resp.~$\varphi_2$) be the corresponding substitution witnessing 
$w_2 \in L(\tau)$ (resp.~$w_2 \in L(\pi)$).

Put $(w_1,+)$ and $(w_2,+)$ into $T$.  Since, for every $\delta \in \Sigma$, 
$w_1$ does not contain the subword $\delta\delta$ while $w_2$ does not contain 
the subword $\delta\overline{\delta}\delta$, and $w_1,w_2$ both start and end 
with different letters, $\tau$ must be of the shape $x_1A_1x_2A_2\ldots A_kx_{k+1}$, 
where $k \leq n-1$ and for all $i \in \{1,\ldots,k\}$, $A_i \in \{0,1,01,10\}$.  
Thus one may assume, without loss of generality, that $\tau$ is a simple block-regular 
pattern.     
For each position $p$ of $w_2$ such that $\tau[\overline{\cI}_{\varphi_1,\tau}(p)] \in \Sigma$ but 
$\pi[\overline{\cI}_{\varphi_2,\pi}(p)] \in X$, note that $p \geq 2$ and $w_2[p-1]$ must be equal to 
$w_2[p]$, and since $\tau$ does not
contain a substring of the shape $\delta\delta$ for any $\delta \in \Sigma$
(as observed 
earlier), it follows that $\tau[\overline{\cI}_{\varphi_1,\tau}(p-1)] \in X$.
Consequently, $\tau(\ve) \sqsubseteq \pi(\ve)$.  One may then conclude
from Lemma \ref{lem:simplebrnegative} that adding $(\wh{\pi(\ve)},-)$ to
$T$ ensures $\tau(\ve) = \pi(\ve)$.  As $\tau$ is simple block-regular, 
we have that $L(\tau) = L(\pi)$, as required. 
~\qed
}

\medskip
\noindent The basic proof idea of Lemma \ref{lem:tdsimplebregpatbinary} -- using
positive examples to exclude certain types of constant segments of the target pattern -- 
can also be generalised to the case $|\Sigma| \geq 3$, although the details of the 
construction are more tedious.

\begin{lem}\label{lem:tdsimplebregpatatleast3}
Suppose $z = |\Sigma| \geq 3$.
If $\pi \in \SR^z$, then $\TD(\pi,R\Pi^z) \leq 3$. 
\end{lem}

\def\prooflemtdsimplebregpatatleast3{
\proof
Suppose $\Sigma = \{a_1,a_2,\ldots,a_k\}$, where $k \geq 3$,
and $\pi = x_1a_{i_1}x_2a_{i_2}\ldots a_{i_{n-1}}x_n$, where 
$x_1,x_2,\ldots,x_n \in X$.  If $n = 2$, then one may verify
directly that for any $b \in \Sigma \sm \{a_{i_1}\}$, 
$\{(a_{i_1},+),$ $(ba_{i_1}b,+),(\ve,-)\}$ is a teaching set for
$\pi$ w.r.t.\ $R\Pi^z$.  We assume in what follows that
$n \geq 3$.  Again, $T = \{(w_1,+),(w_2,+),(w_3,-)\}$ 
will denote a teaching set for $\pi$ w.r.t.\ $R\Pi^z$, where
$w_1,w_2$ and $w_3$ are defined below.  Further, $\tau$ will denote a 
regular pattern that is consistent with $T$.

\begin{description}[leftmargin=0cm]
\item[$w_1$:] For every substring of $\pi$ of the shape
$a_{i_j}x_{j+1}a_{i_{j+1}}$, define $\varphi(x_{j+1})$
according to the following case distinction.
\begin{description}
\item[Case i:] $i_j$ and $i_{j+1}$ have opposite parities.
Set $\varphi(x_{j+1}) = \ve$. 
\item[Case ii:] $i_j$ and $i_{j+1}$ have equal parities.
Fix some $j' \in \{1,\ldots,k\}$ such that $j'$ and $i_j$
have opposite parities (which implies that $j'$ and $i_{j+1}$
also have opposite parities), and set $\varphi(x_{j+1}) = a_{j'}$.
For all other variables $x$ occurring in $\pi$, set $\varphi(x) = \ve$. 
\end{description} 
Set $w_1 = \varphi(\pi)$.

\item[$w_2$:] For every substring of $\pi$ of the shape
$a_{i_j}x_{j+1}a_{i_{j+1}}$, define $\psi(x_{j+1})$
according to the following case distinction.
\begin{description}[leftmargin=0cm]
\item[Case i:] $i_j$ and $i_{j+1}$ have equal parities.
Set $\psi(x_{j+1}) = \ve$.
\item[Case ii:] $i_j$ is even and $i_{j+1}$ is odd.
\begin{description}[leftmargin=0cm]
\item[Case ii.1:] 
$j > 1$ and $i_{j-1}$ is even.
Pick any odd $j' \in \{1,\ldots,k\}$ 
such that 
$a_{j'} \neq a_{j+1}$, and set $\psi(x_{j+1}) = a_{j'}$.
\item[Case ii.2:] $j > 1$ and $i_{j-1}$ is odd, or $j = 1$.
Pick any even $j' \in \{1,\ldots,k\}$ 
and pick any odd $j'' \in \{1,\ldots,k\}$ such that 
$a_{j''} \neq a_{i_{j+1}}$, 
and set $\psi(x_{j+1}) = a_{j'}a_{j''}$.  
\end{description}
\item[Case iii:] $i_j$ is odd and $i_{j+1}$ is even.
Pick any odd $j' \in \{1,\ldots,k\}$ such that 
$a_{j'} \neq a_{i_j}$, and set $\psi(x_{j+1}) = a_{j'}$.
\end{description}
Furthermore, pick $j_1,j_2 \in \{1,\ldots,k\}$ such that
$a_{j_1} \notin \{a_{i_1},a_{i_2}\}$ and $a_{j_2} \notin \{a_{i_{n-1}},a_{i_{n-2}}\}$;
set $\psi(x_1) = a_{j_1}$ and $\psi(x_n) = a_{j_2}$.\footnote{Such $j_1$ and $j_2$ must exist since $|\Sigma| \geq 3$.}
For all other variables $x$ occurring in $\pi$, set $\psi(x) = \ve$.
Set $w_2 = \psi(\pi)$.



\item[$w_3$:] Arguing as in the proof of Lemma \ref{lem:tdsimplebregpatbinary},
the consistency of $\tau$ with $(w_1,+)$ and $(w_2,+)$ implies that $\tau$
is of the shape $x_1A_1x_2A_2\ldots A_{k-1}x_k$, where every maximal 
constant block $A_i$ has length at most $2$; furthermore, if $A_i = a_{\ell}a_{\ell'}$,
then $\ell$ and $\ell'$ have opposite parities.

Note that Lemma \ref{lem:simplebrnegative} cannot be directly applied
here since the consistency of $\tau$ with $(w_1,+)$ and $(w_2,+)$
does not imply that $\tau$ is simple block-regular.
We will, however, give a different construction of $w_3$ by analysing
a decomposition of $w_2$ containing subwords $\beta_1,\beta_2,\ldots,\beta_{n-2}$
such that any maximal constant block of $\tau$ is a subword
of some $\beta_j$ (details are to follow).

For each $j \in \{1,\ldots,n-2\}$, define $\beta_j := a_{i_j} \psi(x_{j+1}) a_{i_{j+1}}$.
The positions of $\beta_1,\ldots,\beta_{n-2}$ are illustrated below.

\begin{equation}\label{eqn:w2decomposebeta}
w_2 = \psi(x_1)\overbrace{a_{i_1}\psi(x_2)a_{i_2}}^{\beta_1} \ldots
\overbrace{a_{i_j}\psi(x_{j+1})a_{i_{j+1}}}^{\beta_j} \ldots
\overbrace{a_{i_{n-2}}\psi(x_{n-1})a_{i_{n-1}}}^{\beta_{n-2}}\psi(x_n).
\end{equation} 

Corresponding to each $\beta_j$, where $j \in \{1,\ldots,n-2\}$, we define
a word $\alpha_j$ based on the following case distinction. 

\begin{description}[leftmargin=0cm]
\item[Case i:] $\beta_j = a_{i_j}a_{i_{j+1}}$, where $i_j$ and $i_{j+1}$
have equal parities. 
\begin{description}[leftmargin=0cm]
\item[Case i.1:] $i_j$ and $i_{j+1}$ are even.
\begin{description}[leftmargin=0cm]
\item[Case i.1.1:] $j-1 \geq 1$ and $i_{j-1}$ is odd, $j+2 \leq n-1$
and $i_{j+2}$ is odd.
Then $\psi(x_j) = a_{j'}$ for some odd $j'$ such that $a_{j'} \neq a_{i_{j-1}}$
and $\psi(x_{j+2}) = a_{j''}$ for some odd $j''$ such that $a_{j''} \neq a_{i_{j+2}}$.  
Set
$$
\alpha_j = \left\{\begin{array}{ll}
a_{i_{j+1}}a_{j''}a_{j'}a_{i_j} & \mbox{if $a_{i_j} \neq a_{i_{j+1}}$;} \\
a_{j'}a_{i_j}a_{j''} & \mbox{if $a_{i_j} = a_{i_{j+1}}$.}\end{array}\right. 
$$ 
\item[Case i.1.2:] $j-1 \geq 1$ and $i_{j-1}$ is odd; either 
$j+2 \leq n-1$ and $i_{j+2}$ is even, or $j+2 > n-1$.
Then $\psi(x_j) = a_{j'}$ for some odd $j'$ such that $a_{j'} \neq a_{i_{j-1}}$.  
If $j+2 \leq n-1$ and $i_{j+2}$ is even, define $\alpha_j$ as in Case i.1.1 but 
with all occurrences of $a_{j''}$ deleted.
If $j+2 > n-1$, define $\alpha_j$ as in Case i.1.1 but with all occurrences of
$a_{j''}$ replaced with $\psi(x_n)$ and $\psi(x_n)$ appended to $\alpha_j$.
\item[Case i.1.3:] $j+2 \leq n-1$ and $i_{j+2}$ is odd; either
$j-1 \geq 1$ and $i_{j-1}$ is even, or $j-1 < 1$.  Then
$\psi(x_{j+2}) = a_{j''}$ for some odd $j''$ such that $a_{j''} \neq a_{i_{j+2}}$.
If $j-1 \geq 1$ and $i_{j-1}$ is even, define $\alpha_j$ as in Case i.1.1 but 
with all occurrences of $a_{j'}$ deleted.  If $j-1 < 1$,
define $\alpha_j$ as in Case i.1.1 but with all occurrences of $a_{j'}$
replaced with $\psi(x_1)$ and $\psi(x_1)$ prepended to $\alpha_j$.
\item[Case i.1.4:] $j-1 \geq 1$ and $i_{j-1}$ is even, or $j-1 < 1$;
$j+2 \leq n-1$ and $i_{j+2}$ is even, or $j+2 > n-1$.
If $j-1 \geq 1, j+2 \leq n-1$ and both $i_{j-1},i_{j+2}$ are even, set
$$
\alpha_j = \left\{\begin{array}{ll}
a_{i_{j+1}}a_{i_j} & \mbox{if $a_{i_j} \neq a_{i_{j+1}}$;} \\
a_{i_j} & \mbox{if $a_{i_j} = a_{i_{j+1}}$.}\end{array}\right.
$$
If $j-1 < 1$, set 
$$
\alpha_j = \left\{\begin{array}{ll}
\psi(x_1)a_{i_{j+1}}\psi(x_1)a_{i_j} & \mbox{if $a_{i_j} \neq a_{i_{j+1}}$;} \\
\psi(x_1)a_{i_j} & \mbox{if $a_{i_j} = a_{i_{j+1}}$.}\end{array}\right.
$$
If $j+2 > n-1$, set
$$
\alpha_j = \left\{\begin{array}{ll}
a_{i_{j+1}}\psi(x_n)a_{i_j}\psi(x_n) & \mbox{if $a_{i_j} \neq a_{i_{j+1}}$;} \\
a_{i_j}\psi(x_n) & \mbox{if $a_{i_j} = a_{i_{j+1}}$.}\end{array}\right.
$$
\end{description}
\item[Case i.2:] $i_j$ and $i_{j+1}$ are odd.
If $j-1 \geq 1$ and $j+2 \leq n-1$, set 
$$
\alpha_j = \left\{\begin{array}{ll}
a_{i_{j+1}}a_{i_j} & \mbox{if $a_{i_j} \neq a_{i_{j+1}}$;} \\
a_{i_j} & \mbox{if $a_{i_j} = a_{i_{j+1}}$.}\end{array}\right.
$$
If $j-1 < 1$, set 
$$
\alpha_j = \left\{\begin{array}{ll}
\psi(x_1)a_{i_{j+1}}\psi(x_1)a_{i_j} & \mbox{if $a_{i_j} \neq a_{i_{j+1}}$;} \\
\psi(x_1)a_{i_j} & \mbox{if $a_{i_j} = a_{i_{j+1}}$.}\end{array}\right.
$$
If $j+2 > n-1$, set 
$$
\alpha_j = \left\{\begin{array}{ll}
a_{i_{j+1}}\psi(x_n)a_{i_j}\psi(x_n) & \mbox{if $a_{i_j} \neq a_{i_{j+1}}$;} \\
a_{i_j}\psi(x_n) & \mbox{if $a_{i_j} = a_{i_{j+1}}$.}\end{array}\right.
$$
\end{description}
\item[Case ii:] $i_j$ is odd and $i_{j+1}$ is even.
\begin{description}[leftmargin=0cm]
\item[Case ii.1:] $j+2 \leq n-1$ and $i_{j+2}$ is odd;
$j-1 \geq 1$ and $i_{j-1}$ is even.
Suppose 
$\beta_j = a_{i_j} a_{j_1} a_{i_{j+1}}$ and $\beta_{j+1} = a_{i_{j+1}}a_{j_2} 
a_{j_3} a_{i_{j+2}}$ for some even $j_2$ and odd $j_1$ and $j_3$,
where $a_{j_1} \neq a_{i_j}$ and $a_{j_3} \neq a_{i_{j+2}}$.
Set 
$$
\alpha_j = \left\{\begin{array}{ll} 
a_{j_1} a_{i_{j+1}} a_{j_2} \psi(x_j) a_{j_2} a_{i_j} a_{j_1} & \mbox{if $a_{j_3} = a_{i_j}$;} \\
a_{j_1} a_{i_{j+1}} a_{j_2} \psi(x_j) a_{j_2} a_{j_3} a_{i_j} a_{j_1} & \mbox{if $a_{j_3} \neq a_{i_j}$.}\end{array}\right.
$$

\item[Case ii.2:] $j+2 \leq n-1$ and $i_{j+2}$ is odd; either
$j-1 \geq 1$ and $i_{j-1}$ is odd, or $j-1 < 1$.
If $j-1 \geq 1$ and $i_{j-1}$ is odd, define $\alpha_j$ as in Case ii.1 
(note that $\psi(x_j) = \ve$ in this case).
If $j-1 < 1$, define $\alpha_j$ as in Case ii.1 but with $\psi(x_1)$
prepended to $\alpha_j$.

\item[Case ii.3:] $j-1 \geq 1$ and $i_{j-1}$ is even; either
$j+2 \leq n-1$ and $i_{j+2}$ is even, or $j+2 > n-1$.
Suppose $\beta_j = a_{i_j}a_{j_1}a_{i_{j+1}}$
for some odd $j_1$ such that $a_{j_1} \neq a_{i_j}$.
If $j+2 \leq n-1$ and $i_{j+2}$ is even, set $\alpha_j = 
a_{j_1}a_{i_{j+1}}\psi(x_j)a_{i_j}a_{j_1}$.  
If $j+2 > n-1$, set $\alpha_j = a_{j_1}a_{i_{j+1}}\psi(x_n)
\psi(x_j)\psi(x_n)a_{i_j}a_{j_1}\psi(x_n)$.

\item[Case ii.4:] $j-1 \geq 1$ and $i_{j-1}$ is odd, or $j-1 < 1$;
$j+2 \leq n-1$ and $i_{j+2}$ is even, or $j+2 > n-1$.
Suppose $\beta_j = a_{i_j}a_{j'}a_{i_{j+1}}$, where $j'$ is
odd and $a_{j'} \neq a_{i_j}$.
If $j-1 \geq 1$, $i_{j-1}$ is odd, $j+2 \leq n-1$
and $i_{j+2}$ is even, set $\alpha_j = a_{j'}a_{i_{j+1}}a_{i_j}a_{j'}$.
If $j-1 < 1$, set $\alpha_j = \psi(x_1)a_{j'}a_{i_{j+1}}\psi(x_1)a_{i_j}a_{j'}$.
If $j+2 > n-1$, set $\alpha_j = a_{j'}a_{i_{j+1}}\psi(x_n)a_{i_j}a_{j'}\psi(x_n)$. 
\end{description}
\item[Case iii:] $i_j$ is even and $i_{j+1}$ is odd.
\begin{description}[leftmargin=0cm]
\item[Case iii.1:] $\beta_j = a_{i_j}a_{j_1}a_{j_2}a_{i_{j+1}}$
for some even $j_1$ and odd $j_2$ such that $a_{j_2} \neq a_{i_{j+1}}$.
Set 
$$
\alpha_j = \left\{\begin{array}{ll}
a_{j_2}a_{i_{j+1}}\psi(x_{j+2})\psi(x_j)a_{i_j}a_{j_1}a_{j_2} & \mbox{if $2 \leq j \leq n-3$;} \\
\psi(x_1)a_{j_2}a_{i_{j+1}}\psi(x_{j+2})\psi(x_1)a_{i_j}a_{j_1}a_{j_2} & \mbox{if $j-1 < 1$;} \\
a_{j_2}a_{i_{j+1}}\psi(x_n)\psi(x_j)a_{i_j}a_{j_1}a_{j_2}\psi(x_n) & \mbox{if $j+2 > n-1$.}\end{array}\right. 
$$

\item[Case iii.2:] $\beta_j = a_{i_j}a_{j_2}a_{i_{j+1}}$ for some
odd $j_2$ such that $a_{j_2} \neq a_{i_{j+1}}$ (note that if $j-1 \geq 1$,
then $i_{j-1}$ is even and so $\psi(x_j) = \ve$).
Define $\alpha_j$ as in Case iii.1, but with all occurrences of $a_{j_1}$ deleted.

\end{description}
\end{description}

\medskip
Set $w_3 := \alpha_1\alpha_2\ldots\alpha_{n-2}$.

\end{description}
By construction, $w_1 \in L(\pi)$ and $w_2 \in L(\pi)$.
Furthermore, induction on $j = 1,\ldots,n-2$ shows that
the longest prefix of $x_1a_{i_1}x_2a_{i_2}x_2\ldots a_{i_{n-1}}x_n$
matching $\alpha_1\ldots\alpha_j$ is $x_1a_{i_1}x_2\ldots a_{i_j}x_{j+1}$.
Hence $w_3 \notin L(\pi)$.  The lemma will follow from the next two claims.


\begin{subclaim}\label{clm:taumorphismw2}
Suppose $h,g:(X \cup \Sigma)^* \mapsto \Sigma^*$ are constant-preserving morphisms witnessing
$w_1 \in L(\tau)$ and $w_2 \in L(\tau)$ respectively, and 
suppose $\pi(\ve) = a_{i_1}a_{i_2}\ldots a_{i_{n-1}} \sqsubseteq \tau(\ve)$.
Let $\langle p_1,p_2,\ldots,$ $p_{n-1}\rangle$ be a sequence of positions of $\tau$
such that $\tau[p_j] = a_{i_j}$ for all $j \in \{1,\ldots,n-1\}$.
For each $j \in \{1,\ldots,n-1\}$, let $q_j$ be the position of $w_1$ occupied by
the specific occurrence of $a_{i_j}$ indicated with braces in Equation (\ref{eqn:w1decomposeqi}).
\begin{equation}\label{eqn:w1decomposeqi}
w_1 := \varphi(x_1)\overbrace{a_{i_1}}^{q_1} \varphi(x_2) \overbrace{a_{i_2}}^{q_2} \ldots 
\varphi(x_{i_j}) \overbrace{a_{i_j}}^{q_j} \varphi(x_{i_{j+1}})  \ldots 
\varphi(x_{n-1}) \overbrace{a_{i_{n-1}}}^{q_{n-1}} \varphi(x_n).
\end{equation}     
Similarly, let $R_j$ be the sequence of positions of $w_2$ indicated with braces in 
Equation (\ref{eqn:w1decomposeri}). 
\begin{equation}\label{eqn:w1decomposeri}
w_2 := \rlap{$\underbrace{\phantom{\psi(x_1)a_{i_1}\psi(x_2)}}_{R_1}$} \psi(x_1)a_{i_1} \overbrace{\psi(x_2) a_{i_2} 
\psi(x_3)}^{R_2} \ldots \overbrace{\psi(x_j) a_{i_j} \psi(x_{j+1})}^{R_j} \ldots \rlap{$\overbrace{\phantom{\psi(x_{n-2}) a_{i_{n-2}}
\psi(x_{n-1})}}^{R_{n-2}}$}\psi(x_{n-2}) a_{i_{n-2}} \underbrace{\psi(x_{n-1})a_{i_{n-1}}\psi(x_n)}_{R_{n-1}}.
\end{equation}     
Let $I^{const}_{h,\tau}$ (resp.~$I^{const}_{g,\tau}$) be the mapping of sequences 
of positions of constants in $\tau$ to sequences of positions of $w_1$ 
(resp.~$w_2$) induced by $h$ (resp.~$g$).  
Then for all $j \in \{1,\ldots,n-1\}$, $I^{const}_{h,\tau}(\langle p_j \rangle) 
= \langle q_j \rangle$ and $I^{const}_{g,\tau}(\langle p_j\rangle)$
is a subsequence of $R_j$.

In particular, if $a_{i_1}a_{i_2}\ldots a_{i_{n-1}} \sqsubseteq \tau(\ve)$, then
$L(\tau) = L(\pi)$.
\end{subclaim}

\begin{subclaim}\label{clm:w3match}
Let $\eta$ be any regular pattern such that $\{w_1,w_2\} \subset L(\eta)$ and 
$a_{i_1}a_{i_2}\ldots a_{i_{n-1}} \not\sqsubseteq \eta(\ve)$.
Then $w_3 \in L(\eta)$.
\end{subclaim} 

\noindent\emph{Proof of Claim \ref{clm:taumorphismw2}.} Let $P_1,P_2,\ldots,P_{n-1}$ denote
the sequences of positions of $w_1$ indicated by braces in Equation (\ref{eqn:morphismscopew1}).
\begin{equation}\label{eqn:morphismscopew1}
w_1 := \rlap{$\underbrace{\phantom{\varphi(x_1)a_{i_1}\varphi(x_2)}}_{P_1}$} \varphi(x_1)a_{i_1} \overbrace{\varphi(x_2) a_{i_2} \varphi(x_3)}^{P_2} \ldots 
\overbrace{\varphi(x_j) a_{i_j} \varphi(x_{j+1})}^{P_j} \ldots \rlap{$\overbrace{\phantom{\varphi(x_{n-2}) a_{i_{n-2}}
\varphi(x_{n-1})}}^{P_{n-2}}$}\varphi(x_{n-2}) a_{i_{n-2}} \underbrace{\varphi(x_{n-1})a_{i_{n-1}}\varphi(x_n)}_{P_{n-1}}.
\end{equation}
It suffices to show that whenever $j \in \{1,\ldots,n-1\}$, $I^{const}_{h,\tau}(\langle p_j\rangle)$ 
is a subsequence of $P_j$; the claim that $I^{const}_{h,\tau}(\langle p_j \rangle) = \langle q_j\rangle$ 
will then follow from the fact that $\varphi(x_j) \notin N(x_j,\pi)$ for all $j \in \{1,\ldots,n-1\}$.  
So assume, by way of contradiction, that there were a least $\ell \in \{1,\ldots,n-1\}$ such that 
$I^{const}_{h,\tau}(\langle p_{\ell} \rangle)$ is not a subsequence
of $P_{\ell}$.  First, suppose that $I^{const}_{h,\tau}(\langle p_{\ell} \rangle)$
were a subsequence of some $P_{\ell'}$ with $\ell' < \ell$.  Then,
since $\varphi(x_{\ell}) \notin \{a_{i_{\ell}},a_{i_{\ell-1}}\}$ and
$\varphi(x_{\ell-1}) \neq a_{i_{\ell-1}}$, $I^{const}_{h,\tau}(\langle p_{\ell-1} \rangle)$
is not a subsequence of $P_{\ell-1}$.  Iterating the preceding argument
then gives that for all $j \leq \ell$, $I^{const}_{h,\tau}(\langle p_j \rangle)$
is not a subsequence of $P_j$, a contradiction. 
A similar argument holds if $I^{const}_{h,\tau}(\langle p_{\ell} \rangle)$ were
a subsequence of some $P_{\ell''}$ with $\ell'' > \ell$.

The proof that $I^{const}_{g,\tau}(\langle p_j\rangle)$ is a subsequence of $R_j$ 
is similar (making crucial use of the definition of $\psi$).  
This establishes the first part of the claim.

Now we establish the second part of the claim.  Note that from the first 
part of the claim, if $i_j$ is odd, then $I^{const}_{g,\tau}(\langle p_j\rangle)$
cannot be a subsequence of the sequence of positions of $w_2$
corresponding to $\psi(x_j)$ (resp.~$\psi(x_{j+1})$).
If $i_j$ is even, then $I^{const}_{g,\tau}(\langle p_j\rangle)$ cannot 
be a subsequence of the sequence of positions of $w_2$ corresponding
to $\psi(x_j)$.  Furthermore, suppose $I^{const}_{g,\tau}(\langle p_j\rangle)$
were a subsequence of the sequence of positions of $w_2$ corresponding
to $\psi(x_{j+1})$; then if $j+1 \leq n-1$, $i_{j+1}$ must be odd and
therefore $I^{const}_{g,\tau}(\langle p_{j+1}\rangle)$ equals $\langle q' \rangle$, 
where $q'$ is the position of $w_2$ occupied by $a_{i_{j+1}}$ in
$R_{j+1}$.  

From the fact that $\{w_1,w_2\} \subset L(\tau)$, we know that
$\tau$ must start as well as end with variables.
For any $\alpha \in (X \cup \Sigma)^*$, let $o(\alpha)$ denote
the number of substrings of $\alpha$ of the shape $b_1xb_2$,
where $x \in X \cup \{\ve\}$, $b_1,b_2 \in \Sigma$ and
$b_1,b_2$ have opposite parities.  Note that $o(w_2) = o(\pi)$. 
Since $I^{const}_{h,\tau}(\langle p_j\rangle) = \langle q_j \rangle$ 
whenever $j \in \{1,\ldots,n-1\}$, it follows that if 
$a_{i_1}\ldots a_{i_{n-1}} \sqsubset \tau(\ve)$, then there is some
position $p'$ of $\tau$ such that for some $j \in \{1,\ldots,n-2\}$,
$p_j < p' < p_{j+1}$ and $\tau[p'] = \varphi(x_{i_{j+1}}) \in \Sigma$.
By the definition of $\varphi$, if $\varphi(x_{i_{j+1}}) = a_{j'}$, then
$j'$ has parity opposite to that of $i_j$ as well as $i_{j+1}$.  Thus   
$o(\tau) > o(\pi)$.  But $w_2 \in L(\tau)$ implies $o(\tau) \leq o(\pi)$,
and therefore $\tau(\ve) = a_{i_1}\ldots a_{i_{n-1}}$.
The fact that $w_1 \in L(\tau)$ (resp.~$w_2 \in L(\tau)$) implies 
that a variable occurs in $\tau$ between every pair $a_{i_j},a_{i_{j+1}}$ such
that $i_j$ and $i_{j+1}$ have equal (resp.~opposite) parities.   
Thus $L(\tau) = L(\pi)$.\qed~(Claim \ref{clm:taumorphismw2})

\medskip
\noindent\emph{Proof of Claim \ref{clm:w3match}.}
Our strategy to show $w_3 \in L(\eta)$ is as follows.
First, fix some constant-preserving morphism $g:(X \cup \Sigma)^* \mapsto \Sigma^*$
such that $g(\eta) = w_2$.  Then $g$ induces a mapping
$\mathcal{I}_{g,\eta}$ of closed intervals of $\{1,\ldots,|\eta|\}$ to
closed intervals of $\{1,\ldots,w_2\}$ such that for all
$[p_1,p_2] \subseteq \{1,\ldots,|\eta|\}$, $g(\eta[p_1]\ldots
\eta[p_2]) = w_2[\mathcal{I}_{g,\eta}([p_1,p_2])]$.  One may take the ``inverse''
$\overline{\cI}_{g,\eta}$ of $\mathcal{I}_{g,\eta}$, where, for all $[q_1,q_2] \subseteq \{1,\ldots,|w_2|\}$,
$\overline{\cI}_{g,\eta}([q_1,q_2]) = [s_1,s_2]$ for some $s_1,s_2 \in \{1,\ldots,|\eta|\}$
such that $[q_1,q_2]$ is a subinterval of $\mathcal{I}_{g,\eta}([s_1,s_2])$
and for all proper subintervals $R$ of $[s_1,s_2]$, $[q_1,q_2]$ is not a subinterval of
$\mathcal{I}_{g,\eta}(R)$. 
Let $r_1,\ldots,r_{n-1}$ be the positions of $a_{i_1},\ldots,a_{i_{n-1}}$
respectively in $w_2$ marked with braces in Equation (\ref{eqn:w2decomposeqi}).
\begin{equation}\label{eqn:w2decomposeqi}
w_2 := \psi(x_1) \overbrace{a_{i_1}}^{r_1} \psi(x_2) \overbrace{a_{i_2}}^{r_2} \psi(x_3) 
\ldots \psi(x_j) \overbrace{a_{i_j}}^{r_j} \psi(x_{j+1}) \ldots \psi(x_{n-2}) 
\overbrace{a_{i_{n-2}}}^{r_{n-2}} \psi(x_{n-1}) \overbrace{a_{i_{n-1}}}^{r_{n-1}}\psi(x_n).
\end{equation}     
By our assumption on $\eta$, there is a least $\ell \in \{1,\ldots,n-1\}$
such that $\eta[\cJ_{g,\eta}([r_{\ell},r_{\ell}])]$ is a variable and
if there is a least $r' > r_{\ell}$ such that $\eta[\cJ_{g,\eta}([r',r'])]$ is a constant,
then $\eta[\cJ_{g,\eta}([r',r'])] \neq a_{i_{\ell}}$.
As argued at the beginning of the construction of $w_3$,
$\eta$ starts and ends with variables, and every maximal constant 
block $A$ of $\eta$ has length at most $2$; furthermore, if the 
length of $A$ is exactly $2$, then $A = a_{j_1}a_{j_2}$
for some $j_1,j_2 \in \{1,\ldots,k\}$ such that $j_1$ and $j_2$ have
opposite parities.  

We define a set $\cC$ 
consisting of all possible intervals of positions of $w_2$ of length at most $2$
such that for every maximal constant block of $\eta$, say $\eta[J]$ for some
closed interval $J \subseteq \{1,\ldots,|\eta|\}$, there is an $I \in \cC$
for which $\cI_{g,\eta}(J) \subseteq I$.

First, suppose $i_{\ell}$ is even and the first letter of $\psi(x_{\ell+1})$ 
equals $a_{i_{\ell}}$.  Then $\cC$ consists of all intervals of 
positions of $w_2$ of the form
\begin{enumerate}[label=\roman*]
\item $[q,q+1]$, where $q < r_{\ell}-1$ and $w_2[q]w_2[q+1] = a_{j_1}a_{j_2}$ 
for some $j_1,j_2 \in \{1,\ldots,k\}$ with opposite parities, or 
\item $[q,q+1]$, where $q > r_{\ell}+1$ 
and $w_2[q]w_2[q+1] = a_{j_1}a_{j_2}$ for some $j_1,j_2 \in \{1,\ldots,k\}$ 
with opposite parities, or 
\item $[q,q]$, where $q < r_{\ell}-1$ and if $q \geq 2$, then $w_2[q-1]w_2[q]w_2[q+1] = 
b a_{j_3}a_{j_4}$ for some $b \in \Sigma$ 
and $j_3,j_4 \in \{1,\ldots,k\}$ such that $j_3$ and $j_4$ have equal parities,
and 
$b = a_{j_5}$ for some $j_5 \in \{1,\ldots,k\}$ such that 
$j_5$ and $j_3$ have equal parities; if $q < 2$, then the same holds with 
$w_2[q-1]$ and $b$ replaced with $\ve$, or
\item $[q,q]$, where $q = r_{\ell}-1$ and if $q \geq 2$, then $w_2[q-1]w_2[q] = ba_{j_6}$ 
for some $j_6 \in \{1,\ldots,k\}$ and $b \in \Sigma$ such that 
if $b = a_{j_7}$ for some $j_7 \in \{1,\ldots,k\}$,
then $j_7$ and $j_6$ have equal parities; if $q < 2$, then the same holds with
$w_2[q-1]$ and $b$ replaced with $\ve$, or 
\item $[q,q]$ for some $q > r_{\ell}+2$ such that if $q+1 \leq |w_2|$, then $w_2[q-1]w_2[q]w_2[q+1] = 
a_{j_8}a_{j_9}b$ for some $j_8,j_9 \in \{1,\ldots,k\}$
with equal parities and $b \in \Sigma$ such that if $b = a_{j_{10}}$ for some
$j_{10} \in \{1,\ldots,k\}$, then $j_{10}$ and $j_9$ have equal parities;
if $q+1 > |w_2|$, then the same holds with $w_2[q+1]$ and $b$ replaced with $\ve$,
or 
\item $[q,q]$, where $q = r_{\ell}+2$ and if $q+1 \leq |w_2$, then $w_2[q]w_2[q+1] = a_{j_{11}}b$
for some $j_{11} \in \{1,\ldots,k\}$ and $b \in \Sigma$
such that if $b = a_{j_{12}}$ for some $j_{12} \in \{1,\ldots,k\}$, then
$j_{12}$ and $j_{11}$ have equal parities; if $q+1 > |w_2|$, then
the same holds with $w_2[q+1]$ and $b$ replaced with $\ve$.    
\end{enumerate}  
Second, suppose $i_{\ell}$ is odd or the first letter of 
$\psi(x_{\ell+1})$ is not equal to $a_{i_{\ell}}$.
Then we define $\cC$ exactly as above but with three differences:
first, $q > \ell+1$ is replaced with $q > \ell$ in (ii); second, $q > \ell+2$ is 
replaced with $q > \ell+1$ in (v); third, $q = \ell+2$ is replaced
with $q = \ell+1$ in (vi).
We next define a one-one mapping $F$ from $\cC$ to the set of all intervals
of positions of $w_3$ satisfying the following conditions for all $[q,q],[q,q+1] \in \cC$:
\begin{itemize}
\item $F([q,q+1]) = [q',q'+1]$ for some $q' \in \{1,\ldots,|w_3|-1\}$ with 
$w_2[q]w_2[q+1] = w_3[q']w_3[q'+1]$.
\item $F([q,q]) = [q',q']$ for some $q' \in \{1,\ldots,|w_3|\}$ with
$w_2[q] = w_3[q']$.
\item Suppose $q_1$ and $q_2$ are the left endpoints of $I_1$ and
$I_2$ respectively, where $I_1,I_2 \in \cC$, $I_1 \neq I_2$ and $q_1 < q_2$ (note that
no two distinct members of $\cC$ intersect).  Let $q'_1$ and $q'_2$
be the left endpoints of $F(I_1)$ and $F(I_2)$ respectively.  
Then $q'_1 < q'_2$ and $F([I_1]) \cap F([I_2]) = \emptyset$.  
\end{itemize} 
Note that the existence of an $F$ satisfying the above three conditions implies
that for any sequence $\langle I_1,I_2,\ldots,I_m \rangle$ of intervals of positions
of $w_2$ such that every $I_i$ corresponds to a maximal constant block
of $\eta$ and for all $i,j \in \{1,\ldots,m\}$ with $i < j$,
$I_i \cap I_j = \emptyset$, and the left endpoint of $I_i$ is smaller than that
of $I_j$, there is a corresponding sequence $\langle I'_1,I'_2,\ldots,I'_m\rangle$
of intervals of positions of $w_3$ such that for all $i,j \in \{1,\ldots,m\}$
with $i < j$, $w_2(I_i) = w_3(I'_i)$, $I'_i \cap I'_j = \emptyset$,
and the left endpoint of $I_{i'}$ is smaller than that of $I_{j'}$. 
Thus, since $\eta$ starts as well as ends with variables, the existence of
such an $F$ will suffice to show that $w_3 \in L(\eta)$.  We consider
a case distinction based on the earlier definition of $\cC$. 
Let $Q_1,\ldots,Q_{n-2}$ be the closed intervals of positions of $w_3$
corresponding to the occurrences of $\alpha_1,\ldots,\alpha_{n-2}$ respectively
as shown in Equation (\ref{eqn:w3decomposealpha}).
\begin{equation}\label{eqn:w3decomposealpha}
w_3 := \overbrace{\alpha_1}^{Q_1} \ldots \overbrace{\alpha_j}^{Q_j} \ldots \overbrace{\alpha_{n-2}}^{Q_{n-2}}.
\end{equation} 
Consider any $I \in \cC$.
\begin{description}[leftmargin=0cm]
\item[Case 1:] $I = [r_j,r_j+1]$ for some $j < \ell$, where, if $w_2[r_j,r_j+1] = a_{j'}a_{j''}$
for some $j',j'' \in \{1,\ldots,k\}$, then $j'$ and $j''$ have opposite parities.
Note that if $i_j$ were odd, then by Cases i and iii in the construction of $w_2$,
$w_2[r_j+1] = a_{j'}$ would imply that $j'$ is odd, which is impossible
by Conditions i and ii in the definition of $\cC$.  Hence $i_j$ is even.  
Furthermore, an inspection of Cases i and ii in the construction
of $w_2$ shows that $r_j+1 \neq r_{j+1}$, and therefore $r_j+1$ is the position of 
the first letter of $\psi(x_{j+1})$ in $w_2$; moreover, $i_{j+1}$ is odd.
Suppose $\beta_j = a_{i_j}a_{j_1}a_{i_{j+1}}$
for some odd $j_1$ such that $a_{j_1} \neq a_{i_{j+1}}$ (the positions
of $\beta_1,\ldots,\beta_{n-2}$ are illustrated in Equation (\ref{eqn:w2decomposebeta})).
From Case iii.2 in the construction of $w_3$, one sees that
$\alpha_j = \ga a_{i_j}a_{j_1}$ for some $\ga \in \Sigma^*$;
fix $\ga$.
Set $F(I) = \left[\sum_{1\leq l < j}|\alpha_l|+|\ga|+1,
\sum_{1\leq l < j}|\alpha_l|+|\ga|+2\right]$. 

\item[Case 2:] $I = [r_j-1,r_j]$ for some $j < \ell$.
First, suppose $j-1 \geq 1$.  Then an argument similar to that in Case 1.1 shows that
$i_j$ must be even and $i_{j-1}$ must be odd.
From Cases i.1.1 and iii in the construction of $w_3$,
one sees that 
$\alpha_j = \ga_1 w_2[r_j-1] a_{i_{j}} \ga_2$
for some $\ga_1,\ga_2 \in \Sigma^*$; fix such $\ga_1$ and $\ga_2$.
Set $F(I) = \left[\sum_{1 \leq l < j}|\alpha_l|+|\ga_1|+1,\sum_{1 \leq l < j}|\alpha_l|+|\ga_1|+2\right]$.        

Second, suppose $j-1 < 1$.  From Cases i.1.3, i.1.4, i.2, ii.4 and iii
in the construction of $w_3$, we deduce that
there are $\ga_1,\ga_2 \in \Sigma^*$ such that
$\alpha_1 = \ga_1 \psi(x_1) a_{i_1} \ga_2$; fix such $\ga_1$ and $\ga_2$.
Set $F(I) = \left[|\ga_1|+1,|\ga_1|+2\right]$. 

\item[Case 3:] $I = [r_j+1,r_j+2]$ for some $j < \ell$
such that $\psi(x_{j+1}) = w_2[r_j+1]w_2[r_j+2]$.
Based on the case distinction in the construction of $w_2$,
one sees that $i_j$ must be even and if $j+2 \leq n-1$,
then $a_{i_{j+1}}$ must be odd.
From Case iii.1, we deduce that $\alpha_j = \ga \psi(x_{j+1})$
for some $\ga \in \Sigma^*$.  Set
$F(I) = \left[\sum_{1 \leq l < j}|\alpha_l|+|\ga|+1,\sum_{1 \leq l < j}
|\alpha_l|+|\ga|+2\right]$.  

\item[Case 4:] $I = [r_j-1,r_j]$ for some $j > \ell$.
Arguing as in the earlier cases, $i_j$ must be even
and $i_{j-1}$ must be odd.  By examining Case ii
in the construction of $w_3$, one sees that
$\alpha_{j-1} = w_2[r_j-1]a_{i_j}\ga$ for some
$\ga \in \Sigma^*$.  Set $F(I) = \left[\sum_{1 \leq l < j-1}
|\alpha_l|+1,\sum_{1 \leq l < j-1}|\alpha_l|+2\right]$.

\item[Case 5:] $I = [r_j,r_j+1]$ for some $j > \ell$.
First, suppose $j+1 \leq n-1$.  Arguing as before, 
$i_j$ and $i_{j-1}$ must be even while $i_{j+1}$ must be odd.
It follows from Case i.1 in the construction of $w_3$
that $\alpha_{j-1} = \ga_1 a_{i_j}w_2[r_j+1] \ga_2$ for some
$\ga_1,\ga_2 \in \Sigma^*$; fix such $\ga_1$ and $\ga_2$.
Set $F(I) = \left[\sum_{1 \leq l < j-1}|\alpha_l|+|\ga_1|+1,
\sum_{1 \leq l < j-1}|\alpha_l|+|\ga_1|+2 \right]$.

Second, suppose $j+1 > n-1$, i.e. $j = n-1$.  It follows from Cases
i.1.2 and i.1.4 in the construction of $w_3$ that for some $\ga_1,\ga_2 \in \Sigma^*$, 
$\alpha_{n-2} = \ga_1 a_{i_{n-1}} \psi(x_n) \ga_2$; fix
such $\ga_1$ and $\ga_2$.  Set $F(I) = \left[\sum_{1 \leq l 
< j-1}|\alpha_l|+|\ga_1|+1,\right.$ $\left.\sum_{1 \leq l < j-1}\right.$ $\left.|\alpha_l|+|\ga_1|+2 
\right]$.   

\item[Case 6:] $I = [r_j+1,r_j+2]$ for some $j > \ell$ such
that $\psi(x_{j+1}) = w_2[r_j+1]w_2[r_j+2]$.
Based on the case distinction in the construction of $w_2$,
we deduce that $i_{j-1}$ and $i_{j+1}$ are odd while 
$i_j$ is even.  It follows from Case ii in the construction of
$w_3$ that $\alpha_{j-1} = \ga_1 a_{i_j} \ga_2 w_2[r_j+1]
w_2[r_j+2] \ga_3$ for some $\ga_1,\ga_2,\ga_3 \in \Sigma^*$;
fix such $\ga_1,\ga_2$ and $\ga_3$.  Set
$F(I) = \left[\sum_{1 \leq l < j-1}|\alpha_l|+|\ga_1|+|\ga_2|+2,
\sum_{1 \leq l < j-1}\right.$ $\left.|\alpha_l|+|\ga_1|+|\ga_2|+3 \right]$.  

\item[Case 7:] $I = [r_j,r_j]$ for some $j < \ell$.
\begin{description} 
\item[Case 7.1:] $i_j$ is even.
First, suppose $j-1 \geq 1$.
Then both $i_{j-1}$ and $i_{j+1}$ must be even.
From Case i.1 in the construction of $w_3$, we deduce
that there exist $\ga_1,\ga_2 \in \Sigma^*$ such that 
$\alpha_j = \ga_1 a_{i_j} \ga_2$ and $|\ga_2| \leq 1$;
fix such $\ga_1$ and $\ga_2$.
Set $F(I) = \left[\sum_{1 \leq l < j}|\alpha_l|+|\ga_1|+1,
\sum_{1 \leq l < j}|\alpha_l|+|\ga_1|+1\right]$. 

Second, suppose $j-1 < 1$.  It follows from Cases i.1
and iii in the construction of $w_3$ that there exist
$\ga_1,\ga_2 \in \Sigma^*$ such that $\alpha_1 = \ga_1 a_{i_1} \ga_2$
and $|\ga_2| \leq 2$; fix such $\ga_1$ and $\ga_2$.
Set $F(I) = \left[|\ga_1|+1,|\ga_1|+1\right]$. 

\item[Case 7.2:] $i_j$ is odd. 
It follows from Cases i.2 and ii in the construction of $w_3$ that
there exist $\ga_1,\ga_2 \in \Sigma^*$ such that
$\alpha_j = \ga_1 a_{i_j} \ga_2$ and $|\ga_2| \leq 1$;
fix such $\ga_1$ and $\ga_2$.  Set $F(I) = \left[\sum_{1 \leq l < j}
|\alpha_l|+|\ga_1|+1,\right.$ $\left.\sum_{1 \leq l < j}|\alpha_l|+|\ga_1|+1 \right]$. 
\end{description}

\item[Case 8:] $I = [r_j,r_j]$ for some $j > \ell$.
From the case distinction in the construction of $w_3$, we deduce
that there exist $\ga_1,\ga_2 \in \Sigma^*$ with $|\ga_1| \leq 1$
such that $\alpha_{j-1} = \ga_1 a_{i_j} \ga_2$; fix such $\ga_1$
and $\ga_2$.  Set $F(I) = \left[\sum_{1 \leq l < j-1}|\alpha_l|+|\ga_1|+1,
\sum_{1 \leq l < j-1}|\alpha_l|+|\ga_1|+1\right]$.

\item[Case 9:] $I = [1,1]$.
Observe from 
the construction
of $w_3$ that $\alpha_1$ starts with $\psi(x_1)$.  Set $F(I) = [1,1]$. 

\item[Case 10:] $I = [|w_2|,|w_2|]$. 
Observe from 
the construction
of $w_3$ that $\alpha_{n-2}$ ends with $\psi(x_n)$.
Set $F(I) = [|w_3|,|w_3|]$.

\end{description}

This completes the definition of $F$. 
By Claim \ref{clm:w3match}, since $\{w_1,w_2\} \subset L(\tau)$ and
$w_3 \notin L(\tau)$, one has that $a_{i_1}a_{i_2}\ldots a_{i_{n-1}} \sqsubseteq \tau(\ve)$.
Thus by Claim \ref{clm:taumorphismw2}, $L(\tau) = L(\pi)$.
Therefore $T = \{(w_1,+),(w_2,+),(w_3,-)\}$ is indeed a teaching set
for $\pi$ w.r.t.\ $R\Pi^z$.~\qed
}

\medskip
\noindent
The next result determines upper (for $|\Sigma| \in \{1,\infty\}$) and lower (for $|\Sigma| \in  \natnum 
\cup \{\infty\}$) bounds 
for the 
\TD\ of any given simple block-regular 
pattern w.r.t.\ the whole class of patterns.  
It turns out that these bounds vary with the alphabet size.

\begin{thm}\label{thm:tdsimplebregclassofpatterns}
Suppose $z \in \natnum \cup \{\infty\}$ and $\pi = x_1 c_1 x_2 \ldots c_{n-1} x_n$
for some $c_1,\ldots,$ $c_{n-1} \in \Sigma$ and $n \geq 2$.
(i) If $z \in \{1,\infty\}$, then $\TD(\pi,\Pi^z) \in \{1,3\}$.
(ii) If $2 \leq z < \infty$, then $\TD(\pi,\Pi^z) = \Omega(|\pi|)$.
\end{thm} 

\def\proofthmtdsimplebregclassofpatterns{
\proof
Note that for any $0 \in \Sigma$, $x_1$ has the teaching set $\{(\ve,+),(0,+)\}$.
Now suppose $\pi$ contains at least one constant symbol.

\noindent\emph{Assertion (i).} 
If $\Sigma = \{0\}$, then there is some $m \geq 1$ such that $\pi$ is equivalent 
to the pattern $0^mx_1$, and so $\pi$ may be taught with the examples $(0^m,+),
(0^{m+1},+)$ and $(0^{m-1},-)$. 

If $|\Sigma| = \infty$, then one can choose distinct constants $a_1,a_2,\ldots,a_n
\in \Sigma \sm \{c_1,\ldots,c_{n-1}\}$.
Any pattern $\tau$ consistent with the examples $(\pi(\ve),+)$ and
$(\pi[x_1 \ra a_1, x_2 \ra a_2,\ldots, x_n \ra a_n],+)$ must be simple block-regular
and satisfy $\tau(\ve) \sqsubseteq \pi(\ve)$.  By Lemma \ref{lem:simplebrnegative},
the example $(\wh{\pi(\ve)},-)$ will ensure, in addition, 
that $\tau(\ve) \not\sqsubset \pi(\ve)$.

Finally, note that any simple block-regular pattern not equivalent to $x_1$
must be taught using at least $3$ examples (for a similar proof, 
see \cite[Theorem 12.1]{bayeh17}. 

\noindent\emph{Assertion (ii).} 
First, suppose $\Sigma = \{a_1,a_2,\ldots,a_{\ell}\}$ for some $\ell \geq 3$.
We show that any teaching set for $\pi$ w.r.t.\ $\Pi^{\ell}$ must contain at least
$\left\lfloor \displaystyle\frac{n}{\ell} \right\rfloor$ positive examples.
Assume 
that some teaching set $T$ for $\pi$ w.r.t.\ $\Pi^{\ell}$ contains $k$ positive examples $(w_1,+),\ldots,
(w_k,+)$ for some $k \geq 1$. 
For each $i \in \{1,\ldots,k\}$, fix a substitution $h_i: X \mapsto \Sigma^*$ such 
that $h_i(\pi) = w_i$.  Let $\{z^i_j:i,j \in \natnum\}$ be a subset of $X$ such that 
$z^i_j \neq z^{i'}_{j'}$ whenever $(i,j) \neq (i',j')$.
For each $i \in \{1,\ldots,k\}$, let $g_i:\Sigma^* \mapsto X^*$
be a morphism such that $g_i(a_j) = z^i_j$ for all $j \in \{1,\ldots,\ell\}$.
Let $\pi'$ be the pattern derived from $\pi$ by replacing each 
$x \in \Var(\pi)$ with the string $g_1(h_1(x))g_2(h_2(x))\ldots g_k(h_k(x))$;
$\pi'$ can be written in the form $A_1 c_1 A_2 \ldots c_{n-1} A_n$,
where $A_1,A_2,\ldots,A_n \in \{z^i_j: 1 \leq i \leq k \wedge 1 \leq j \leq \ell\}^*$.
By construction, $w_i \in L(\pi')$ for all $i \in \{1,\ldots,k\}$.
In particular, note that if $\pi'_i$ is the restriction of $\pi'$ to 
$\{z^i_1,\ldots,z^i_l\} \cup \Sigma$, then $w_i \in L(\pi'_i)$. 
Furthermore, since $\pi'$ is similar to $\pi$,
one has $L(\pi') \subseteq L(\pi)$
and so $\pi'$ is consistent with $T$.  As $T$ is a teaching set for
$\pi$ w.r.t.\ $\Pi^{\ell}$, $L(\pi') = L(\pi)$ and therefore every $A_i$ 
contains at least one free variable.  Hence 
\begin{equation}\label{eqn:setfreevarpiprimepi}
\left|\{x:\mbox{$x$ is a free variable of $\pi'$}\}\right|
\geq |\Var(\pi)| = n.
\end{equation}  
On the other hand, since $\left|\{x:\mbox{$x$ is a free variable of $\pi'$}\}\right| \subseteq
\{z^i_j: 1 \leq i \leq k \wedge 1 \leq j \leq \ell\}$,
we have
\begin{equation}\label{eqn:setfreevarpiprimezk}
\left|\{x:\mbox{$x$ is a free variable of $\pi'$}\}\right| 
\leq \ell k.
\end{equation} 
It now follows from Equations (\ref{eqn:setfreevarpiprimepi}) and (\ref{eqn:setfreevarpiprimezk})
that $n \leq \ell k$, and therefore $k \geq \left\lfloor \displaystyle\frac{n}{\ell} \right\rfloor$,
as required.   

The proof for binary alphabets is similar.  Suppose $\Sigma = \{0,1\}$.
Define an operation $\mathcal{O}$ on any $\tau \in R\Pi^2$ as follows: 
pick the first occurrence of a substring of $\tau$ of the shape $x \delta x' \overline{\delta} x''$, 
where $x \in X$ and $\delta \in \Sigma$ and delete $x'$.  If no such
substring occurs in $\tau$, set $\mathcal{O}(\tau) = \tau$.
Then for all $\tau \in R\Pi^2$, one has $\mathcal{O}(\tau) = \tau'$
for some $\tau' \in R\Pi^2$ with $L(\tau') = L(\tau)$ \cite[Lemma 2]{Nessel05}. 
 
We iteratively apply $\mathcal{O}$ to $\pi$ until no new regular pattern is 
produced; that is to say, we find the least $k$ such that $\mathcal{O}^{k+1}(\pi) = 
\mathcal{O}^k(\pi)$.  Setting $\tau' = \mathcal{O}^k(\pi)$, notice that for
all $\eta \in \Pi^2$ with $\eta$ similar to $\tau'$ and $L(\eta) = L(\tau')$,
every maximal variable block of $\eta$ must contain a free variable.
To see this, let $\eta = A_1 c_1 \ldots c_{n-1} A_n$ and
$\tau' = x_1 c_1 \ldots c_{n-1} x_n$ (after normalisation of $\tau'$), 
where $A_1,\ldots,A_n \in X^*$ and $c_1,\ldots,c_{n-1} \in \{0,1,01,10\}$.
Choose some $\delta_1 \in \Sigma$ that differs from the first symbol of
$c_1$, and set $w_1 = \tau'[x_1 \ra \delta_1]$.  Since $L(\eta) = L(\tau')$,
we have $w_1 \in \tau'$ and therefore $A_1$ must contain a free variable.
A similar argument shows that $A_n$ contains a free variable.
Now consider any $i \in \{2,\ldots,n-1\}$.  If $\N(x_i,\tau') = \{\delta\}$
for some $\delta \in \Sigma$, then setting $w_i = \tau'[x_i \ra \overline{\delta}]$
gives $w_i \in L(\tau') = L(\eta)$ and so $A_i$ must contain a free
variable.  If $\N(x_i,\tau') = \{0,1\}$, then at least one of $c_{i-1}$ and $c_i$,
say $c_{i-1}$, equals $\delta\overline{\delta}$ for some $\delta \in \Sigma$.
Pick $\delta' \in \Sigma$ that differs from the first symbol of $c_i$
(if $c_i = \delta\overline{\delta}$ instead, let $\delta' \in \Sigma$ be a
letter that differs from the last symbol of $c_{i-1}$). 
Setting $w_i = \tau'[x_i \ra \delta']$ then gives $w_i \in L(\eta)$, and
so $A_i$ contains a free variable.

The proof for the case $|\Sigma| \geq 3$ may now be applied to $\tau'$.
Note that $|\Var(\tau')| \geq \left\lfloor\displaystyle\frac{2n}{3}\right\rfloor$,
and so the earlier proof gives that every teaching set for $\tau'$ w.r.t.\
$\Pi^2$ must contain at least $\left\lfloor\displaystyle\frac{n}{3}\right\rfloor$ 
positive examples.~\qed
}

\medskip
\noindent
We do not know whether the lower bound given in Assertion (ii) of Theorem 
\ref{thm:tdsimplebregclassofpatterns} is also an upper bound (up to numerical
constant factors).  In the proof of \cite[Proposition 4]{bayeh17}, 
it was shown that the \TD\ of every simple block-regular pattern $\pi$ is $O(2^{|\pi|})$.

\section{Finite Distinguishability of $m$-Quasi-Regular, Non-Cross $m$-Regular and $m$-Regular Patterns}

This section studies the problem of determining the subclass
of finitely distinguishable patterns w.r.t.\ three classes: the $m$-quasi-regular patterns,
the non-cross $m$-regular patterns, and the $m$-regular patterns.  The first two classes are 
interesting from an algorithmic learning perspective as they provide natural examples
of pattern language families that are learnable in the limit\footnote{Roughly speaking, a class of languages
is learnable in the limit if there is a learning algorithm such that, given any infinite sequence of all positive 
examples for any language $L$ in the class, 
the algorithm outputs a corresponding sequence of guesses for the target language (based on a representation 
system for the languages in the class) that converges to a fixed representation for $L$;
this model is due to Gold \cite{Gol67}.} \cite{Mit98,Reidenbach06}.  
The $m$-regular patterns are a fairly natural generalisation
of the $m$-quasi-regular patterns; as will be seen later, the class of constant-free
$4$-regular patterns is not identifiable in the limit over binary alphabets, and
in particular, not all $m$-regular patterns are finitely distinguishable over
binary alphabets.

\begin{nota}\label{nota:finiterepetitions}
Fix any $\ell \geq 0$ and $z,m \geq 1$.  
An \emph{$\ell$-variable} pattern is one that has at most $\ell$ distinct variables.  
Let $\Pi^z_{\ell,m}$ denote the class of $\ell$-variable patterns $\pi$ such that every
variable occurs at most $m$ times in $\pi$; if $\ell = \infty$, 
then there is no uniform upper bound on the number of distinct variables occurring 
in any $\pi \in \Pi^z_{\ell,m}$; if $m = \infty$, then there is no uniform upper
bound on the number of times any variable can occur. 
We call every $\pi \in \Pi^z_{\infty,m}$ an \emph{$m$-regular pattern}.
$\Pi^z_{\infty,m,cf}$ denotes the class of all constant-free $m$-regular patterns. 

Let $\QR^z_{\ell,m}$ denote the class of all $\ell$-variable patterns $\pi$ such
that every variable of $\pi$ occurs exactly $m$ times; again, if $\ell = \infty$, then there
is no uniform upper bound on the number of distinct variables occurring in
any $\pi \in \QR^z_{\ell,m}$.  Every $\pi \in \QR^z_{\infty,m}$ is known as an 
$m$-quasi-regular pattern \cite{Mit98}.  We denote the class of constant-free $m$-quasi-regular patterns 
by $\QR^z_{\infty,m,cf}$.


\end{nota}

\medskip
\noindent  
Mitchell \cite{Mit98} showed that for any $m \geq 1$, the class of $m$-quasi-regular
pattern languages is learnable in the limit.  The next theorem shows that for all
$z \geq 1$, every $m$-quasi-regular pattern even has a finite teaching
set w.r.t.\ $\QR^z_{\infty,m}$.  Thus, at least as far as $m$-quasi-regular
patterns are concerned, version space learning with a helpful teacher is just as powerful 
as learning in the limit.  We begin with a lemma, which states that for any given
$m$-quasi-regular pattern $\pi$ and 
every $m$-quasi-regular pattern $\tau$ with $L(\tau) \not\subseteq L(\pi)$,
there is some $S \subseteq \Var(\tau)$ of size at most linear in $|\Var(\pi)|$ 
for which $L\left(\tau{\big|}_{\Sigma\cup S}\right) \not\subseteq L(\pi)$;
for any $S' \subseteq X \cup \Sigma$, $\tau{\big|}_{S'}$ is the subsequence of $\tau$
obtained by deleting symbols not in $S'$. 

\begin{lem}\label{lem:mquasiregnotsubsetbound}
Fix $\Sigma$ with $z = |\Sigma| \geq 2$ and $\{0,1\} \subseteq \Sigma$.
Suppose $m \geq 1$ and $\pi,\tau \in \QR^z_{\infty,m}$.
If $\tau(\ve) = \pi(\ve)$ and $L(\tau) \not\subseteq L(\pi)$,
then there is some $S \subseteq \Var(\tau)$ with $|S| \leq 1 + \left(|\pi(\ve)| + m +4\right) \cdot |\Var(\pi)|$
such that $L\left(\tau{\big|}_{\Sigma\cup S}\right) \not\subseteq L(\pi)$.
\end{lem}

\def\prooflemmquasiregnotsubsetbound{
\proof
Given that $L(\tau) \not\subseteq L(\pi)$ and $\tau(\ve) = \pi(\ve)$,
both $\tau$ and $\pi$ contain at least one variable, and so there is
some $S \subseteq \Var(\tau)$ of minimum possible size such that
$L\left(\tau{\big|}_{\Sigma \cup S}\right) \not\subseteq L(\pi)$.  Fix such an $S$.
By the choice of $S$, one has $L\left(\tau{\big|}_{\Sigma \cup (S \sm \{y'\})}\right) \subseteq
L(\pi)$ for all $y' \in S$.  Fix any $y \in S$, and set $S' := S \sm \{y\}$.
Without loss of generality, assume $S' = \{x_1,\ldots,x_{\ell}\}$ ($S'$ may
also be empty).  As noted earlier, $L\left(\tau{\big|}_{\Sigma \cup S'}\right) \subseteq L(\pi)$.

Now suppose, by way of contradiction, that $|S| > 1 + \left(|\pi(\ve)| + m +4\right) \cdot |\Var(\pi)|$.
Let $\varphi: X \mapsto \Sigma^*$ be the substitution defined by $\varphi(x_i) =
0 1^{2i\cdot|\tau|}  0$ for all $i \in \{1,\ldots,\ell\}$ and $\varphi(z) = \ve$
for all $z \in X \sm \{x_1,\ldots,x_{\ell}\}$.  Set $w := \varphi(\tau)$.
We first establish the following claim.
\begin{subclaim}\label{clm:noofoccurrencesofvarphixi}
For all $i \in \{1,\ldots,\ell\}$, $w$ contains exactly $m$ occurrences of
$\varphi(x_i) = 0 1^{2i\cdot|\tau|} 0$.  Furthermore, all $m$ occurrences
of $\varphi(x_i)$ are disjoint.
\end{subclaim}

\noindent\emph{Proof of Claim \ref{clm:noofoccurrencesofvarphixi}.}
Fix any $i \in \{1,\ldots,\ell\}$.  Since $\tau \in \QR^z_{\infty,m}$, there are at least 
$m$ occurrences of $\varphi(x_i)$ in $w$.  We show that there
cannot be any occurrence of $\varphi(x_i)$ that overlaps with (i) a
constant part of $\tau$, or (ii) an occurrence of $\varphi(x_j)$ for some $j \in 
\{1,\ldots,\ell\}$ such that $\varphi(x_j)$ and $\varphi(x_i)$ occupy different
intervals of positions of $w$. 

Assume otherwise.  Consider any $j \in \{1,\ldots,\ell\}$.  Since
$\varphi(x_j)$ starts and ends with $0$, the occurrences of $\varphi(x_j)$
and $\varphi(x_i)$ coincide or $\varphi(x_j)$ overlaps with $\varphi(x_i)$
only at the first or last position of $\varphi(x_i)$.

First, suppose an occurrence of $\varphi(x_i)$ overlaps with a constant part
of $\tau$.  Since $|01^{2i\cdot|\tau|}0| > |\tau|$, this occurrence of $\varphi(x_i)$
must overlap with an occurrence of $\varphi(x_j)$ that is generated by a variable of 
$\tau$ for some $j \in \{1,\ldots,\ell\}$. 
By the observation in the preceding paragraph, since the occurrences of $\varphi(x_j)$
and $\varphi(x_i)$ must be different, $\varphi(x_j)$ can overlap with $\varphi(x_i)$
only at the first or last position of $\varphi(x_i)$.  It follows that
each of the $2i\cdot|\tau|$ occurrences of $1$ in $\varphi(x_i)$ must overlap
with a constant part of $\tau$, which is impossible as $2i\cdot|\tau| > |\tau|$. 

Second, suppose an occurrence of $\varphi(x_i)$ overlaps with an occurrence of
$\varphi(x_j)$ for some $j \in \{1,\ldots,\ell\}$ such that $\varphi(x_j)$ and
$\varphi(x_i)$ occupy different intervals of positions of $w$.
An argument similar to that in the preceding paragraph shows that 
each of the $2i\cdot|\tau|$ occurrences of $1$ in $\varphi(x_i)$ must
overlap with a constant part of $\tau$, which is impossible.~\qed~(Claim \ref{clm:noofoccurrencesofvarphixi})

\medskip
\noindent Let the variable part of $\tau{\big|}_{\Sigma \cup S'}$ (i.e.\ $\tau{\big|}_{S'}$) be $x_{i_1}\ldots x_{i_{m\ell}}$ (since
$\tau{\big|}_{\Sigma \cup S'}$ has $\ell$ distinct variables, it has $m\ell$ variable
occurrences).
Set $c = |\pi(\ve)|$, and write $w$ as
\begin{equation}\label{eqn:wvarphitaudecompose}
w := \underbrace{\gamma_1}_{J_1}\underbrace{\varphi(x_{i_1})}_{H_1} \underbrace{\gamma_2}_{J_2}
\underbrace{\varphi(x_{i_2})}_{H_2} \ldots \underbrace{\gamma_j}_{J_j} \underbrace{\varphi(x_{i_j})}_{H_j} \underbrace{\gamma_{j+1}}_{J_{j+1}} \ldots
\underbrace{\varphi(x_{i_{m\ell}})}_{H_{m\ell}} \underbrace{\gamma_{m\ell+1}}_{J_{m\ell+1}},
\end{equation}
where $\gamma_1,\ldots,\gamma_{m\ell+1} \in \Sigma^*$, $\tau(\ve) = \gamma_1\gamma_2\ldots\gamma_{m\ell+1}$
and $J_1,H_1,\ldots,J_{m\ell},H_{m\ell},J_{m\ell+1}$ are the intervals of positions of $w$
corresponding to the subwords marked in Equation (\ref{eqn:wvarphitaudecompose}).
Since $L\left(\tau{\big|}_{\Sigma \cup S'}\right) \subseteq L(\pi)$, there is a morphism 
$\theta: X^* \mapsto X^*$ such that $\theta(\pi) = w$.

We claim that for all $j \in \{1,\ldots,m\ell-m-c-3\}$,
$\cI_{\theta,\pi}$ maps the positions of at least two variable occurrences 
of $\pi$ to intervals of positions of $w$ that overlap with the interval
corresponding to
$$
\gamma_j \varphi(x_{i_j}) \gamma_{j+1} \ldots \gamma_{j+c+m+3}\varphi(x_{i_{j+c+m+3}}). 
$$
Formally, this means there are at least two positions of $\pi$ occupied by variables,
say $p_1$ and $p_2$, such that
\begin{equation}\label{eqn:atleast2varscover}
\cI_{\theta,\pi}(p_k) \cap \left(J_j \cup H_j \cup \ldots \cup J_{j+c+m+3} \cup H_{j+c+m+3}\right) \neq \emptyset
\end{equation}
for $k \in \{1,2\}$.
Suppose the latter statement does not hold.  For all $i \in \{1,\ldots,\ell\}$, since $|\varphi(x_i)| 
> |\tau| \geq |S| + c > m\cdot|\Var(\pi)|+c = |\pi|$, no constant part of $\pi$
can cover $\varphi(x_i)$, and so there must be some 
$q \in \{1,\ldots,|\pi|\}$ such that $\pi[q]$ is a variable and $\cI_{\theta,\pi}(q)$
covers $J_{j+1} \cup H_{j+1} \cup \ldots \cup H_{j+c+m+2} \cup J_{j+c+m+3}$, 
i.e.\ 
\begin{equation}\label{eqn:cithetapiqcovers1}
J_{j+1} \cup H_{j+1} \cup \ldots \cup H_{j+c+m+2} \cup J_{j+c+m+3} \subseteq \cI_{\theta,\pi}(q)
\end{equation}
Since every variable of $\pi$ occurs exactly
$m$ times, there must be at least $m$ occurrences of 
$$
w' := \gamma_{j+1}\varphi(x_{i_{j+1}}) \ldots \varphi(x_{i_{j+c+m+2}})\gamma_{j+c+m+3}
$$
in $w$.  According to Claim \ref{clm:noofoccurrencesofvarphixi}, $\varphi(x_i)$ occurs
exactly $m$ times in $w$ for all $i \in \{1,\ldots,\ell\}$, and all its $m$ occurrences
are disjoint.  Thus for all distinct $j_1,j_2 \in \{j+1,\ldots,j+c+m+2\}$, $i_{j_1} \neq 
i_{j_2}$.  Furthermore, since there are at most $c$ indices $i$ with $\gamma_i \neq \ve$, 
$w'$ contains at least $c+m+1 - c = m+1$ subwords of the shape $\varphi(x_{i_{j_1}})\varphi(x_{i_{j_2}})$, 
where $j_1 \neq j_2$ and $j_1,j_2 \in \{1,\ldots,\ell\}$.  This means there are at least $m+1$ pairs
$(j_1,j_2)$ with $j_1 \neq j_2$ and $j_1,j_2 \in \{1,\ldots,\ell\}$ such that
$\tau{\big|}_{\Sigma \cup S'}$ contains exactly $m$ occurrences of the substring $x_{j_1}x_{j_2}$.
Since $y$ occurs exactly $m$ times in $\tau{\big|}_{\Sigma \cup S}$ (we recall that $S = S' \cup \{y\}$),
there is at least one pair $(k_1,k_2)$ with $k_1 \neq k_2$ and $k_1,k_2 \in \{1,\ldots,\ell\}$
such that $x_{k_1}x_{k_2}$ occurs exactly $m$ times in $\tau{\big|}_{\Sigma \cup S}$.
But by Theorem \ref{thm:jiang94suffcondsubset}, $\tau{\big|}_{\Sigma \cup S}$ would then be equivalent to 
$\tau{\big|}_{\Sigma \cup (S \sm \{x_{k_2}\})}$,
contradicting the minimality of $|S|$.  Thus there are indeed at least $2$ positions
of variables in $\pi$, say $p_1$ and $p_2$, such that (\ref{eqn:atleast2varscover}) holds. 

Arguing inductively, it follows that the number of variable occurrences of $\pi$ (including
variable repetitions) is at least $\displaystyle\frac{m\ell}{c+m+4}$.  Consequently,
\begin{equation*}
\begin{aligned}
|\Var(\pi)| \geq \displaystyle\frac{1}{m} \cdot \displaystyle\frac{m\ell}{c+m+4} 
= \displaystyle\frac{\ell}{c+m+4},
\end{aligned}  
\end{equation*}
and so 
\begin{equation*}
\begin{aligned}
|S| = |S'| + 1 
= \ell + 1 
\leq 1 + (c+m+4)\cdot|\Var(\pi)|,
\end{aligned}
\end{equation*}
as desired.~\qed
}

\begin{thm}\label{thm:cfqrplfinitetd}
If $z = 1$, then $\TD(\QR^z_{\infty,m}) = 3$.  If
$z \geq 2$, then for every $\pi \in \QR^z_{\infty,m}$, $\TD(\pi,\QR^z_{\infty,m}) 
= O(2^{|\pi(\ve)|}+D\cdot(|\pi(\ve)|+D\cdot m)^{D\cdot m})$,
where $D := \max(\{(1/m)\cdot(2\cdot|\pi|-|\pi(\ve)|),1+(|\pi(\ve)|+m+4)\cdot|\Var(\pi)|\})$. 
\end{thm} 

\def\proofthmcfqrplfinitetd{
\proof
We first consider the case $z = 1$.  Suppose $\Sigma = \{0\}$.  Every language generated by a pattern in 
$\QR^1_{\infty,m}$ is equivalent to a pattern of the shape $0^k x^m$ or $0^{k'}$, where $k \in \natnum_0$
and $k' \in \natnum$.
Let $\pi := 0^k x^m$.  If $k \geq m$, then $\pi$ can be taught using the sample $\{(0^k,+),(0^{k+m},+)(0^{k-m},-)\}$:
the two examples $(0^k,+)$ and $(0^{k-m},-)$ uniquely identify the constant part of $\pi$, while
$(0^{k+m},+)$ distinguishes $\pi$ from the constant pattern $0^k$.
If $k < m$, then $\{(0^k,+),(0^{k+m},+)\}$ is a teaching set for $\pi$: since $k < m$, $(0^k,+)$ already
uniquely identifies the constant part of $\pi$, while as before $(0^{k+m},+)$ ensures that $\pi$ is not a constant
pattern. 
Let $\pi' := 0^{k'}$.  Then $\{(0^{k'},+),(0^{k'+m},-)\}$ is a teaching set for $\pi'$: the constant part
of any pattern $\tau$ consistent with $(0^{k'},+)$ is equal to $0^{k''}$ for some $k'' \leq k'$; if $L(\tau) \neq L(\pi)$,
then $\tau$ contains a variable $x$ such that for some $i \geq 1$ with $k''+mi = k'$, $0^{k'}$ is obtained from $\tau$ by 
substituting $0^{i}$ for $x$.  Replacing $x$ with $0^{i+1}$ yields $0^{k'+m} \in L(\tau)$,
and so $\tau$ is inconsistent with $(0^{k'+m},-)$.       
In any one of the above cases, one has $\TD(\pi,\QR^1_{\infty,m}) \leq 3$.
Furthermore, suppose $\eta := 0^m x^m$.  Any teaching set for $\eta$ must contain
at least one positive and one negative example since $L(0^m) \subset L(\eta)$ and
$L(\eta) \subset L(x^m)$; an additional positive example is needed to distinguish $\eta$
from all constant patterns.  Hence $\TD(\eta,\QR^1_{\infty,m}) \geq 3$.   

Now suppose $z \geq 2$.  Fix any $\pi \in \QR^z_{k,m}$.  We build a teaching set $T$ for $\pi$ w.r.t.\ $\QR^z_{\infty,m}$.  
Let $\eta$ denote any pattern in $\QR^z_{\infty,m}$ that is consistent with $T$.  
First, put $(\pi(\ve),+)$ into $T$.  Next, for every $w \sqsubset \pi(\ve)$,
put $(w,-)$ into $T$.  The $O(2^{|\pi(\ve)|})$ examples added to $T$ up to the present stage ensure that 
$\eta(\ve) = \pi(\ve)$.    
By \cite{Mit98}, there is a finite tell-tale set for $\pi$ w.r.t.\ $\QR^z_{\infty,m}$, 
that is, a finite set $S \subseteq L(\pi)$ such that for all $\tau \in \QR^z_{\infty,m}$, 
one has $S \subseteq L(\tau) \subseteq L(\pi)\Rightarrow L(\tau) = L(\pi)$; furthermore,
\cite[Lemma 9]{Mit98} implies that this set $S$ has size $O(\lceil D_1 \cdot (|\pi(\ve)|+D_1\cdot m)^{D_1\cdot m}\rceil)$, 
where $D_1 := (1/m)\cdot(2|\pi|-|\pi(\ve)|)$.  
Put $\{(w',+): w' \in S\}$ into $T$.  The examples in $T$ now ensure that 
$\eta(\ve) = \pi(\ve)$ and $L(\eta) \not\subset L(\pi)$. 
Thus if $L(\eta) \neq L(\pi)$, then $L(\eta) \not\subseteq L(\pi)$.  
Next, for each $\tau \in \QR^z_{1 + \left(|\pi(\ve)| + m +4\right) \cdot \left|\Var(\pi)\right|,m}$ such that 
$L(\tau) \not\subseteq L(\pi)$ and $\tau(\ve) = \pi(\ve)$, pick some $v_{\tau} \in L(\tau) \sm L(\pi)$ and put $(v_{\tau},-)$ into 
$T$; note that there are 
$O(D_2 \cdot (|\pi(\ve)|+D_2\cdot m)^{D_2\cdot m})$ many such $\tau$ (up to equivalence), where
$D_2 := 1+(|\pi(\ve)|+m+4)\cdot |\Var(\pi)|$.  As was observed earlier, 
if $L(\eta) \neq L(\pi)$, then $L(\eta) \not\subseteq L(\pi)$, and so by Lemma \ref{lem:mquasiregnotsubsetbound}, 
$\eta(\ve) = \pi(\ve)$ implies there is some $\tau' \in \QR^z_{1 + \left(|\pi(\ve)| + m +4\right) \cdot 
\left|\Var(\pi)\right|,m}$ with $L(\tau') 
\subseteq L(\eta)$ and $L(\tau') \not\subseteq L(\pi)$; the negative example $(v_{\tau'},-)$ would therefore 
ensure that $\eta$ is inconsistent with $T$.  At this stage, $T$ has altogether
$O(2^{|\pi(\ve)|}+D\cdot(|\pi(\ve)|+D\cdot m)^{D\cdot m})$ examples,
where $D := \max(\{(1/m)\cdot(2\cdot|\pi|-|\pi(\ve)|),1+(|\pi(\ve)|+m+4)\cdot|\Var(\pi)|\})$.~\qed 
}

\medskip
\noindent
Next, we show that the \PBTD\ of the class of constant-free $m$-quasi-regular pattern languages
is exactly $1$ for large enough alphabet sizes.  We establish this value by observing that
if the adjacency graph of a constant-free $m$-quasi-regular pattern $\pi$ 
\cite[Chapter 3]{Lothaire02} has a 
colouring satisfying certain conditions, where each colour corresponds to a letter in the alphabet,
then such a colouring can be used to construct a positive example for $\pi$ that
distinguishes it from all shorter constant-free $m$-quasi-regular patterns.  

\begin{thm}\label{thm:pbtdmquasiregularcf}
For any $z \geq 1$, $\TD(\QR^z_{\infty,1,cf}) = \PBTD(\QR^z_{\infty,1,cf}) = 0$.
Suppose $m \geq 2$.  If $z = |\Sigma| \geq 4m^2+1$, then $\PBTD(\QR^z_{\infty,m,cf}) = 1$. 
\end{thm}

\def\proofthmpbtdmquasiregularcf{
\noindent\emph{Proof of Theorem \ref{thm:pbtdmquasiregularcf}.}
If $m = 1$, then $\QR^z_{\infty,m,cf}$ contains only the pattern $x$ and so
$\TD(\QR^z_{\infty,1,cf}) = \PBTD(\QR^z_{\infty,1,cf})$ $= 0$.  Suppose $m \geq 2$.
Given $\pi,\tau \in \QR^z_{\infty,m,cf}$ that are succinct, define $\pi \prec \tau$
iff $|\tau| < |\pi|$.  For any succinct pattern $\pi \in \QR^z_{\infty,m,cf}$ with
$\Var(\pi) = \{x_1,\ldots,x_n\}$, define the \emph{adjacency graph of $\pi$}, denoted
$\AG(\pi)$, to be the bipartite graph whose vertex set comprises two copies of $\Var(\pi)$,
one denoted $\Var(\pi)^L := \{x_1^L,\ldots,x_n^L\}$ and the other denoted
$\Var(\pi)^R := \{x_1^R,\ldots,x_n^R\}$, such that an edge connects
$x_i^L$ and $x_j^R$ iff $x_ix_j$ is a substring of $\pi$ \cite[Chapter 3]{Lothaire02}.
We find the least $k$ such that some $k$-colouring $c:\Var(\pi)^L \cup \Var(\pi)^R
\mapsto \{1,\ldots,k\}$ of $\AG(\pi)$ satisfies the following conditions.
\begin{enumerate}
\item For all $i \in \{1,\ldots,n\}$, $c(x_i^L) = c(x_i^R)$.
\item For any distinct $j_1,j_2 \in \{1,\ldots,n\}$, if $(x_i^L,x_{j_1}^R) \in E(\AG(\pi))$ 
and $(x_i^L,x_{j_2}^R) \in E(\AG(\pi))$
(resp.~$(x_{j_1}^L,x_i^R) \in E(\AG(\pi))$ and $(x_{j_2}^L,x_i^R) \in E(\AG(\pi))$), 
then $c(x_{j_1}^R) \neq c(x_{j_2}^R)$
(resp.~$c(x_{j_1}^L) \neq c(x_{j_2}^L)$).
\end{enumerate}
We show that $k \leq 4m^2+1$.  Let $G$ be the graph obtained from $\AG(\pi)$
by contracting the pair $(x_i^L,x_i^R)$ of vertices for all $i \in \{1,\ldots,n\}$
(i.e.\ the vertices $x_i^L$ and $x_i^R$ are replaced with a single vertex $x_i$ such that $x_i$ is
adjacent to any vertex to which $x_i^L$ and $x_i^R$ were originally adjacent) and
deleting all loops.
Choose the minimum $k'$ such that some
colouring $c':V(G) \mapsto \{1,\ldots,k'\}$ is a $2$-distance colouring of $G$.
Let $c'': \Var(\pi)^L \cup \Var(\pi)^R \mapsto \{1,\ldots,k'\}$ be the colouring
of $\AG(\pi)$ defined by $c''(x_i^L) = c''(x_i^R) = c'(x_i)$ for all
$i \in \{1,\ldots,n\}$.
Note that for any distinct $j_1,j_2
\in \{1,\ldots,n\}$, $(x_i^L,x_{j_1}^R) \in E(\AG(\pi))$ and $(x_i^L,x_{j_2}^R) \in E(\AG(\pi))$
(resp.~$(x_{j_1}^L,x_i^R) \in E(\AG(\pi))$ and $(x_{j_2}^L,x_i^R) \in E(\AG(\pi))$) together imply 
that $d_G(x_{j_1},x_{j_2}) \leq d_{\AG(\pi)}(x_{j_1}^R,x_{j_2}^R) \leq 2$ (resp.~$d_G(x_{j_1},x_{j_2})
\leq d_{\AG(\pi)}(x_{j_1}^L,x_{j_2}^L) \leq 2$); hence $c''$ satisfies Conditions 1 and 2 with $c''$ in place
of $c$, and therefore $k \leq k'$.  Furthermore, $\Delta(G)$ is equal to the maximum, over all 
$i \in \{1,\ldots,n\}$,
of the number of substrings of $\pi$ of the shape $x_jx_i$ or $x_ix_{j'}$ (where 
$j \neq i$ and $j' \neq i$); this is bounded above by $2m$ because every variable of 
$\pi$ occurs exactly $m$ times.  Thus by Lemma \ref{lem:graphcolor2distance},
$k \leq k' \leq 4m^2+1$.

Fix distinct letters $a_1,\ldots,a_k \in \Sigma$ and any strictly increasing sequence
$2 < p_1 < \ldots < p_n$ of positive integers.  
For each $i \in \{1,\ldots,n\}$, fix some $\xi_i \in \{1,\ldots,k\}$ such that $\xi_i \neq 
c(x_i)$. 
Let $\varphi: X \mapsto \Sigma^*$ be the substitution 
defined by $\varphi(x_i) = a_{c(x_i)} a_{\xi_i}^{p_i} a_{c(x_i)}$ for all 
$i \in \{1,\ldots,n\}$ and $\varphi(x') = \varepsilon$ for all $x' \in X \sm \Var(\pi)$.
Set $w := \varphi(\pi)$.  Thus if $\pi = x_{l_1} x_{l_2} \ldots x_{l_{n'}}$,
\begin{equation}\label{eqn:quasiregularpositiveeg}
w := \underbrace{a_{c(x_{l_1})} a_{\xi_{l_1}}^{p_{l_1}} a_{c(x_{l_1})}}_{I_1} \ldots
\underbrace{a_{c(x_{l_i})} a_{\xi_{l_i}}^{p_{l_i}} a_{c(x_{l_i})}}_{I_i} \ldots 
\underbrace{a_{c(x_{l_{n'}})} a_{\xi_{l_{n'}}}^{p_{l_{n'}}} a_{c(x_{l_{n'}})}}_{I_{n'}}.
\end{equation} 
Let $\tau$ be any succinct pattern in $\QR^z_{\infty,m,cf}$ such that
$w \in L(\tau)$ and $\tau \not\prec \pi$.  It will be argued that
$L(\tau) = L(\pi)$.  Suppose $\psi: X^* \mapsto \Sigma^*$ is a morphism witnessing
$w \in L(\tau)$.  Let $I_1,\ldots,I_{n'}$ be the closed intervals corresponding
to the positions of the subwords of $w$ marked with braces in (\ref{eqn:quasiregularpositiveeg}).
We show that for each $j \in \{1,\ldots,n'\}$, there is some $j' \in \{1,\ldots,|\tau|\}$
such that $I_j \subseteq \cI_{\psi,\tau}(j')$, i.e.\ there is a single position
of $\tau$ that is mapped under $\psi$ to a subword of $w$ covering 
$a_{c(x_{l_j})} a_{\xi_{l_j}}^{p_{l_j}} a_{c(x_{l_j})}$. 
Assume otherwise; let $i_0 \in \{1,\ldots,n'\}$ be the least integer for
which the latter statement is false.
It follows that $I_{i_0}$ contains a cut-point of $w$ relative to $(\psi,\tau)$.
Further, one observes that $\cI_{\psi,\tau}$ cannot map any single position
of $\tau$ to a proper superset of $I_i$ for any given $i \in \{1,\ldots,n'\}$:

\begin{subclaim}\label{clm:substringrestrictpsix}
Fix any $x \in \Var(\tau)$.  For all $i \in \{1,\ldots,n\}$ and
$j \in \{1,\ldots,k\}$, neither $a_j a_{c(x_i)} a_{\xi_i}^{p_i}
a_{c(x_i)}$ nor $a_{c(x_i)} a_{\xi_i}^{p_i}
a_{c(x_i)} a_j$ is a subword of $\psi(x)$.
\end{subclaim}

\noindent\emph{Proof of Claim \ref{clm:substringrestrictpsix}.}
Suppose, by way of contradiction, that $a_j a_{c(x_i)} a_{\xi_i}^{p_i}
a_{c(x_i)}$ were a subword of $\psi(x)$ for some $x \in \Var(\tau)$.  
Since $a_{c(x_i)} \neq a_{\xi_i}$ (by the choice of $\xi_i$), 
$p_i \geq 3$ and $p_{i'} \neq p_{j'}$ for all distinct $i',j' \in \{1,\ldots,n\}$, 
there are exactly $m$ (non-overlapping) occurrences of the word 
$a_{c(x_i)} a_{\xi_i}^{p_i} a_{c(x_i)}$ in $w$.  Suppose
these occurrences are represented by the intervals $I_{j_1},\ldots,I_{j_m}$
of positions of $w$, where $j_1 < \ldots < j_m$.  Hence if $x$ occupies positions 
$q_1,\ldots,q_m$ of $\tau$, where $q_1 < \ldots < q_m$, then $I_{j_{\ell}}
\subset \cI_{\psi,\tau}(q_{\ell})$ for all $\ell \in \{1,\ldots,m\}$.
As $a_j$ occupies the position just before the leftmost point of $I_{j_{\ell}}$ in 
$w$ for all $\ell \in \{1,\ldots,m\}$, $x$ cannot be the first symbol of $\tau$.  
Thus there is some $x_{j'} \in \Var(\tau)$
with $j' \neq i$ such that $j = c(x_{j'})$, which means that
$a_{c(x_{j'})} a_{c(x_i)} a_{\xi_i}^{p_i} a_{c(x_i)}$
occurs exactly $m$ times in $w$.  Now there cannot be exactly
$m$ occurrences of the substring $x_{j'}x$ in $\tau$; otherwise, the subpattern
obtained from $\tau$ by deleting all occurrences of $x_{j'}$ would be equivalent
to $\tau$, contradicting the succinctness of $\tau$.
Therefore there must be some $x_{j''} \in \Var(\tau)$ (possibly equal to $x$) with
$j'' \neq j'$ such that $x_{j''}x$ is a substring of $\tau$,
and so $a_{c(x_{j''})} a_{c(x_i)} a_{\xi_i}^{p_i}
a_{c(x_i)}$ must be a subword of $w$.  However, by the choice of $c$ -- in particular,
Condition 2, $c(x_{j'}) \neq c(x_{j''})$ and thus $a_{c(x_{j'})} a_{c(x_i)} a_{\xi_i}^{p_i}
a_{c(x_i)}$ cannot occur exactly $m$ times in $w$, a contradiction.
An analogous proof shows that $a_{c(x_i)} a_{\xi_i}^{p_i}
a_{c(x_i)} a_j$ cannot be a subword of $\psi(x)$ for any given
$j \in \{1,\ldots,k\}$.~\qed~(Claim \ref{clm:substringrestrictpsix})

By Claim \ref{clm:substringrestrictpsix} and the choice of $i_0$ (which
implies, in particular, that $I_{i_0}$ contains a cut-point), 
$\left|\overline{\cI}_{\psi,\tau}\left(\bigcup_{\ell \leq i_0} I_{\ell}\right)\right| \geq i_0+1$.     
By applying Claim \ref{clm:substringrestrictpsix} successively to
$w(I_{i_0}),w(I_{i_0+1}),\ldots,w(I_{n'})$, it follows that for $j = i_0+1,i_0+2,\ldots,n'$,
$\left|\overline{\cI}_{\psi,\tau}\left(\bigcup_{\ell \leq j} I_{\ell}\right)\right| \geq j+1$
and so $|\tau| \geq n'+1$, implying that $\tau \prec \pi$, contrary
to assumption.

Consequently, for each $j \in \{1,\ldots,n'\}$, there is some $j' \in \{1,\ldots,|\tau|\}$
such that $I_j \subseteq \cI_{\psi,\tau}(j')$; by Claim \ref{clm:substringrestrictpsix},
one also has $\cI_{\psi,\tau}(j') \subseteq I_j$.  Thus, since
$a_{c(x_{l_i})} a_{\ell_{l_i}}^{p_{l_i}} a_{c(x_{l_i})}$ occurs
exactly $m$ times in $w$ for all $i \in \{1,\ldots,n'\}$ and
the subword of $w$ corresponding to the interval $I_{i'}$ is different from that 
corresponding to $I_{i''}$ whenever $l_{i'} \neq l_{i''}$, one has (after normalising $\tau$ and $\pi$)
$\pi \sqsubseteq \tau$.  As $|\tau| \leq |\pi|$, it follows that
$L(\tau) = L(\pi)$, as required.~\qed
}

\medskip
\noindent
While the \PBTD\ of the class of $m$-quasi-regular patterns remains open in full generality, we observe
that over unary alphabets, the \PBTD\ of this class is exactly $2$ for any $m \geq 1$. 

\begin{prop}\label{prop:pbtdqrunary}
For any $m \geq 1$, $\PBTD(\QR^1_{\infty,m}) = 2$.
If $z \geq 2$, then $\PBTD(\QR^z_{\infty,m}) \geq 2$.
\end{prop}    

\def\proofproppbtdqrunary{
\proof
We first note that over a unary alphabet $\Sigma = \{0\}$, any pattern of the shape
$0^k x_1^m \ldots x_n^m$, where $k \geq 0$ and $n \geq 1$, is equivalent to $0^k x^m$.  
Given any patterns $\pi$ and $\pi'$ of the shape $0^k x^m$ or $0^k$, define
$\pi \prec \pi'$ iff 
\begin{enumerate}
\item $\pi'$ is a constant pattern and $\pi$ contains at least one variable, or
\item both $\pi$ and $\pi'$ are non-constant patterns and $|\pi(\ve)| < |\pi'(\ve)|$. 
\end{enumerate}  
For any constant pattern $\pi$, a teaching set for $\pi$ w.r.t.\ $(\QR^1_{\infty,m},\prec)$
is $\{(\pi,+)\}$: $\pi$ is preferred to all non-constant patterns while any constant pattern
different from $\pi$ cannot be consistent with $(\pi,+)$.
For any pattern $\tau := 0^k x^m$, where $k \geq 0$, a teaching set for $\tau$
w.r.t.\ $(\QR^1_{\infty,m},\prec)$ is $\{(0^k,+),(0^{k+m},+)\}$:
no constant pattern can be consistent with this sample; furthermore, since the constant
part of any pattern consistent with this sample has length at most $|\tau|$, it follows
from Condition 2 above that $\tau$ is preferred to all $\tau'$ such that $\tau'$ is consistent with
the sample and $L(\tau') \neq L(\tau)$.

To see that $\PBTD(\QR^z_{\infty,m}) \geq 2$ for all $z \geq 1$, one may apply \cite[Theorem 34]{GRSZ2016}; 
according to this theorem, $\PBTD(\QR^1_{\infty,m}) > 1$ because $\QR^1_{\infty,m}$ contains all constant patterns 
as well as infinitely many patterns that generate infinite languages.~\qed
}

\medskip
\noindent
A \emph{non-cross pattern} $\pi$ is a constant-free pattern of the shape
$x_0^{n_0}x_1^{n_1}\ldots x_k^{n_k}$, where $n_0,n_1,\ldots,n_k$ $\in \natnum$.
Let $\NC\Pi^z_{\infty,m}$ denote the class of all non-cross patterns $\pi$ over any $\Sigma$
with $|\Sigma| = z$ such that every variable of $\pi$ occurs at most $m$ times.
$\NC\Pi^z_{\infty,\infty}$ coincides with $\NC\Pi^z$, the class of
all non-cross patterns.
The next main result shows that for any fixed $m$, the \TD\ of every pattern in $\NC\Pi^z_{\infty,m}$ 
is not only finite, but also has a uniform upper bound depending only on $m$.
Slightly more interestingly, the teaching complexity of $\NC\Pi^z_{\infty,m}$ in the preference-based
teaching model varies with the alphabet size when $m \geq 2$: over unary alphabets, the $\PBTD$ of this class is
exactly linear in $m$, while over alphabets of size at least $2$, the $\PBTD$ is exactly $1$. 
In the following lemma, we observe certain properties of an ``unambiguous'' word that
was constructed in \cite[Lemma 13]{Reidenbach06}.

\begin{lem}\label{lem:noncrossstructure}(Based on \cite[Lemma 13]{Reidenbach06})
Suppose $\{0,1\} \subseteq \Sigma$.  Fix any $m \geq 2$, and let $\pi = x_0^{n_0}\ldots x_k^{n_k}$, 
where $n_0,\ldots,n_k \in \{2,\ldots,m\}$.  Suppose there are positive numbers $\ell$ and $i_1,\ldots,i_{\ell}$ such that
\begin{equation}\label{eqn:widef}
w := \underbrace{(01)^{i_1}}_{I_1} \underbrace{(001)^{i_2}}_{I_2} \ldots \underbrace{(0^j1)^{i_j}}_{I_j} 
\ldots \underbrace{(0^{\ell-1}1)^{i_{\ell-1}}}_{I_{\ell-1}} \underbrace{(0^{\ell}1)^{i_{\ell}}}_{I_{\ell}} \in L(\pi),
\end{equation}
where, for each $j \in \{1,\ldots,\ell\}$, $I_j$ is the closed interval of positions of $w$ occupied
by the subword $(0^j1)^{i_j}$ as indicated with braces in Equation (\ref{eqn:widef}).
For each $j \in \{0,\ldots,k\}$, let $J_j$ denote the closed interval of positions of $\pi$ occupied by $x_j^{n_j}$.
Let $h$ be any substitution such that $h(\pi) = w$ and $h(x_i) \neq \ve$ for all $i \in \{0,\ldots,k\}$.
Then the following hold.
\begin{enumerate}[label=(\roman*)]
\item  For all $j \in \{0,\ldots,k\}$, $h(x_j)$ is of the shape $(0^{j'}1)^{i'}$ for some 
$j' \in \{1,\ldots,\ell\}$ and $i' \in \{1,\ldots,i_{j'}\}$.
\item For each $j \in \{1,\ldots,\ell\}$, there are $g_j \in \{0,\ldots,k\}$ and $h_j \in \{0,\ldots,k-g_j\}$ such that $I_j =
\coprod_{l=0}^{h_j}\cI_{h,\pi}(J_{g_j+l})$.
\end{enumerate} 
\end{lem}

\def\prooflemnoncrossstructure{
\proof
\noindent\emph{Assertion (i).}
Assume, by way of contradiction, that there is a least $j_0$ such that 
$h(x_{j_0})$ does not satisfy the claim.  It will be shown by induction that for every variable $x$ of
$\pi$ that does not lie to the left of $x_{j_0}^{n_{j_0}}$, $h(x)$ ends with $0$; since $w$ ends with
$1$, this would contradict the fact that $h(\pi) = w$.

By the choice of $x_{j_0}$, $h(x_{j_0})$ has one of the following shapes: (1) $0^p$ for some $p \in \{1,\ldots,\ell\}$,
(2) $0^{p'} 1 \ldots 1 0^{p''} 1$ for some $p',p'' \in \{1,\ldots,\ell\}$ with $p'' > p'$, or (3) $0^{p'''}1\ldots 1 0^{p''''}$ for some
$p''',p'''' \in \{1,\ldots,\ell\}$.  If $h(x_{j_0})$ has the shape given in (2), then, since $x_{j_0}$ occurs at least
twice in $\pi$, $w$ must contain a subword of the shape $0^{p''} 1 0^{p'} 1$ for some $p',p'' \in \{1,\ldots,\ell\}$
with $p' < p''$, which is impossible (as seen from the shape of $w$ in Equation (\ref{eqn:widef})).
Hence (1) or (3) holds, so the induction statement (i.e.\ that for every variable $x$ of
$\pi$ that does not lie to the left of $x_{j_0}^{n_{j_0}}$, $h(x)$ ends with $0$) holds for $x = x_{j_0}$. 

Now consider any variable $x$ of $\pi$ that lies to the right of $x_{j_0}^{n_{j_0}}$.
By the induction hypothesis, it may be assumed that for every variable $x'$ of $\pi$ lying to the right of $x_{j_0}^{n_{j_0}}$
and to the left of $x$, $h(x')$ ends with $0$.  If $h(x)$ starts with $1$, then, since $x$ is repeated at least
once in $\pi$ and every occurrence of $1$ in $w$ is preceded by $0$, $h(x)$ must end with $0$.
Suppose $h(x)$ starts with $0$ and ends with $1$.  If $x'$ is the variable immediately preceding $x$ in $\pi$, then
by the induction hypothesis, $h(x')$ is of the shape $\alpha 0$ for some $\alpha \in \{0,1\}^*$; thus, since $x$ occurs 
at least twice in $\pi$, 
if $\pi'$ denotes the suffix of $\pi$ starting at the first occurrence of $x$, then $h(\pi')$ is of the shape $0^{p_0}1
0^{p_1}1 \beta$ for some $p_0,p_1 \in \{1,\ldots,\ell\}$ with $p_1 > p_0$ and some $\beta \in \{0,1\}^*$.
As $p_1 > p_0$, $h(x)$ cannot be equal to $0^{p_0}1$, and therefore $h(x)$ must be of the shape 
$0^{p_0}1 \ldots 0^{p_2} 1$ for some $p_2 \in \{1,\ldots,\ell\}$ with $p_2 > p_0$.  But $w$ does not contain
any subword of the shape $0^{p_2} 1 0^{p_0} 1$ with $p_2 \in \{1,\ldots,\ell\}$ and $p_0 < p_2$.
The latter contradiction implies that if $h(x)$ starts with $0$, then it must also end with $0$.  This completes
the induction step and establishes the claim.

\noindent\emph{Assertion (ii).}
It suffices to show that for all $j \in \{0,\ldots,k\}$, there is some
$j' \in \{1,\ldots,\ell\}$ such that $\cI_{h,\pi}(J_j) \subseteq I_{j'}$.  By Assertion (i), there are $j'' \in \{1,\ldots,\ell\}$
and $i'' \in \{1,\ldots,i_{j''}\}$ such that $h(x_j^{n_j}) = (0^{j''}1)^{i''}$.  Furthermore, if $j \geq 1$, then
$h(x_{j-1}^{n_{j-1}})$ ends with $1$.  One observes from Equation (\ref{eqn:widef})
that any occurrence of $0^{j''}1$ in $w$ that starts after an occurrence of $1$ or 
is a prefix of $w$ must belong to the interval $I_{j''}$.  Consequently, $\cI_{h,\pi}(J_j) \subseteq I_{j''}$, as was to be
shown.~\qed
}


\begin{thm}\label{thm:tdnoncrossbounded}
For all $z \in \natnum \cup \{\infty\}$, $\TD(\NC\Pi^z_{\infty,1}) = \PBTD(\NC\Pi^z_{\infty,1}) = 0$.
Suppose $m \geq 2$. 
\begin{enumerate}[label=(\roman*)]
\item If $z = 1$, then $\TD(\NC\Pi^z_{\infty,m}) = \Theta(m)$ and
$\PBTD(\NC\Pi^z_{\infty,m}) = \Theta(m)$.  
\item 
For any $n \in \natnum_0$, let $\omega(n)$ denote the number of distinct prime factors of $n$
and let $\Pi(n)$ denote the number of prime powers not exceeding $n$.
If $z \geq 2$, then $\max(\{\omega(n): n \leq m\}) \leq \TD(\NC\Pi^z_{\infty,m}) \leq 2 + \Pi(m-1)$ 
and $\PBTD(\NC\Pi^z_{\infty,m})  = \PBTD(\NC\Pi^z) = 1$.  In particular, 
$\max(\{\omega(n): n \leq m\}) \leq 
\TD(\NC\Pi^z_{\infty,m}) <  O\left((m-1)^{\frac{1}{2}}\log(m-1)\right)+\displaystyle\frac{1.25506(m-1)}{\log(m-1)}$. 
\end{enumerate}
\end{thm}

\def\proofthmtdnoncrossbounded{
\proof
If $m = 1$, then $\NC\Pi^z_{\infty,m}$ contains only the pattern $x_1$ (up to equivalence) and thus
$\TD(\NC\Pi^z_{\infty,1}) = \PBTD(\NC\Pi^z_{\infty,1}) = 0$.  Suppose $m \geq 2$.

\noindent\emph{Assertion (i).}
Suppose $\Sigma = \{0\}$.  We identify every pattern language $L(\pi)$ such that $\pi = x_0^{n_0}\ldots x_k^{n_k}$
with its 
Parikh image $\{\vec{v} \cdot \vec{x}: \vec{x} \in \nats^{k+1}\}$, where $\vec{v} = (n_0,\ldots,n_k)$.
Thus teaching $\NC\Pi^z_{\infty,m}$ is equivalent to teaching the class $\cC_m :=\{\{\vec{v} \cdot \vec{x}: x \in \nats^k\}:
\vec{v} \in \{1,\ldots,m\}^k \wedge k \in \pnats\}$.  Since the \PBTD\ is a lower bound for the \TD,
it suffices to show that $\TD(L,\cC_m) = O(m)$ for all $L \in \cC_m$ and $\PBTD(\cC_m) = \Omega(m)$.  
Let $L = \{\vec{v} \cdot \vec{x}: \vec{x} \in \nats^{k+1}\}$, 
where $\vec{v} = (n_0,\ldots,n_k) \in \{1,\ldots,m\}^{k+1}$;
without loss of generality, it may be assumed that for all distinct $i$ and $j$, $n_i$ does not divide $n_j$
(otherwise, if $n_i \mid n_j$, then the linear set $L'$ obtained from $L$ by deleting the entry $n_j$ from $\vec{v}$
in the definition of $L$ would be equal to $L$).  It is shown that $L$ can be taught w.r.t.\ $\cC_m$ using at most $m$
examples.  Let $T$ be the sample consisting of all pairs $(p,\ell_p)$ such that $p \leq m$ and $\ell_p = +$
if $p \in L$ and $\ell_p = -$ if $p \notin L$ (that is, $T$ consists of all examples for $L$ in the 
domain $\{0,1,2,\ldots,m\}$).  Consider any $H \in \cC_m$ that is consistent with
$T$.  Since $\{n_0,\ldots,n_k\} \subseteq L$, the linearity of
$H$ (resp.~$L$) implies that $L \subseteq H$.  Furthermore, pick $\{n'_0,\ldots,n'_{k'}\} 
\subseteq \{1,\ldots,m\}$ so that $H$ is equal to $\{\vec{w} \cdot \vec{x}: x \in \nats^{k'+1}\}$
for $w = (n'_0,\ldots,n'_{k'})$.
The consistency of $H$ with $T$ implies that $\{n'_0,\ldots,n'_{k'}\} \subseteq \{n_0,\ldots,n_k\}$
and hence $H \subseteq L$.  Therefore $H = L$ and so $T$ is indeed a teaching set
for $L$ w.r.t.\ $\cC_m$.

Now it is shown that $\PBTD(\cC_m) = \Omega(m)$.
We reuse the construction in the proof
of \cite[Lemma 29]{gao15}.  Assume that $m \geq 6$, and set $m' = \left\lfloor \displaystyle\frac{m}{3} \right\rfloor$. 
Let $\cF$ be the class $\{\spn{\{m'\} \cup \{p_i: 1 \leq i \leq m'-1\}}:(\forall i \in \{1,\ldots,m'-1\})[ p_i \in \{m'+i,2m'+i\}]\}$.
Note that $\cF \subseteq \cC_m$.  Furthermore, every member of $\cF$ is of the shape $\{0,m'\} \cup \{p_i: 1 \leq i \leq m'-1\}
\cup \{x: x \geq 2m'\}$, where $p_i \in \{m'+i,2m'+i\}$ for all $i \in \{1,\ldots,m'-1\}$. 
Thus the \TD\ of every member of $\cF$ is at least $m'-1$, and therefore $\PBTD(\cC_m) \geq
\PBTD(\cF) \geq m'-1$.  This establishes that $\TD(\cC_m) = \Theta(m)$ and $\PBTD(\cC_m) = \Theta(m)$.       

\noindent\emph{Assertion (ii).}
Suppose $\{0,1\} \subseteq \Sigma$.  We first show that $\PBTD(\NC\Pi^z_{\infty,m}) = 1$.  
Let $\prec$ be the preference relation on $\NC\Pi^z_{\infty,m}$
defined according to the following hierarchy, in order of decreasing priority.  Suppose 
$\pi$ and $\tau$ are non-cross patterns in canonical form belonging to $\NC\Pi^z_{\infty,m}$.
(Here ``prefer $\pi$ to $\tau$'' means $\tau \prec \pi$.) 
\begin{description}
\item[Rule 1:] With highest priority: prefer $\pi$ to $\tau$ if $L(\pi) \neq L(x_0)$ and $L(\tau) = L(x_0)$.
\item[Rule 2:] With second highest priority: suppose both $\pi$ and $\tau$ contain at least two distinct variables;  
prefer $\pi$ to $\tau$ if $\pi$ has fewer variables than $\tau$.
\item[Rule 3:] With third highest priority: prefer $\pi$ to $\tau$ if $L(\pi) \subset L(\tau)$.   
\end{description} 
Suppose $\pi = x_0^{n_0}\ldots x_k^{n_k}$, 
where $n_0,\ldots,n_k \in \pnats$.  If there is some $i$ with $n_i = 1$,
then $\pi$ has the teaching set $\{(0,+)\}$ w.r.t.\ $\NC\Pi^z_{\infty,m}$.  
Suppose now that $n_i \geq 2$ for all $i$.  
Let $T = \{(w_1,+)\}$, where
$$
w_1 := (01)^{n_0} (001)^{n_1} \ldots (0^j1)^{n_{i+1}} \ldots (0^{k+1}1)^{n_k}.
$$
 Let $\tau := y_0^{m_0}\ldots y_{\ell}^{m_{\ell}}$ denote any 
pattern in $\NC\Pi^z_{\infty,m}$ that is consistent with $T$ and $\tau \not\prec \pi$.
By Rule 1, $m_i \geq 2$ for all $i \in \{0,\ldots,\ell\}$, that is, $L(\tau) \neq L(x_0)$. 
By Lemma \ref{lem:noncrossstructure}, the consistency of $\tau$ with $(w_1,+)$ implies that 
$\tau$ is equivalent to $x_0$ or every variable of $\tau$ occurs at least twice and for each
$j \in \{0,\ldots,k\}$, 
there are nonnegative integers $s_{j,0},\ldots,s_{j,l_j}$ and 
$i_{j,0},i_{j,1},\ldots,i_{j,l_j} \in \{0,\ldots,\ell\}$ with
$i_{j,h} < i_{j',h'}$ whenever $j < j'$ or $j = j' \wedge h < h'$ such that
$\sum_{h=0}^{l_j} s_{j,h} m_{i_{j,h}} = n_j$.  In particular, since $L(\tau) \neq L(x_0)$,
$\tau$ contains at least $k+1$ variables.  By Rule 2, $\tau$ must contain exactly $k+1$
variables.          
It follows that $\tau$ is equivalent to $x_0^{n'_0}x_1^{n'_1}\ldots x_k^{n'_k}$, where,
for each $i \in \{0,\ldots,k\}$, $n'_i \mid n_i$.  If there were a least $i' \in \{0,\ldots,k\}$ such
that $n'_i < n_i$ (that is, $n'_i$ properly divides $n_i$), then $L(\pi) \subset L(\tau)$ and
so $\tau \prec \pi$ by Rule 3, contradicting the choice of $\tau$.  Thus $n'_i = n_i$ for all $i \in \{0,\ldots,k\}$
and therefore $L(\tau) = L(\pi)$, as required.

Next, it will be shown that $\TD(\NC\Pi^z_{\infty,m})$ is at most $2$ plus the number of prime powers 
(including primes) less than $m$; this is equal to $2+\sum_{i=1}^{\lfloor\log(m-1)\rfloor} \varrho\left((m\right.$ $\left.-1)^{\frac{1}{i}}\right)$,
where $\varrho(x)$ denotes the number of primes less than or equal to $x$.
As observed earlier, the pattern $x_0$ can be taught with the single
example $(0,+)$.  Suppose $\pi = x_0^{n_0}\ldots x_k^{n_k}$, 
where $n_i \geq 2$ for all $i \in \{0,\ldots,k\}$.  We build a teaching set $T$ consisting of the following
examples; $\eta := y_0^{m_0} \ldots y_{\ell}^{m_{\ell}}$ will denote any non-cross pattern 
(in canonical form) in $\NC\Pi^z_{\infty,m}$ that is consistent
with $T$.  First, put $(v_1,+)$ into $T$, where
$$
v_1 := (01)^{n_0} (001)^{n_1} \ldots (0^j1)^{n_{i+1}} \ldots (0^{k+1}1)^{n_k}.
$$
According to Lemma \ref{lem:noncrossstructure}, the consistency of $\eta$ with $(v_1,+)$
implies that for each $j \in \{0,\ldots,k\}$, there are nonnegative integers $s_{j,0},\ldots,s_{j,l_j}$
and $i_{j,0},\ldots,i_{j,l_j} \in \{0,\ldots,\ell\}$ such that $i_{j,h} < i_{j',h'}$ 
iff $j < j'$ or $j = j' \wedge h < h'$, and $\sum_{r=0}^{l_j} s_{j,r} m_{i_{j,r}} = n_j$.   
Second, define
$$
v_2 := (01)^{m!} (001)^{m!} \ldots (0^j1)^{m!} \ldots (0^{k+1}1)^{m!} (0^{k+2}1)^{m!}, 
$$ 
and put $(v_2,-)$ into $T$.  Note that by Lemma \ref{lem:noncrossstructure}, $v_2$ is indeed
a negative example for $\pi$ because any pattern $\pi'$ with $v_2 \in L(\pi')$ is equivalent
to $x_0$ or it contains at least $k+2$ variables that occur at least twice.  Furthermore, Lemma 
\ref{lem:noncrossstructure} also implies that $\eta$ is not equivalent to $x_0$ and
that $\eta$ contains at most $k+1$ variables.  Since the consistency of $\eta$ with $(v_1,+)$
implies that $\eta$ contains at least $k+1$ distinct variables, it follows that $\eta$ contains exactly
$k+1$ variables, each of which occurs at least twice.  That is to say, $\eta$ is of the shape
$x_0^{n'_0}x_1^{n'_1}\ldots x_k^{n'_k}$, where, for each $i \in \{0,\ldots,k\}$, $n'_i \mid n_i$.
It remains to ensure that $n'_i$ does not properly divide $n_i$ for any $i \in \{0,\ldots,k\}$.

Let $\{q_0^{r_0},\ldots,q_{\ell'}^{r_{\ell'}}\}$ be the set of all prime powers 
that are maximal proper prime power factors of the $n_i$'s; in other words, for every $j \in \{0,\ldots,\ell'\}$, 
there is some $j_0 \in \{0,\ldots,k\}$ with $q_j^{r_j} \mid n_{j_0}$ and $q_j^{r_j} \neq n_{j_0}$ but
$q_j^{r_j+1} \nmid n_{j_0}$.  For each $j \in \{0,\ldots,\ell'\}$, let $d_j$ be the number of $i$'s between
$0$ and $k$ (inclusive) such that $q_j^{r_j}$ does not divide $n_i$, and set $e_j = q_j^{r_j-1} \cdot
\prod_{p~\mbox{is prime}\wedge q_j \neq p \leq m} p^{\left\lfloor\frac{\log(m)}{\log(p)}\right\rfloor}$.  
Now define 
$$
t_j := (01)^{e_j} (001)^{e_j} \ldots (0^l)^{e_j} \ldots (0^{d_j+1}1)^{e_j}
$$     
for every $j \in \{0,\ldots,\ell'\}$, and put $(t_j,-)$ into $T$.

We first show that $t_j \notin L(\pi)$ for every $j \in \{0,\ldots,\ell'\}$.  This will be achieved by means of
a proof by contradiction; assuming that $t_j \in L(\pi)$, one can construct
a one-one mapping $F$ from $\{1,\ldots,d_j+1\}$ to $\{i \in \{0,\ldots,k\}: q_j^{r_j} \nmid n_i\}$ as follows.  
Given any $i \in \{1,\ldots,d_j+1\}$, it follows from
Lemma \ref{lem:noncrossstructure} that there are nonnegative integers $s_{i,0},\ldots,
s_{i,l_i}$ and $u_{i,0},\ldots,u_{i,l_i} \in \{0,\ldots,k\}$ such that $u_{i,g} < u_{i',g'}$ iff $i < i'$
or $i = i'$ and $g < g'$, and $\sum_{k=0}^{l_i} s_{i,k} n_{u_{i,k}} = e_j$.
Note that since $q_j^{r_j} \nmid e_j$, there must exist a least $h_i$
such that $q_j^{r_j} \nmid n_{u_{i,h_i}}$.  Define $F(i) = u_{i,h_i}$.
Then $range(F) \subseteq \{i \in \{0,\ldots,k\}: q_j^{r_j} \nmid n_i\}$; furthermore,
$i < i' \Rightarrow u_{i,h_i} < u_{i',h_{i'}} \Leftrightarrow F(i) < F(i')$.
Thus $F$ is indeed a one-one mapping, so that 
\begin{equation*}
\begin{aligned}
d_j+1 &= \left\vert\{1,\ldots,d_j+1\}\right\vert \\ 
&\leq \left\vert range(F)\right\vert ~(\mbox{by the one-one property of $F$}) \\
&\leq \left\vert\{i \in \{0,\ldots,k\}: q_j^{r_j} \nmid n_i\}\right\vert ~(\mbox{since $range(F) \subseteq \{i \in \{0,\ldots,k\}: q_j^{r_j} \nmid n_i\}$}) \\
&= d_j,
\end{aligned}
\end{equation*}
a contradiction.

To complete the proof, it will be shown that if there were a least $i'' \in \{0,\ldots,k\}$ such
that $n'_{i''}$ properly divides $n_{i''}$ (as noted above, $\eta$ is of the shape
$x_0^{n'_0}x_1^{n'_1}\ldots x_k^{n'_k}$, where $n'_i \mid n_i$ for all $i \in \{0,\ldots,k\}$),
then there would be a least $j' \in \{0,\ldots,\ell'\}$ such that $t_{j'} \in L(\eta)$.
Suppose such an $i''$ did exist.  Then there must be a least $j'' \in \{0,\ldots,\ell'\}$
for which $q_{j''}^{r_{j''}} \mid n_{i''}$ and $n'_{i''} \mid n_{i''}q_{j''}^{-1}$.  Hence
the number of $i$'s between $0$ and $k$ (inclusive) such that $q_{j''}^{r_{j''}} \nmid n'_i$
is at least $1$ more than the number of $i$'s between $0$ and $k$ (inclusive) such that
$q_{j''}^{r_{j''}} \nmid n_i$, and the number of $j_1$'s between $0$ and $k$ inclusive 
such that $n'_{j_1} \mid e_{j''}$ is at least $d_{j''}+1$.  
Consequently, $t_{j''} \in L(\eta)$, which is the desired contradiction.
In conclusion, $n'_i = n_i$ for all $i \in \{0,\ldots,k\}$ and thus $\eta$ is equivalent to $\pi$;
this establishes that $T$ is a teaching set for $\pi$ w.r.t.\ $\NC\Pi^z_{\infty,m}$.

To prove that $\TD(\NC\Pi^z_{\infty,m}) \geq \max(\{\omega(n): n \leq m\})$, pick any $n \leq m$
such that $\omega(n) \geq \omega(n')$ for all $n' \leq m$.  Let $q_1,\ldots,q_{\omega(n)}$ be all the
prime factors of $n$, and consider the non-cross pattern $\theta := x_1^{\prod_{i=1}^{\omega(n)} p_i}$. 
For each $i \in \omega(n)$, set $\theta_i := x_1^{\prod_{j \neq i} p_j}$.
We note that $\theta \in \NC\Pi^z_{\infty,m}$ and for all $i \in \{1,\ldots,\omega(n)\}$, $\theta_i \in 
\NC\Pi^z_{\infty,m}$.  Furthermore, whenever $i \neq j$, $L(\theta_i) \cap L(\theta_j) \subseteq L(\theta)$.  
It follows that $\TD(\theta,\NC\Pi^z_{\infty,m}) \geq \omega(n)$.~\qed
}

\medskip
\noindent
It is possible that neither the lower bound nor the upper bound on $\TD(\NC\Pi^z_{\infty,m})$ given in 
Theorem \ref{thm:tdnoncrossbounded} is tight for almost all $m$.
The proof of Theorem \ref{thm:tdnoncrossbounded} (c.f.\ Appendix \ref{appen:proofthmtdnoncrossbounded}) shows that 
the \TD\ of any general 
non-cross pattern $\pi$ w.r.t.\ $\NC\Pi^z_{\infty,m}$ (for any fixed $z \geq 2$ and
$m \geq 2$) is at most $2$ plus the number of maximal proper prime factors of the variable frequencies
of $\pi$, but as the following example shows, this upper bound is not always sharp even for non-cross
succinct patterns with three variables; a pattern $\pi$ is \emph{succinct} \cite{Mit98,Rei08} iff there is no  
pattern $\tau$ such that $L(\tau) = L(\pi)$ and $|\tau| < |\pi|$.

\begin{exmp}\label{exmp:noncrosslowerbound}
Suppose $\{0,1\} \subseteq \Sigma$.  Let $\pi = x_1^4 x_2^8 x_3^9$.
There are $3$ maximal proper prime power factors of $4,8$ and $9$, namely,
$2,4$ and $3$, and so by the proof of Theorem \ref{thm:tdnoncrossbounded},
the \TD\ 
of $\pi$ w.r.t.\ $\NC\Pi^{|\Sigma|}_{\infty,9}$ is at most
$2+3 = 5$.  However, $\pi$ has a teaching set of size $4$ (further details are given in
Appendix \ref{appen:noncrosslowerbound}). 
\end{exmp}

\medskip
\noindent
The next result exemplifies the general observation that a larger alphabet allows pattern languages to be distinguished 
using a relatively smaller number of labelled examples. 


\begin{thm}\label{thm:pbtdgeneralinfinite}
$\PBTD(\Pi^{\infty}) = 2$ and for any $m \geq 1$, $\PBTD(\Pi^1_{\infty,m}) = \Theta(m)$.
\end{thm}

\def\proofthmpbtdgeneralinfinite{
\proof
We first compute $\PBTD(\Pi^{\infty})$. 
It will be assumed that every pattern $\pi$ in the present proof is succinct, i.e.\ $|\pi'| \geq |\pi|$
for all $\pi'$ such that $L(\pi') = L(\pi)$.  Define a preference relation $\prec$ on $\Pi^z$ based
on the following preference hierarchy, 
where $\pi$ and $\tau$ are any two given succinct patterns:
\begin{description}
\item[Rule 1:] With highest priority, prefer $\pi$ to $\tau$ (i.e.\ $\tau \prec \pi$) if $|\pi(\ve)| > |\tau(\ve)|$. 
\item[Rule 2:] With second highest priority, prefer $\pi$ to $\tau$ (i.e.\ $\tau \prec \pi$) if $L(\pi) \subseteq L(\tau)$.
\end{description}
Given any $\pi \in \Pi^{\infty}$, one can construct a teaching set $T$ of size at most $2$ for $\pi$ w.r.t.\
$(\Pi^z,\prec)$ as follows; $\tau$ will denote any pattern in $\Pi^z$ that is consistent with $T$ and
$\tau \not\prec \pi$.  First, put 
$(\pi(\ve),+)$ into $T$.  Since $\tau(\ve) \sqsubseteq \pi(\ve)$, Rule 1 will ensure that $\tau(\ve) = \pi(\ve)$,
that is, $\pi$ and $\tau$ have identical constant parts.  Second, suppose $\Var(\pi) = \{x_0,\ldots,x_{k-1}\}$.  Choose
a set $\{a_0,\ldots,a_{k-1}\}$ of $k$ distinct letters such that $\{a_0,\ldots,a_{k-1}\} \cap \Const(\pi) = \emptyset$,         
and put $(\pi[x_i \ra a_i, 0 \leq i \leq k-1],+)$ into $T$.
By Theorem \ref{thm:jiang94suffcondsubset}, the fact that $\pi[x_i \ra a_i, 0 \leq i \leq k-1] \in L(\tau)$
implies $L(\pi) \subseteq L(\tau)$.  By Rule 2, one has $L(\tau) = L(\pi)$, as required.

To see that $\PBTD(\Pi^{\infty}) \geq 2$, one may apply \cite[Theorem 34]{GRSZ2016}; according to this theorem, 
$\PBTD(\Pi^{\infty}) > 1$ because $\Pi^{\infty}$ contains all constant patterns as well as infinitely many patterns that 
generate infinite languages.

Next, it is shown that $\PBTD(\Pi^1_{\infty,m}) = \Theta(m)$.  
Suppose $\Sigma = \{0\}$.  It follows from Theorem \ref{thm:tdnoncrossbounded}
and the monotonicity of the \PBTD\ \cite[Lemma 6]{GRSZ2016} that 
$\PBTD(\Pi^1_{\infty,m}) \geq \PBTD(\NC\Pi^1_{\infty,m}) = \Theta(m)$.
For the upper bound, we observe that every pattern in $\Pi^1_{\infty,m}$ is equivalent
to a pattern of the shape $0^k x_1^{n_1}\ldots, x_{\ell}^{n_{\ell}}$, where $k + \ell \geq 1$
and $1 \leq n_1 < \ldots < n_{\ell} \leq m$ (this follows from the fact that over unary alphabets,
equivalence of two patterns is preserved under permutations of the patterns' symbols and that 
any two terms of the shape $x_i^n x_j^n$ can be combined into a single term $x_i^n$).
Define the preference relation $\prec$ on $\Pi^1_{\infty,m}$ as follows:
for any $\pi,\pi' \in \Pi^1_{\infty,m}$, $\pi \prec \pi'$ iff 
\begin{itemize}
\item $|\pi'(\ve)| > |\pi(\ve)|$, or
\item $\pi'(\ve) = \pi(\ve)$ and $L(\pi') \subset L(\pi)$. 
\end{itemize} 
Suppose $\pi \in \Pi^1_{\infty,m}$.  If $\pi = 0^k$ for some $k \geq 1$,
then $\pi$ can be taught w.r.t.\ $(\Pi^1_{\infty,m},\prec)$ using the single positive
example $(0^k,+)$ since all patterns containing $0^k$ must have a constant part of length
at least $k = |\pi|$ and $\pi$ is preferred to all patterns with a constant part of
length less than $k$.  Suppose $\pi = 0^k x_1^{n_1} \ldots x_{\ell}^{n_{\ell}}$ for
some $\ell \geq 1$ such that $n_i < n_j$ whenever $i < j$.
A teaching set for $\pi$ is $T := \{(0^k,+)\} \cup \{(0^{k+n_i},+): i \in \{1,\ldots,\ell\} \}$.
Let $\tau$ denote any pattern in $\Pi^1_{\infty,m}$ that is consistent with $T$.
The positive example $(0^k,+)$ ensures that $\tau(\ve) = \pi(\ve)$.
Furthermore, since $0^{k+n_i} \in L(\tau)$ for all $i \in \{1,\ldots,\ell\}$,
it follows that $L(\tau) \subseteq L(\pi)$, and so by the definition of $\prec$, 
$L(\tau) = L(\pi)$. 
\hfill\qed
}

\medskip
\noindent The next series of results deal with the finite distinguishability problem
for the general class of $m$-regular patterns.  We begin with a few preparatory results.  
The first part of Theorem \ref{thm:jiang94suffcondsubset}
gives a sufficient criterion for the inclusion of pattern languages, and it was observed
by Jiang, Kinber, Salomaa and Yu \cite{Jiang94}; the second part, due to
Ohlebusch and Ukkonen \cite{Ohlebusch97}, states that the existence of a constant-preserving
morphism from $\pi$ to $\tau$ (where $\pi$ and $\tau$ are similar) also implies 
$L(\tau) \subseteq L(\pi)$ if $\Sigma$ contains at least two letters that do not
occur in $\pi$ or $\tau$.  The second result is based on a few lemmas due to
Reidenbach \cite[Lemmas 4--6]{Rei08}, adapted to the 
case of general patterns over an infinite alphabet. 

\begin{thm}\cite{Jiang94,Ohlebusch97}\label{thm:jiang94suffcondsubset}
Let $\Sigma$ be an alphabet, and let $\pi,\tau \in \Pi^{|\Sigma|}$.
Then $L(\pi) \subseteq L(\tau)$ if there exists a constant-preserving
morphism $g:(X \cup \Sigma)^* \mapsto (X \cup \Sigma)^*$ with $g(\tau) = \pi$.
If $|\Sigma| \geq |\Const(\pi)| + 2,|\Sigma| \geq |\Const(\tau)|+2$ and
$\pi$ is similar to $\tau$, then $L(\pi) \subseteq L(\tau)$ only if there exists a 
constant-preserving morphism $g:(X \cup \Sigma)^* \mapsto (X \cup 
\Sigma)^*$ with $g(\tau) = \pi$.
\end{thm}

\begin{lem}(Based on \cite{Rei08})\label{lem:succinctinfinitealphabet}
Suppose $|\Sigma| = \infty$.  Fix any $\pi \in \Pi^{\infty}$ such that
$\pi$ is succinct.  Let $Y = \{y_1,y_2,\ldots\}$ be an infinite set
of variables such that $Y \cap \Var(\pi) = \emptyset$.  Suppose $\tau 
\in \pi \shuffle Y^*$.  Then $L(\tau) = L(\pi)$ iff
\begin{enumerate}[label=(\roman*)]
\item For all $Y' \in Y^+$ and $\delta,\delta'\in\Const(\pi)$, the following hold: 
(a) $Y'\delta$ is not a prefix of $\tau$, (b) $\delta Y'$ is not a suffix of $\tau$, 
(c) $\delta Y' \delta'$ is not a substring of $\tau$;  
\item There is a constant-preserving morphism $g:(X \cup \Sigma)^* \mapsto (X \cup \Sigma)^*$ 
such that $g(\pi) = \tau$;
\item For all constant-preserving morphisms $h:(X \cup \Sigma)^* \mapsto (X \cup \Sigma)^*$
with $h(\pi) = \tau$ and for all $x \in \Var(\pi)$, if there exist $Y_1,Y_2 
\in Y^*$ such that $Y_1 x Y_2$ is a substring of $\tau$ and $Y_1$ (resp.~$Y_2$)
is not immediately preceded (resp.~succeeded) by any $y \in Y$ w.r.t.\ $\tau$,
then there are splittings $Y^1_1Y^2_1$ and $Y^1_2Y^2_2$ of $Y_1$
and $Y_2$ respectively for which $h(x) = Y^2_1 x Y^1_2$. 
\end{enumerate}
\end{lem}

\def\prooflemsuccinctinfinitealphabet{
\proof
The ``if'' direction of the lemma follows from Condition (ii), Theorem \ref{thm:jiang94suffcondsubset}
and the fact that $L(\pi) \subseteq L(\tau)$ (which is in turn implied by $\tau \in \pi \shuffle
Y^*$ and $Y \cap \Var(\tau) = \emptyset$).  We prove the ``only if'' direction of the
lemma.  
\begin{description}[leftmargin=0cm]
\item[Condition (i):] Assume, by way of contradiction, that $Y'\delta$ were a prefix of $\tau$.
Fix some $\omega \in \Sigma \sm \Const(\pi)$ and $y \in \Var(Y')$.  Set $w = \tau[y \ra \omega]$.
Then $w \in L(\tau)$ by construction; on the other hand, since $\tau$ starts with $\delta$
but $w$ starts with $\omega \neq \delta$, $w \notin L(\pi)$.  The proofs that $\delta Y'$ is not
a suffix of $\tau$ and $\delta Y' \delta'$ is not a substring of $\tau$ are similar.  
\item[Condition (ii):] Note that Condition (i) implies $\pi$ is similar to $\tau$.  If $L(\tau) \subseteq L(\pi)$, 
then (a) $|\Sigma| = \infty$, (b) $\pi$ is similar to $\tau$ and (c) the second part of Theorem 
\ref{thm:jiang94suffcondsubset} together imply Condition (ii).  
\item[Condition (iii):] Let $h:(X \cup \Sigma)^* \mapsto (X \cup Y \cup \Sigma)^*$ be any 
constant-preserving morphism such that $h(\pi) = \tau$.  Let $p_1,p_2,\ldots,p_n$
be all the positions of $\pi$ that are occupied by variables, where $p_1 < p_2 < \ldots < p_n$, 
and for all $j \in \{1,\ldots,n\}$, let $x_{i_j}$ denote the variable at the $p_j^{th}$ position of 
$\pi$. 
(For example, if $\pi = x_1 0 x_1 0 x_2 x_1 x_3 x_3$, then $i_1 = i_2 = 1, i_3 = 2, i_4 = 1$ and 
$i_5 = i_6 = 3$.)  

Suppose there is a least $j \in \{1,\ldots,n\}$ such that $h(x_{i_j})$ is not of the shape 
$Y_1 x_{i_j} Y_2$, where $Y_1,Y_2 \in Y^*$.  Note that $\Const(h(x)) = \emptyset$ for all
$x \in \Var(\pi)$, for otherwise $h(\pi)$ would have more occurrences of constants than $\tau$ 
(by Condition (ii), $\tau$ is similar to $\pi$). 
Since 
$x_{i_1},x_{i_2},\ldots,x_{i_n}$ occur in $\tau$ in the same order as their appearance in $\pi$, 
there exists some $j_1 \in \{1,\ldots,n\}$ such that $j_1 \geq j$ and $h(x_{i_{j_1}}) \in Y^*$.  Now let $\pi'$ be 
the pattern obtained from $\pi$ by
deleting all occurrences of $x_{i_{j_1}}$. 
Let $h':(X \cup \Sigma)^* \mapsto (X \cup \Sigma)^*$ be a 
constant-preserving morphism such that $h'(x) = (h(x)){\big|}_{\Sigma \cup \Var(\pi)}$ for all
$x \in \Var(\pi)$.  Then one has 
\begin{equation*}
\begin{split}
h'(\pi') &= (h(\pi')){\big|}_{\Sigma \cup \Var(\pi)} \\ 
~ &= (h(\pi)){\big|}_{\Sigma \cup \Var(\pi)} ~ (\mbox{since $h(x_{i_{j_1}}) \in Y^*$ and $\pi' = \pi{\big|}_{\Sigma \cup (\Var(\pi)\sm \{x_{i_{j_1}}\})}$}) \\
~ &= \tau{\big|}_{\Sigma \cup \Var(\pi)} ~ (\mbox{since $h(\pi) = \tau$}) \\
~ &= \pi ~ (\mbox{since $\tau \in \pi \shuffle Y^*$ and $Y \cap \Var(\pi) = \emptyset$}).  
\end{split}
\end{equation*}
Consequently, by Theorem \ref{thm:jiang94suffcondsubset}, $L(\pi) \subseteq L(\pi')$.  
By construction, $L(\pi') \subseteq L(\pi)$ and so $L(\pi) = L(\pi')$.  But $\pi'$ 
is a pattern shorter than $\pi$ that generates the same language as $\pi$, contrary to 
the hypothesis that $\pi$ is succinct.
\end{description} 
\hfill\qed
}

\medskip
\noindent
The next crucial lemma shows that for any fixed $m \geq 1$, only finitely 
many negative examples are needed to distinguish a succinct pattern 
$\pi$ from all patterns $\pi' \in \Pi^{\infty}_{\infty,m}$ obtained by 
shuffling $\pi$ with an infinite set $Y$ of variables such that $Y$ and 
$\Var(\pi)$ are disjoint.

\begin{lem}\label{lem:subsetrestrictedvarwitness}
Fix $\Sigma$ with $|\Sigma| = \infty$.
Suppose $k \geq 0$, $m \geq 1$ and $\pi \in \Pi^{\infty}_{k,m}$.
Let $Y = \{y_1,y_2,\ldots\}$ be an infinite set of variables such that 
$Y \cap \Var(\pi) = \emptyset$.  Suppose $\tau \in \left(\pi \shuffle
Y^*\right) \cap \Pi^{\infty}_{\infty,m}$.  There is some
$\tau' \in \Pi^{\infty}_{4mk+|\pi|+2,m}$ such that $\tau' = 
\tau{\big|}_{\Sigma \cup \Var(\pi) \cup S}$ for some finite $S \subset Y$, 
and if $L(\pi) \subset L(\tau)$, then $L(\pi) \subset L(\tau')$.    
\end{lem}

\def\prooflemsubsetrestrictedvarwitness{
\proof
We split the analysis into two cases.
\begin{description}[leftmargin=0cm]
\item[Case 1:] There are $Y' \in Y^+$ and $\delta,\delta'\in\Sigma$ such that 
at least one of the following holds: (i) $Y'\delta$ is a prefix of $\tau$, 
(ii) $\delta Y'$ is a suffix of $\tau$ or (iii) $\delta Y' \delta'$ is a 
substring of $\tau$.    
Suppose (i) holds.  Pick some $y \in \Var(Y')$ and let $\tau'$ be the
restriction of $\tau$ to $\Sigma \cup \Var(\pi) \cup \{y\}$.  We
show $L(\tau') \supset L(\pi)$.
By construction, $L(\tau') \supseteq L(\pi)$.  
Fix some $\omega \in \Sigma \sm \Const(\pi)$, and set $w = \tau'[y \ra \omega]$.
Then $w \in L(\tau')$.  Further, since $\pi$ starts with $\delta$ but $w$ starts with 
$\omega \neq \delta$, one has $w \notin L(\pi)$.   
A similar proof applies if (ii) or (iii) holds.

\item[Case 2:] Not Case 1.  Let $x_1$ (resp.~$x_n$) be the leftmost 
(resp.~rightmost) variable of $\pi$.  
For each $x \in \Var(\pi)$, let $Y^x_{\ell}$ (resp.~$Y^x_r$) be the longest 
substring $Z$ in $Y^*$ such that every occurrence of $x$ in $\tau$ is immediately
preceded (resp.~succeeded) by $Z$.  For each occurrence of $x \in \Var(\pi)$,
identify the unique $y \in Y$ such that $y$ immediately precedes the
corresponding occurrence of $Y^x_{\ell}x$, and put $y$ into $S^x$ (if no such
$y$ exists, then nothing needs to be done).  Similarly, for each occurrence 
of $x \in \Var(\pi)$, identify the unique $z \in Y$ such that $z$ immediately
succeeds the corresponding occurrence of $xY^x_r$, and put $z$ into
$S^x$ (again, nothing needs to be done if no such $z$ exists).
Further, if the last (resp.~first) symbol occurring in $\tau$ is some $y \in Y$,
put $y$ into $S^{x_n}$ (resp.~$S^{x_1}$).  Lastly, for every substring of $\tau$ of the shape 
$x Y' \delta$ (resp.~$\delta Y' x$), where $\delta \in \Sigma, Y' \in Y^+$ and $x \in \Var(\pi)$,
put the last (resp.~first) symbol of $Y'$ into $S^x$.       

Let $\tau'$ be the restriction of $\tau$ to $\Sigma \cup \Var(\pi) \cup \bigcup_{x \in \Var(\pi)}S^x$.
Note that $\tau' \in \Pi^{\infty}_{4mk+|\pi|+2,m}$
and $\tau' = \tau{\big|}_{\Sigma \cup \Var(\pi) \cup S}$ for some finite
$S \subset Y$.  
Suppose there is a constant-preserving morphism $g:(X \cup \Sigma)^* 
\mapsto (X \cup \Sigma)^*$ such that $g(\pi) = \tau'$. 
We show that this implies the existence of a constant-preserving morphism $g':(X \cup \Sigma)^* 
\mapsto (X \cup \Sigma)^*$ such that $g'(\pi) = \tau$.  It will then follow that 
whenever $L(\pi) \subset L(\tau)$, one has $L(\pi) \subset L(\tau')$,
as required. 
By Lemma \ref{lem:succinctinfinitealphabet}, every occurrence of any $y \in Y$
in $\tau'$ is contained in a substring of $\tau'$ of the shape $x Y'$ or $Y'x$
for some $Y' \in Y^+$, and for every $x \in \Var(\pi)$, there are $Z^x_{\ell},Z^x_r 
\in Y^*$ for which $I_{g,\pi}$ maps the position $p_x$ 
of %
the $t^{th}$ 
occurrence of $x$ in $\pi$ (for any $t \leq \left|\pi{\big|}_x\right|$) 
to an interval $J_{p_x}$ of positions of $\tau'$ corresponding to an occurrence 
of $Z^x_{\ell} x Z^x_r$ in $\tau'$ such that the position of the $t^{th}$ occurrence
of $x$ in $\tau'$ belongs to $J_{p_x}$. 
Suppose $\tau = \rho_1 x_1 \cdots x_n \rho_n$ and $\tau' = \rho'_1 x_1
\cdots x_n \rho'_n$, where $\rho_1,\rho'_1,\rho_n,\rho'_n \in Y^*$.  

Our first step is to show $Y^{x_1}_{\ell} = \rho_1$ and $Y^{x_n}_r = \rho_n$.
So assume, by way of contradiction, that at least one of the following holds: 
(i) $Y^{x_1}_{\ell} \neq \rho_1$ or (ii) $Y^{x_n}_r \neq \rho_n$.  
Suppose (i) holds. 
Since $Y^{x_1}_{\ell} \neq \rho_1$, there is a unique $y \in Y$ immediately preceding
the first occurrence of $Y^{x_1}_{\ell}x_1$ in $\tau$.  Note that $Z^{x_1}_{\ell} = \rho'_1 
= \rho_1{\big|}_{\bigcup_{x\in\Var(\pi)}S^x}$.
Furthermore, there is another substring of $\tau$ of the shape $s Y^{x_1}_{\ell}x_1$, 
where $s \in \left(Y \sm \{y\}\right) \cup \Var(\pi) \cup \Sigma$.
If $s \in Y \sm \{y\}$, then $s \in S^{x_1}$ and so $Z^{x_1}_{\ell} \neq 
\rho_1{\big|}_{\bigcup_{x\in\Var(\pi)}S^x} = \rho'_1$, a contradiction.
If $s \in \Var(\pi) \cup \Sigma$, then 
one has 
\begin{equation*}
\begin{split}
\#(y)[Z^{x_1}_{\ell}] &\leq \#(y)[Y^{x_1}_{\ell}] \\ 
& < \#(y)[\rho_1] ~~ \mbox{(since $y$ precedes the first occurrence of $Y^{x_1}_{\ell}x_1$ in $\tau$)} \\ 
&= \#(y)\left[\rho_1{\big|}_{\bigcup_{x\in\Var(\pi)}S^x}\right] ~~ \mbox{(since $y \in S^{x_1}$)} \\
&= \#(y)[\rho'_1],
\end{split}
\end{equation*} 
which again shows $Z^{x_1}_{\ell} \neq \rho'_1$, a contradiction.  
A similar proof shows that (ii) contradicts the definition of $Y^{x_n}_r$. 

The next step is to show that for every substring of $\tau$ of the shape
$\delta Y'x$ (resp.~$xY'\delta$), where $\delta \in \Sigma, Y' \in Y^*$
and $x \in \Var(\pi)$, one has $Y^x_{\ell} = Y'$ (resp.~$Y^x_r = Y'$).
The proof is similar to that in the preceding paragraph.
Suppose there is a substring of $\tau$ of the shape $\delta Y' x$ and
$Y^x_{\ell} \neq Y'$.  There is a unique $y \in Y$ at the $\left(|Y'| - |Y^x_{\ell}|
\right)^{th}$ position of $Y'$, and so by the definition of $S^x$ one
has $y \in S^x$.  
Then, as argued in the preceding paragraph, one has $Z^x_{\ell} \neq Y'{\big|}_{\bigcup_{x
\in\Var(\pi)}S^x}$, and so such a $g$ as described earlier cannot exist.  The proof for 
substrings of $\tau$ of the shape $xY'\delta$ is similar.
       
Thus one may safely assume that (i) $Y^{x_1}_{\ell} = \rho_1$, (ii) $Y^{x_n}_r = \rho_n$,
and (iii) for all substrings of $\tau$ of the shape $\delta Y' x$ (resp.~$xY'\delta$),
where $\delta \in \Sigma, Y' \in Y^*$ and $x \in \Var(\pi)$, we have $Y^x_{\ell} = Y'$
(resp.~$Y^x_r = Y'$).  

We next observe that for any $x_i \in \Var(\pi)$ and $y \in \bigcup_{x\in\Var(\pi)}S^x$,
$\#(y)[Z^{x_i}_{\ell}] \leq \#(y)[Y^{x_i}_{\ell}]$.  To see this, suppose first
that there is an occurrence of $Y^{x_i}_{\ell}x_i$ that is not immediately preceded
by any $y \in Y$.  Then $Z^{x_i}_{\ell} = Y^{x_i}_{\ell}{\big|}_{\bigcup_{x\in\Var(\pi)}S^x}$
and thus for all $y \in \bigcup_{x\in\Var(\pi)}S^x$, $\#(y)[Z^{x_i}_{\ell}] \leq \#(y)[Y^{x_i}_{\ell}]$.
 Second, suppose that every occurrence of $Y^{x_i}_{\ell}$ is immediately preceded
by some $y \in Y$.  Thus, by the choice of $Y^{x_i}_{\ell}$, there must exist distinct
$y',y'' \in Y$ such that $y' Y^{x_i}_{\ell}x_i$ and $y'' Y^{x_i}_{\ell}x_i$ are substrings
of $\tau$.  By the definition of $S^{x_i}$, $y',y'' \in S^{x_i}$.  Hence both
$y' Y^{x_i}_{\ell}{\big|}_{\bigcup_{x\in\Var(\pi)}S^x}x_i$ and
$y'' Y^{x_i}_{\ell}{\big|}_{\bigcup_{x\in\Var(\pi)}S^x}x_i$ are substrings of $\tau'$,
and therefore $Z^{x_i}_{\ell}$ is a suffix of $Y^{x_i}_{\ell}{\big|}_{\bigcup_{x\in\Var(\pi)}S^x}$.  
Consequently, $\#(y)[Z^{x_i}_{\ell}] \leq \#(y)[Y^{x_i}_{\ell}]$ for all $y \in \bigcup_{x\in\Var(\pi)}S^x$, 
as required.   
Similarly, for any $x_i \in \Var(\pi)$ and $y \in \bigcup_{x\in\Var(\pi)}S^x$,
$\#(y)[Z^{x_i}_r] \leq \#(y)[Y^{x_i}_r]$.

For every $x_i \in \Var(\pi)$, let $\alpha_i$ be the \emph{longest} suffix
of $Y^{x_i}_{\ell}$ such that $\alpha_i{\big|}_{\bigcup_{x \in \Var(\pi)}S^x}$
$= Z^{x_i}_{\ell}$ and let $\beta_i$ be the \emph{shortest} prefix of $Y^{x_i}_r$
such that $\beta_i{\big|}_{\bigcup_{x \in \Var(\pi)}S^x} = Z^{x_i}_r$
(by the remarks in the preceding paragraph, such $\alpha_i$ and $\beta_i$
exist).  Set $g'(x_i) = \alpha_i x_i \beta_i$.  For example, suppose
$Y^{x_i}_{\ell} = y_1y_2^2y_3y_1y_3y_4y_1$, $Y^{x_i}_r = y_1y_2y_3y_1y_3y_2y_4$,
$Z^{x_i}_{\ell} = y_1^2$, $Z^{x_i}_r = y_1y_2y_1$ and $\bigcup_{x \in \Var(\pi)}S^x
= \{y_1,y_2\}$.  Then $\alpha_i = y_3y_1y_3y_4y_1$ and $\beta_i = y_1y_2y_3y_1$.

It remains to verify that $g'(\pi) = \tau$.  By the present case assumption, 
every occurrence of any substring $Z \in Y^*$ of $\tau$ is contained in a
substring $\theta$ of $\tau$ satisfying 
at least one of the following: (a) $\theta = Z x_1 $ and $\theta$ is a prefix of $\tau$; 
(b) $\theta = x_n Z$ and $\theta$ is a suffix of $\tau$; (c) $\theta = x_i Z x_j$,
for some $x_i,x_j \in \Var(\pi)$; (d) $\theta = \delta Z x_i$ for some $\delta \in \Sigma$ 
and $x_i \in \Var(\pi)$; (e) $\theta = x_i Z \delta$ for some $\delta \in \Sigma$ and $x_i \in \Var(\pi)$. 
Thus, since $g'(x_i) = \alpha_i x_i \beta_i$ for all $x_i \in \Var(\pi)$, it suffices to 
show: (a) $\alpha_1 = \rho_1$; (b) $\beta_n = \rho_n$; (c) if $x_iZx_j$ is a substring
of $\tau$ for some $Z \in Y^*$, then $\beta_i\alpha_j = Z$; (d) if $\delta Z x_i$ is a substring of $\tau$
for some $Z \in Y^*$ and $\delta\in\Sigma$, then $\alpha_i = Z$; (e) if $x_i Z \delta$ is a substring of $\tau$
for some $Z \in Y^*$ and $\delta\in\Sigma$, then $\beta_i = Z$.   

\begin{description}[leftmargin=0cm]
\item[Assertion (a):] Note that since $Y^{x_1}_{\ell} = \rho_1$ and $Z^{x_1}_{\ell} = \rho'_1 = 
\rho_1{\big|}_{\bigcup_{x \in \Var(\pi)}S^x}$, we have $Z^{x_1}_{\ell}=Y^{x_1}_{\ell}
{\big|}_{\bigcup_{x \in \Var(\pi)}S^x}$.
Consequently, $\alpha_1 = Y^{x_1}_{\ell} = \rho_1$.
\item[Assertion (b):] An argument similar to that in the proof of Assertion (a) 
yields $Z^{x_n}_r = Y^{x_n}_r{\big|}_{\bigcup_{x \in \Var(\pi)}S^x}$.
Furthermore, since, if $\rho_n \neq \ve$, $S^{x_n}$ must contain the last
variable occurring in $\rho_n$ ($=Y^{x_n}_r$), the shortest prefix of $Y^{x_n}_r$
whose restriction to $\bigcup_{x \in \Var(\pi)}S^x$ equals $Z^{x_n}_r$
is $Y^{x_n}_r$.  Therefore $\beta_n = Y^{x_n}_r = \rho_n$.
\item[Assertion (c):] Suppose that for some $x_i,x_j \in \Var(\pi)$ and $Z \in Y^*$, $x_i Z x_j$ 
is a substring of $\tau$. 
One must show $\beta_i \alpha_j = Z$.      

First, suppose $Z^{x_i}_r = \ve$.  Then $Z^{x_j}_{\ell} = Z{\big|}_{x\in\Var(\pi)S^x}$.
Since $Z^{x_j}_{\ell}$ is a suffix of $Y^{x_j}_{\ell}{\big|}_{x \in \Var(\pi) S^x}$, it follows that
$Z{\big|}_{x\in\Var(\pi)S^x}$ is a suffix of $Y^{x_j}_{\ell}{\big|}_{x \in \Var(\pi) S^x}$.  As 
$Y^{x_j}_{\ell}$ is a suffix of $Z$, one also has that
$Y^{x_j}_{\ell}{\big|}_{x \in \Var(\pi) S^x}$ is a suffix of $Z{\big|}_{x\in\Var(\pi)S^x}$,
and therefore $Y^{x_j}_{\ell}{\big|}_{x \in \Var(\pi) S^x} = Z{\big|}_{x\in\Var(\pi)S^x}$.
If $Y^{x_j}_{\ell} \neq Z$, then there is some $y \in Y$ immediately preceding $Y^{x_j}_{\ell}$ in $Z$ such that 
$y \in S^{x_j}$, implying $Z{\big|}_{x\in\Var(\pi)S^x} \neq Y^{x_j}_{\ell}{\big|}_{x \in \Var(\pi) S^x}$.  
Hence $Y^{x_j}_{\ell} = Z$.  Since $\alpha_j$ is the longest suffix of $Y^{x_j}_{\ell}$ whose restriction to
$\bigcup_{x\in\Var(\pi)}S^x$ equals $Z^{x_j}_{\ell}$ and 
$$
Y^{x_j}_{\ell}{\big|}_{\bigcup_{x\in\Var(\pi)}S^x} = Z{\big|}_{x\in\Var(\pi)S^x} = Z^{x_j}_{\ell},
$$
we have $\alpha_i = Y^{x_j}_{\ell}$.
Furthermore, since $\beta_i$ is the shortest prefix of 
$Y^{x_i}_r$ whose restriction to $\bigcup_{x\in\Var(\pi)}S^x$
equals $Z^{x_i}_r$, one has $\beta_i = \ve$, and so
$\beta_i \alpha_j = Y^{x_j}_{\ell} = Z$. 

Second, suppose $Z^{x_i}_r \neq \ve$.  Then $Y^{x_i}_r \neq \ve$.
Recall that $\alpha_j$ is the longest suffix of $Y^{x_j}_{\ell}$ whose
restriction to $\bigcup_{x \in \Var(\pi)}S^x$ equals $Z^{x_j}_{\ell}$,
and that $Y^{x_j}_{\ell}$ is a suffix of $Z$.  Let $Z = \ga\alpha_j$, where 
$\ga \in Y^*$.  Since $Z^{x_i}_r \neq \ve$, $\ga \neq \ve$.
In particular, note that $\ga[|\ga|] \in \bigcup_{x\in\Var(\pi)}S^x$ due to
the following reasons: if $\alpha_j = Y^{x_j}_{\ell}$, then $\ga[|\ga|] \in S^{x_j}$ 
by the definition of $S^{x_j}$; if $\alpha_j$ were a proper suffix of $Y^{x_j}_{\ell}$
and $\ga[|\ga|] \notin \bigcup_{x\in\Var(\pi)}S^x$, then $\ga[|\ga|]\alpha_j$
would be a suffix of $Y^{x_j}_{\ell}$ longer than $\alpha_j$ whose restriction
to $\bigcup_{x\in\Var(\pi)}S^x$ equals $Z^{x_j}_{\ell}$.   
Thus $\ga[|\ga|]$ is equal to the last symbol of $Z^{x_i}_r$; denote this symbol by $y$.
One has $\#(y)[\alpha_j] = \#(y)[Z^{x_j}_{\ell}]$ and thus $\#(y)[\ga] =
\#(y)[Z] - \#(y)[\alpha_j] = \#(y)[Z^{x_i}_r] + \#(y)[Z^{x_j}_{\ell}] - \#(y)[\alpha_j] = \#(y)[Z^{x_i}_r]$.
It follows that $\ga$ is the shortest prefix of $Y^{x_i}_r$ whose restriction
to $\bigcup_{x\in\Var(\pi)}S^x$ equals $Z^{x_i}_r$, which means that
$\ga = \beta_i$.    
\item[Assertion (d):] Suppose that for some $\delta \in \Sigma$, $Z \in Y^*$ and $x_i \in \Var(\pi)$,
$\delta Z x_i$ is a substring of $\tau$.  One must show $\alpha_i = Z$.
As was proven earlier, $Y^{x_i}_{\ell} = Z$.  Hence $Z^{x_i}_{\ell} = Z{\big|}_{\bigcup_{x\in\Var(\pi)}S^x} 
= Y^{x_i}_{\ell}{\big|}_{\bigcup_{x\in\Var(\pi)}S^x}$.  Since $\alpha_i$ is the 
longest suffix of $Y^{x_i}_{\ell}$ with $\alpha_i{\big|}_{\bigcup_{x\in\Var(\pi)}S^x}
= Z^{x_i}_{\ell}$, one has $\alpha_i = Y^{x_i}_{\ell} = Z$.     
\item[Assertion (e):] Suppose that for some $\delta \in \Sigma$, $Z \in Y^*$ and $x_i \in \Var(\pi)$,
$x_iZ\delta$ is a substring of $\tau$.  One must show $\beta_i = Z$.  First,
$Y^{x_i}_r = Z$ was proven earlier.  As in the proof of Assertion (d), $Z^{x_i}_r = Z{\big|}_{\bigcup_{x\in\Var(\pi)}S^x} 
= Y^{x_i}_r{\big|}_{\bigcup_{x\in\Var(\pi)}S^x}$.  Furthermore, since $S^{x_i}$ contains
the last symbol of $Z$, $Y^{x_i}_r$ is the shortest prefix of $Y^{x_i}_r$ ($=Z$) whose restriction
to $\bigcup_{x\in\Var(\pi)}S^x$ equals $Z^{x_i}_r$.
Hence $\beta_i = Y^{x_i}_r = Z$.~~~~~~~~~~~~~~~\qed  
\end{description}
\end{description}
}

\begin{thm}\label{thm:subclassfinitetd}
Suppose $m \geq 1$.
\begin{enumerate}[label=(\roman*)]
\item $\TD(\Pi^1_{\infty,m}) \leq 2^m + m + 1$
and for all $\pi \in \Pi^{\infty}_{k,m}$ with $k \geq 1$, $\TD(\pi,\Pi^{\infty}_{\infty,m}) = O((D+1)^D)$,
where $D := (4mk+|\pi|+2)\cdot m$.
\item 
Let $1\Pi^z_m$ denote the class of patterns $\pi$ over any alphabet of size $z$ such
that $\pi$ contains at most one variable that occurs more than $m$ times.  Suppose
$\pi \in 1\Pi^z_m$.
If $z \geq 4$, then 
$\TD(\pi,1\Pi^z_m) < \infty$ only if $\pi$ contains a variable that occurs more than
$m$ times or $\pi \in \SR^z$. 
If $z = \infty$, then 
$\TD(\pi,1\Pi^z_m) < \infty$ if $\pi$ contains a variable that occurs more than
$m$ times or $\pi \in \SR^z$. 
\end{enumerate}
\end{thm}

\def\proofthmsubclassfinitetd{
\proof
\noindent\emph{Assertion (i).}  Suppose $\Sigma = \{0\}$.  Then every $\pi \in \Pi^1_{\infty,m}$
is equivalent to a pattern of the shape $0^k x_1^{p_1}\ldots x_n^{p_n}$,
where $k \geq 0$ and $0 \leq p_1 < \ldots < p_n \leq m$.  The constant part of
$\pi$ may be taught using the sample $\{(0^k,+)\} \cup \{(0^{k-i},-): 1 \leq i \leq \min(\{k,m\})\}$.
Furthermore, for each $k$, there are at most $\sum_{i=0}^m {m \choose i} =2^m$ many patterns $\pi'$ of the shape
$0^k x_1^{p'_1} \ldots x_{\ell}^{p'_{\ell}}$, where $0 \leq p'_1 < \ldots < p'_{\ell} \leq m$.
For each such pattern $\pi'$ with $L(\pi') \neq L(\pi)$, $\pi'$ can be
distinguished from $\pi$ using a word in the symmetric difference of $L(\pi)$
and $L(\pi')$.  It follows that $\pi$ has a 
teaching set of size at most $2^m+m+1$, as required.

Now suppose $|\Sigma| = \infty$ and $\Sigma \sm \Const(\pi) = \{a_1,a_2,a_3,\ldots\}$.  
Let $k$ be the number of distinct variables in $\pi$.
We build a teaching set $T$ for $\pi$ w.r.t.\ $\Pi^{\infty}_{\infty,m}$.  
Let $\tau$ denote any pattern in $\Pi^{\infty}_{\infty,m}$ that is consistent with $T$.  
Given $\pi = X_1c_1X_2c_2\ldots c_{n-1}X_n \in \Pi^{\infty}_{\infty,m}$, 
where $X_1,X_2,\ldots,X_n \in X^*$ and $c_1,c_2,\ldots,c_{n-1}$ $\in \Sigma^+$,
put all $O(2^{|\pi(\ve)|})$ elements of $\{(\pi(\ve),+)\} \cup \{(v,-): v \sqsubset \pi(\ve)\}$ into $T$; 
these examples ensure that $\tau(\ve) = \pi(\ve)$.  Next, set $w = \pi[x_i \ra a_i: x_i 
\in \Var(\pi)]$ and put $(w,+)$ into $T$.
Then $w \in L(\tau)$ implies there is a substitution $g:X \mapsto \Sigma^*$
such that for some $S \subseteq \Var(\tau)$, $g(\tau{\big|}_{S}) = w$
and $g(x) \neq \ve$ for all $x \in \tau{\big|}_S$.  Fix such an $S$.
Let $g'$ be a morphism such that $g'(a_i) = x_i$ for all $i$; one has 
$(g'\circ g)(\tau{\big|}_S) = \pi$, and
so $L(\pi) \subseteq L(\tau{\big|}_S) \subseteq L(\tau)$.    
There are at most 
$O((1+|\pi|)^{|\pi|})$ patterns $\tau'$ (up to equivalence) such that for 
some substitution $h:X \mapsto \Sigma^*$, $h(\tau') = w$ and $h(x) \neq \ve$ 
for all $x \in \Var(\tau')$; note that each such $\tau'$ satisfies 
$L(\pi) \subseteq L(\tau')$.  For each such $\tau'$ with $L(\tau') \supset L(\pi)$,
pick $w_{\tau'} \in L(\tau') \sm L(\pi)$ and put $(w_{\tau'},-)$ into $T$.
The latter negative examples ensure that $L(\pi) = L(\tau{\big|}_S)$.  Moreover, 
since $\pi$ is succinct and $L(\pi) = L(\tau{\big|}_S)$, it follows from Lemma 
\ref{lem:succinctinfinitealphabet} that $\tau{\big|}_S$ is equal to $\pi$ up to 
a renaming of variables.  Thus, up to a renaming of variables, $\tau \in \pi \shuffle Y^*$
for some infinite set $Y$ of variables with $Y \cap \Var(\pi) = \emptyset$.        
By Lemma \ref{lem:subsetrestrictedvarwitness}, there exists some $\tau' \in 
\Pi^{\infty}_{4mk+|\pi|+2,m}$ such that $\tau' = \tau{\big|}_{S'}$ for some 
finite $S' \subseteq Y$, and if $L(\pi) \subset L(\tau)$, then $L(\pi) \subset L(\tau')$.
For every $\tau'' \in \left(\Pi^{\infty}_{4mk+|\pi|+2,m}\right) \cap
\pi \shuffle Y^*$ with $\tau''(\ve) = \pi(\ve)$ and $L(\tau'') \supset L(\pi)$, pick some 
$w_{\tau''} \in L(\tau'')\sm L(\pi)$
and put $(w_{\tau''},-)$ into $T$; there are at most $O((D+1)^D)$ many 
such $\tau''$ (up to equivalence), where $D := (4mk+|\pi|+2)\cdot m$.  These negative examples ensure that
$L(\tau) \not\supset L(\pi)$.  Therefore $L(\tau) = L(\pi)$, which proves that
$T$ is indeed a teaching set of size $O((D+1)^D)$ for $\pi$ w.r.t.\ $\Pi^{\infty}_{\infty,m}$,
where $D:= (4mk+|\pi|+2)\cdot m$.
}

\medskip
\noindent
The next result shows that over binary alphabets, even the class
of constant-free $4$-regular pattern languages contains patterns with infinite \TD. 
We prove this by modifying Reidenbach's \cite{Reidenbach06} proof of the non-learnability 
of $x_1^2x_2^2x_3^2$ so that every pattern constructed in the proof has variable frequency 
at most $4$.

\begin{thm}(Based on \cite[Theorem 5]{Reidenbach06})\label{thm:sizetwoboundedvarfreq}
Suppose $\pi = x_1^2 x_2^2 x_3^2$.  For any $m \geq 4$, $\TD(\pi,\Pi^2_{\infty,m,cf}) = \infty$.
\end{thm} 

\def\proofthmsizetwoboundedvarfreq{
\proof
Suppose $\Sigma = \{0,1\}$.  Let $\{x_{i,j}: i,j \in \natnum_0\}$ and $\{y_{i,j}: i,j \in \natnum_0\}$
be two disjoint infinite sets of variables.
It suffices to show that $\pi$ does not possess a finite tell-tale w.r.t.\ $\Pi^2_{\infty,4,cf}$ (i.e.\
a finite set $S \subseteq L(\pi)$ such that for all $\tau \in \Pi^2_{\infty,4,cf}$, 
one has $S \subseteq L(\tau) \subseteq L(\pi)\Rightarrow L(\tau) = L(\pi)$).
Following the proof in \cite{Reidenbach06}, assume, by way of contradiction,
that $\pi$ has a finite tell-tale $\{w_1,\ldots,w_n\}$ for some $n \geq 1$.
Without loss of generality, assume that $w_i \neq \ve$ for all $i \in \{1,\ldots,n\}$.
For each $i \in \{1,\ldots,n\}$, there is a substitution $\sigma_i: X \mapsto \Sigma^*$
witnessing $\sigma_i(\pi) = w_i$.
Set $\tilde{\sigma}_i(\pi) := \sigma_i(x_1)\sigma_i(x_2)\sigma_i(x_3)$.
We define, for each $i \in \{1,\ldots,n\}$, patterns $\ga_{i,1},\ga_{i,2}$ and $\ga_{i,3}$
according to the following case distinction.
\begin{description}[leftmargin=0cm]
\item[Case 1:] There is some $\delta \in \Const(w_i)$ such that the last occurrence
of $\delta$ in $\tilde{\sigma}_i(\pi)$ is strictly before the $(|\sigma_i(x_1)|+1)$-st
position of $\tilde{\sigma}_i(\pi)$.  Set $\ga_{i,1} = \ve$.
Let $\ell_1 := \left|\tilde{\sigma}_i(\pi){\big|}_{\{\delta\}}\right|$
$\left(\mbox{resp.~}\ell_2 := \left|\tilde{\sigma}_i(\pi){\big|}_{\{\overline{\delta}\}}\right|\right)$, 
i.e.\ $\ell_1$ (resp.~$\ell_2$) is the number of occurrences of $\delta$
(resp.~$\overline{\delta}$) in $\tilde{\sigma}_i(\pi)$.  Note that the case
assumption implies $\sigma_i(x_2)\sigma_i(x_3) \in \{\overline{\delta}\}^*$.     
Suppose $\ell_1 = 2p_1+r_1$ and $\ell_2 = 2p_2+r_2$ for some $p_1,p_2 \geq 0$
and $r_1,r_2 \in \{0,1\}$.  Let $\tau_i$ be the pattern derived from $\tilde{\sigma}_i(\pi)$
as follows: if $p_1 \geq 1$ (resp.~$p_2 \geq 1$), then for all $j \in \{0,\ldots,p_1-1\}$
(resp.~$j \in \{0,\ldots,p_2-1\}$), substitute $x_{i,j}$ (resp.~$y_{i,j}$) for
the $(2j+1)$-st and $(2j+2)$-nd occurrences of $\delta$ (resp.~$\overline{\delta}$)
in $\tilde{\sigma}_i(\pi)$, and if $r_1 = 1$ (resp.~$r_2 = 1$), then substitute $x_{i,p_1}$ 
(resp.~$y_{i,p_2}$) for the $(2p_1+1)$-st (resp.~$(2p_2+1)$-st) occurrence of $\delta$ 
(resp.~$\overline{\delta}$) in $\tilde{\sigma}_i(\pi)$.  

Define $\ga_{i,2}$ to be the prefix of $\tau_i$ of length $|\sigma_i(x_1)|$ and
define $\ga_{i,3}$ to be the suffix of $\tau_i$ of length $|\sigma_i(x_2)\sigma_i(x_3)|$.

\item[Case 2:] Not Case 1.  Then for all $\delta \in \Const(w_i)$, the position of
the last occurrence of $\delta$ in $\tilde{\sigma}_i(\pi)$ is greater than 
$|\tilde{\sigma}_i(x_1)|$.
Let $\ell'_1 := \left|\tilde{\sigma}_i(\pi){\big|}_{\{0\}}\right|$
$\left(\mbox{resp.~}\ell'_2 :=\right.$ $\left.\left|\tilde{\sigma}_i(\pi){\big|}_{\{1\}}\right|\right)$.
Suppose $\ell'_1 = 2q_1+s_1$ and $\ell'_2 = 2q_2+s_2$ for some $q_1,q_2 \geq 0$
and $s_1,s_2 \in \{0,1\}$.  As in Case 1, let $\tau_i$ be the pattern derived from
$\tilde{\sigma}_i(\pi)$ as follows: if $q_1 \geq 1$ (resp.~$q_2 \geq 1$), then for all 
$j \in \{0,\ldots,q_1-1\}$ (resp.~$j \in \{0,\ldots,q_2-1\}$), substitute $x_{i,j}$ (resp.~$y_{i,j}$) for
the $(2j+1)$-st and $(2j+2)$-nd occurrences of $0$ (resp.~$1$)
in $\tilde{\sigma}_i(\pi)$, and if $s_1 = 1$ (resp.~$s_2 = 1$), then substitute $x_{i,q_1}$ 
(resp.~$x_{i,q_2}$) for the $(2q_1+1)$-st (resp.~$(2q_2+1)$-st) occurrence of $0$ 
(resp.~$1$) in $\tilde{\sigma}_i(\pi)$.  

Define $\ga_{i,1}$ to be the prefix of $\tau_i$ of length $|\sigma_i(x_1)|$,
define $\ga_{i,2}$ to be the substring of $\tau_i$ of length $|\sigma_i(x_2)|$
that starts at the $(|\sigma_i(x_1)|+1)$-st position of $\tau_i$, and
define $\ga_{i,3}$ to be the suffix of $\tau_i$ of length $|\sigma_i(x_3)|$.     
\end{description}
Set 
$$
\tau := (\ga_{1,1}\ldots\ga_{n,1})^2 (\ga_{1,2}\ldots\ga_{n,2})^2 (\ga_{1,3}\ldots\ga_{n,3})^2. 
$$
In order to derive a contradiction, it will be shown that $(a) \{w_1,\ldots,w_n\}
\subseteq L(\tau)$ and (b) $L(\tau) \subset L(\pi)$.\footnote{In \cite{Reidenbach06},
$\tau$ is known as a passe-partout for $\pi$ and $\{w_1,\ldots,w_n\}$.}

\smallskip
\noindent\emph{Proof of (a).} For $i \in \{1,\ldots,n\}$, let
$\varphi_i: X \mapsto \Sigma^*$ be the morphism defined as follows.
If $\sigma_i$ falls into Case 1, let $\delta$ be a letter as defined in Case 1
for $\sigma_i$.  For all $j \in \natnum_0$, set $\varphi_i(x_{i,j}) = \delta$ and 
$\varphi_i(y_{i,j}) = \overline{\delta}$.  For all $i' \neq i$ and $j \in \natnum_0$,
set $\varphi_i(x_{i',j}) = \varphi_i(y_{i',j}) = \ve$.  It may be directly
verified that $\varphi_i(\tau) = w_i$.

Suppose $\sigma_i$ falls into Case 2.  For all $j \in \natnum_0$, set $\varphi_i(x_{i,j}) = 0$ and 
$\varphi_i(y_{i,j}) = 1$.  For all $i' \neq i$ and $j \in \natnum_0$,
set $\varphi_i(x_{i',j}) = \varphi_i(y_{i',j}) = \ve$.  Then $\varphi_i(\tau) = w_i$.  

\smallskip
\noindent\emph{Proof of (b).} By Theorem \ref{thm:jiang94suffcondsubset},
it is enough to show that there is a morphism $\psi: X^* \mapsto X^*$
such that $\psi(\pi) = \tau$ but there does not exist any morphism
$\theta: X^* \mapsto X^*$ for which $\theta(\tau) = \psi$.
For the first part, define, for each $i \in \{1,2,3\}$, the substitution
$\psi(x_i) = \ga_{1,i} \ldots \ga_{n,i}$.  It follows that $\psi(\pi) = \tau$.
For the second part, we first note that by construction, every variable
of $\tau$ that occurs exactly twice must belong to $\Var(\ga_{1,2}\ldots\ga_{n,2}
\ga_{1,3}\ldots\ga_{n,3})$.  Consequently, for all morphisms $\theta: X^*
\mapsto X^*$, if $\theta(\tau)$ contains exactly three variables, each of 
which occurs exactly twice, then $\theta(\tau)$ is equivalent to one of the
following patterns: $x_1x_2x_1x_2x_3^2$, or $x_1x_2x_3x_1x_2x_3$,
or $x_1^2 x_2 x_3 x_2 x_3$. 
Thus $\theta(\tau)$ cannot be equivalent to $\pi$.

We conclude from (a) and (b) that $\{w_1,\ldots,w_n\}$ cannot be a tell-tale 
for $\pi$ w.r.t.\ $\Pi^2_{\infty,4,cf}$, contrary to assumption.
~\qed
}

\begin{rem}\label{rem:noncrosslowerbound}
The lower bound $4$ on $m$ in Theorem \ref{thm:sizetwoboundedvarfreq} is tight
in the sense that the \TD\ 
of $\pi := x_1^2 x_2^2 x_3^2$ 
w.r.t.\ $\Pi^2_{\infty,3}$ is finite.  
In fact, $T := \{(\ve,+),(0^2 1^2 0^2,+),(0,-),$ $(01^20,$ $-),(0^3,-),((01)^2(0^21)^2(0^31)^2(0^41)^2,-)\}$ 
is a teaching set for $\pi$ w.r.t.\ $\Pi^2_{\infty,3}$ (further details are
given in Appendix \ref{appen:binarynoncrosslowerbound}). 
\end{rem}

\section{Conclusion}

Table \ref{tbl:summ} summarises some of the main results of this paper.
For three types of pattern classes studied -- the simple block-regular,
$m$-quasi-regular and $m$-regular non-cross patterns -- it was found
that over any alphabet size, every pattern in the class is finitely distinguishable;
in the case of simple block-regular and $m$-regular non-cross patterns, one 
also has an upper bound on the \TD\ of the class of such patterns that is,
depending on the alphabet size, constant, linear or sublinear in $m$.
The most delicate questions appear to be those concerning the $m$-regular patterns for finite alphabets 
of size at least $2$; we only know that for all $m \geq 4$, there are patterns in 
$\Pi^2_{\infty,m,cf}$ that are not finitely distinguishable (and even not learnable in the limit).
We note that the class of non-cross patterns over any
alphabet and the class of all patterns over infinite alphabets are 
learnable in the limit\footnote{This implies that for every pattern $\pi$ belonging to any one of these classes,
$L(\pi)$ contains a finite set that distinguishes $\pi$ from all $\pi'$ in the class such that 
$L(\pi') \subset L(\pi)$ \cite[Theorem 1]{Ang80}.} \cite{Reidenbach06,Mit98}, but they have relatively restricted subclasses
of finitely distinguishable patterns \cite[Theorems 3,10]{bayeh17}.  Thus the
fact that every pattern in the $m$-regular versions of these classes has a finite
\TD\ suggests that the variable frequency of a pattern class may play
a role in determining whether any given pattern $\pi$ can be finitely distinguished
from all $\pi'$ such that $L(\pi') \not\subseteq L(\pi)$. 
On the other hand, we have seen in Theorem \ref{thm:subclassfinitetd}(ii) that
even constant patterns cannot be finitely distinguished w.r.t.\ the class of patterns with at 
most one variable (but no uniform upper bound on the number of variable occurrences). 
It might be interesting to know whether there is a `natural' class $\Pi$ of $m$-regular patterns such that $\Pi$ is
learnable in the limit but $\TD(\pi,\Pi) = \infty$ for some $\pi \in \Pi$.
We also suspect that $\TD(\Pi^{\infty}_{\infty,m}) = \infty$ for some $m \geq 2$ and $\TD(\QR^z_{\infty,m}) = \infty$
for some finite $z \geq 2$ and $m \geq 1$, but as yet do not know how to prove this.

\medskip
\noindent\textbf{Acknowledgements.} The author
was supported (as RF) by the Singapore Ministry
of Education Academic Research Fund grant MOE2016-T2-1-019 / R146-000-234-112.
I sincerely thank Fahimeh Bayeh, Sanjay Jain and
Sandra Zilles for proofreading the manuscript; their numerous suggestions for corrections
and improvements are gratefully acknowledged.  I also thank Fahimeh Bayeh very much for her
suggestion to look at the \PBTD\ of $m$-quasi-regular patterns over unary 
alphabets.   

\captionsetup[table]{skip=10pt}
\begin{center}
\begin{table}
\centering
\begin{tabular}{|c||c|c|c|}
\hline
\multirow{1}{*}{~}&{\scriptsize{$z = 1$}}&\makebox[1.5cm]{\scriptsize{$2 \le z <\infty$}}&\makebox[1.5cm]{\scriptsize{$z=\infty$}}\\\hline\hline
\multirow{2}{*}{\makebox[1.5cm]{\scriptsize{$\SR^z$}}}&\scriptsize{$\TD = 2$,} 
&\scriptsize{$\TD = 2$,} 
&\scriptsize{$\TD = 2$,} 
\\
&\scriptsize{$\PBTD = 1$ (Thm \ref{thm:tdsimpleblockregular})}
&\scriptsize{$\PBTD = 1$ (Thm \ref{thm:tdsimpleblockregular})}
&\scriptsize{$\PBTD = 1$ (Thm \ref{thm:tdsimpleblockregular})}
\\\hline
\multirow{4}{*}{\makebox[1.5cm]{\scriptsize{$\QR^z_{\infty,m}$}}}&\scriptsize{$\TD = 3$}& 
\scriptsize{$(\forall\pi)[\TD(\pi,\Pi)<\infty]$}&\scriptsize{$(\forall\pi)[\TD(\pi,\Pi)<\infty]$}
\\
& \scriptsize{(Thm \ref{thm:cfqrplfinitetd})}
& \scriptsize{(Thm \ref{thm:cfqrplfinitetd})}
& \scriptsize{(Thm \ref{thm:cfqrplfinitetd})}
\\
&\scriptsize{$\PBTD= 2$ (Prop \ref{prop:pbtdqrunary})}&\scriptsize{$\PBTD \geq 2$ (Prop \ref{prop:pbtdqrunary})}&
\scriptsize{$\PBTD= 2$ (Prop \ref{prop:pbtdqrunary},} \\
& ~
& ~
& \scriptsize{Thm \ref{thm:pbtdgeneralinfinite})}
\\\hline
\multirow{4}{*}{\makebox[1.5cm]{\scriptsize{$\NC\Pi^z_{\infty,m}$}}}&\scriptsize{$\TD/\PBTD = \Theta(m), m \geq 2$,} & 
\scriptsize{$\TD = o(m)$,} 
&\scriptsize{$\TD = o(m)$,}\\
&\scriptsize{$\TD/\PBTD = 0, m = 1$}& 
\scriptsize{$\PBTD= 1, m \geq 2$,}&\scriptsize{$\PBTD = 0, m = 1$}\\ 
&\scriptsize{(Thm \ref{thm:tdnoncrossbounded})}&\scriptsize{$\PBTD = 0, m = 1$}
&\scriptsize{(Thm \ref{thm:tdnoncrossbounded})} \\
&~&\scriptsize{(Thm \ref{thm:tdnoncrossbounded})}&~\\
\hline
\multirow{4}{*}{\makebox[1.5cm]{\scriptsize{$\Pi^z_{\infty,m}$}}}&
\scriptsize{$\TD = O(2^m)$} & 
\scriptsize{$(\exists\pi)[\TD(\pi,\Pi^2_{\infty,4,cf}) = \infty]$}&
\scriptsize{$(\forall\pi)[\TD(\pi,\Pi)<\infty]$}
\\
& \scriptsize{(Thm \ref{thm:subclassfinitetd}(i))}
& \scriptsize{(Thm \ref{thm:sizetwoboundedvarfreq})}
& \scriptsize{(Thm \ref{thm:subclassfinitetd}(i))}
\\
&\scriptsize{$\PBTD = \Theta(m)$ (Thm \ref{thm:pbtdgeneralinfinite})}&
\scriptsize{$\PBTD \geq 2$ (Prop \ref{prop:pbtdqrunary})}&
\scriptsize{$\PBTD= 2$ (Prop \ref{prop:pbtdqrunary},}\\
&~&~&\scriptsize{Thm \ref{thm:pbtdgeneralinfinite})}\\\hline
\end{tabular}
\caption{\scriptsize{\TD\ and \PBTD\ of various pattern classes.  In each entry, $m \ge 1$, the universal (resp.~existential) quantifier is taken over all
patterns belonging to the class in the corresponding row and $\Pi$ refers to the class in the corresponding row.}}
\label{tbl:summ}
\end{table}
\end{center}


%

\vspace*{-0.5in}

\bibliographystyle{plain}


\section*{Appendix}

\newtheorem{exmps}{Example} [subsection]
\renewcommand{\theexmps}{\thesubsection.\arabic{exmps}}

\newtheorem{lems}[exmps]{Lemma} 
\renewcommand{\thelems}{\thesubsection.\arabic{lems}}

\newtheorem{rems}[exmps]{Remark} 
\renewcommand{\therems}{\thesubsection.\arabic{rems}}

\newtheorem{notas}[exmps]{Notation} 
\renewcommand{\thenotas}{\thesubsection.\arabic{notas}}

\newtheorem{subclaim}[exmps]{Claim} 
\renewcommand{\thesubclaim}{\thesubsection.\arabic{subclaim}}

\addcontentsline{toc}{section}{Appendices}
\renewcommand{\thesubsection}{\Alph{subsection}}

\noindent
This appendix contains the proofs not presented in the main part of the paper
as well as additional definitions/notation and examples.

\subsection{Additional Definitions and Notation}

In this section, we introduce additional definitions and notation needed
for the proofs in the appendix.
 
Given any $x \in X$, let $\N(x,\pi)$ denote the set of all $s \in X \cup \Sigma$
such that $s$ is adjacent to an occurrence of $x$ in $\pi$; call $\N(x,\pi)$
the \emph{neighbourhood} of $x$ in $\pi$.
For each $\delta \in \Sigma$ and $w \in (X \cup \Sigma)^*$,  
$\#(\delta)[w]$ denotes the number of occurrences of $\delta$ in $w$.  

If $|\Sigma| = 2$ and
$\delta \in \Sigma$, then $\overline{\delta}$ denotes the unique element
of $\Sigma \sm \{\delta\}$. 
For any $\pi \in (X \cup \Sigma)^+$ and
variables $x_{i_1},\ldots,x_{i_n}$ occurring in $\pi$, let
$\pi[x_{i_1} \ra \alpha_1, \ldots, x_{i_n} \ra \alpha_n]$ denote
the word obtained from $\pi$ by substituting $\alpha_j$ for $x_{i_j}$
whenever $j \in \{1,\ldots,n\}$ and substituting $\ve$ for every other
variable. 
We will often assume that a pattern $\pi \in \Pi^z$ is \emph{normalised} in the sense
that the $k$ variables occurring in $\pi$ are
named $x_1,\ldots,x_k$ in order of their first occurrences from left to right (or $x$ if $k = 1$). 

Given any pattern $\pi$ and substitution $h:X \mapsto 
\Sigma^*$, $h$ induces a mapping of closed intervals of positions of 
$\pi$ to closed intervals of positions of $h(\pi)$.  This mapping will 
be denoted by $\cI_{h,\pi}$.  For any position $p$ of $\pi$, 
$\cI_{h,\pi}(\{p\})$ will simply be written as $\cI_{h,\pi}(p)$.  We 
define the \emph{inverse} of $\cI_{h,\pi}$, denoted $\overline{\cI}_{h,\pi}$, 
to be the mapping of closed intervals of positions of $h(\pi)$ to closed 
intervals of positions of $\pi$ such that for all closed intervals $J \subseteq 
\{1,\ldots,|h(\pi)|\}$, $\overline{\cI}_{h,\pi}(J)$ is the smallest closed 
interval $I \subseteq \{1,\ldots,|\pi|\}$ such that $J \subseteq 
\cI_{h,\pi}(I)$ (in other words, $J \subseteq \cI_{h,\pi}(I)$ and for all 
$I' \subset I$, $J \not\subseteq \cI_{h,\pi}(I')$).  For any position $q$ of 
$h(\pi)$, $\overline{\cI}_{h,\pi}(\{q\})$ will be abbreviated to 
$\overline{\cI}_{h,\pi}(q)$.

Fix any $z = |\Sigma| \geq 1$ and $\pi \in \Pi^z$.
Suppose that $\ga \in L(\pi)$ for some $\ga \in \Sigma^*$,
as witnessed by the substitution $h:X \mapsto \Sigma^*$. 
We define a \emph{cut} of $\ga$ \emph{relative to $(h,\pi)$} to be 
any pair $(I_1,I_2)$ of disjoint nonempty closed intervals of positions 
of $\ga$ such that $I_1 = [r_1,r_2]$ and $I_2 = [r_2+1,r_3]$ for some 
$r_1,r_2,r_3 \in \{1,\ldots,|\ga|\}$, and there exists 
$q \in \{1,\ldots,|\pi|\}$ with $\mathcal{I}_{h,\pi}(q) = I_1$ and
$\mathcal{I}_{h,\pi}(q+1) = I_2$.  If $(I_1,I_2)$ is a cut of $\ga$
relative to $(h,\pi)$, then the right endpoint of $I_1$ (which is one less 
than the left endpoint of $I_2$) will be called a \emph{cut-point} of $\ga$
\emph{relative to $(h,\pi)$}.  If the choice of $(h,\pi)$ is clear 
from the context, then $(I_1,I_2)$ (resp.~the right endpoint of $I_1$) will 
simply be called a \emph{cut} of $\ga$ (resp.~\emph{cut-point} of $\ga$).    

\begin{exmps}\cite{bayeh18}\label{exmp:defncut}
Let $\pi = x_1 x_2 x_1 x_2 x_1$ and $\ga = 01 11 01 11 01$.
Then $h:X \mapsto \Sigma^*$, defined by $h(x_1) = 01$ and $h(x_2) = 11$,
witnesses $\ga \in L(\pi)$.  One has that
\begin{equation}\label{eqn:gacutexamp}
\ga = \underbrace{\overbrace{01}^{I_1}}_{h(x_1)} \underbrace{\overbrace{11}^{I_2}}_{h(x_2)} 
\underbrace{01}_{h(x_1)} \underbrace{11}_{h(x_2)} \underbrace{01}_{h(x_1)},
\end{equation} 
and $(I_1,I_2)$ (where the positions of $\ga$ occupied by $I_1$ and $I_2$ are 
illustrated in Equation (\ref{eqn:gacutexamp})) is a cut of $\ga$ relative
to $(h,\pi)$; the corresponding cut-point of $\ga$ relative to $(h,\pi)$ is $2$. 
\end{exmps}

\medskip
\noindent
The following basic lemma elucidates the connection between the number
of cuts of $h(\pi)$ and the length of $\pi$.  It will be useful
in subsequent results for showing that $L(\pi)$ cannot contain certain 
words.     

\begin{lems}\cite{bayeh18}\label{lem:numberofcutsandpositionsrelation}
If $\ga$ has $d$ distinct cuts relative to $(h,\pi)$, then $\left|\pi\right| \geq d+1$.
\end{lems}

\proof
Given any two consecutive cuts $(I_1,I_2)$ and $(J_1,J_2)$ of $\ga$ such that
the left endpoint of $I_1$ is smaller than the left endpoint of $J_1$, 
$I_1 \neq I_2$ and $I_2 \neq J_2$ together imply that $I_2 \neq J_2$.  Hence $I_1$,
$I_2$ and $J_2$ correspond to three different positions of $\pi$.~\qed 

\subsection{Example of the Mappings $\cI$ and $\overline{\cI}$}

\begin{exmps}\cite{bayeh18}\label{exmp:inducedmappositions}
Suppose $\Sigma = \{a,b\}$ and $\pi = x_1 x_2 x_1 x_2 x_1$.  
Let $h:X \mapsto \Sigma^*$ be the substitution defined by $h(x_1) = ab$ 
and $h(x_2) = bb$.
Then $\ga := h(\pi) \in L(\pi)$ and one has that $\cI_{h,\pi}([1,2]) 
= [1,4]$, $\cI_{h,\pi}([4,5]) = [7,10]$ and $\overline{\cI}_{h,\pi}(5) = \{3\}$.  
\begin{equation*}\label{eqn:inducedmappositions}
\ga = \overbrace{\underbrace{abbb}_{\begin{subarray}{c} \cI_{h,\pi}([1,2]) \\ = ~ [1,4]\end{subarray}}}^{h(x_1x_2)} 
\overbrace{\underbrace{a}_{\begin{subarray}{c} \overline{\cI}_{h,\pi}(5) \\ = ~ \{3\} \end{subarray}} b}^{h(x_1)}
\overbrace{\underbrace{bbab}_{\begin{subarray}{c} \cI_{h,\pi}([4,5]) \\ = ~ [7,10]\end{subarray}}}^{h(x_2x_1)}.
\end{equation*} 
\end{exmps}

\subsection{Proof of Lemma \ref{lem:tdsimplebregpatbinary}}\label{appen:prooflemtdsimplebregpatbinary}

\prooflemtdsimplebregpatbinary

\subsection{Example for Lemma \ref{lem:tdsimplebregpatbinary}}

We illustrate the construction of the teaching set in the proof of Lemma 
\ref{lem:tdsimplebregpatbinary} with the following example.

\begin{exmps}\label{exmp:simpleblockregulartdbinary}
Suppose $\Sigma = \{0,1\}$.  Let $\pi = x_1 0 x_2 0 x_3 1 x_4 1 x_5$.
According to the construction in the proof of Lemma \ref{lem:tdsimplebregpatbinary},
$\pi$ has the teaching set $\{(w_1,+),(w_2,+),$ $(w_3,-)\}$ w.r.t.\ $R\Pi^2$, where 
$w_1,w_2$ and $w_3$ are defined as follows:
($\theta_1$ and $\theta_2$ are substitutions witnessing $w_1 \in L(\pi)$ and 
$w_2 \in L(\pi)$ respectively):  
\begin{itemize}
\item $w_1 = \underbrace{1}_{\theta_1(x_1)}0\underbrace{1}_{\theta_1(x_2)}01\underbrace{0}_{\theta_1(x_4)}1
\underbrace{0}_{\theta_1(x_5)}$;
\item $w_2 = 00 \underbrace{0}_{\theta_2(x_3)} 11 $;
\item $w_3 = \underbrace{0110}_{} \underbrace{1}_{}$.
\end{itemize}   
\end{exmps}

\subsection{Proof of Lemma \ref{lem:tdsimplebregpatatleast3}}\label{appen:tdsimplebregpatatleast3}

\proof
Suppose $\Sigma = \{a_1,a_2,\ldots,a_k\}$, where $k \geq 3$,
and $\pi = x_1a_{i_1}x_2a_{i_2}\ldots a_{i_{n-1}}$ $x_n$, where 
$x_1,x_2,\ldots,x_n \in X$.  If $n = 2$, then one may verify
directly that for any $b \in \Sigma \sm \{a_{i_1}\}$, 
$\{(a_{i_1},+),$ $(ba_{i_1}b,+),(\ve,-)\}$ is a teaching set for
$\pi$ w.r.t.\ $R\Pi^z$.  We assume in what follows that
$n \geq 3$.  Again, $T = \{(w_1,+),(w_2,+),(w_3,-)\}$ 
will denote a teaching set for $\pi$ w.r.t.\ $R\Pi^z$, where
$w_1,w_2$ and $w_3$ are defined below.  Further, $\tau$ will denote a 
regular pattern that is consistent with $T$.

\begin{description}[leftmargin=0cm]
\item[$w_1$:] For every substring of $\pi$ of the shape
$a_{i_j}x_{j+1}a_{i_{j+1}}$, define $\varphi(x_{j+1})$
according to the following case distinction.
\begin{description}
\item[Case i:] $i_j$ and $i_{j+1}$ have opposite parities.
Set $\varphi(x_{j+1}) = \ve$. 
\item[Case ii:] $i_j$ and $i_{j+1}$ have equal parities.
Fix some $j' \in \{1,\ldots,k\}$ such that $j'$ and $i_j$
have opposite parities (which implies that $j'$ and $i_{j+1}$
also have opposite parities), and set $\varphi(x_{j+1}) = a_{j'}$.
For all other variables $x$ occurring in $\pi$, set $\varphi(x) = \ve$. 
\end{description} 
Set $w_1 = \varphi(\pi)$.

\item[$w_2$:] For every substring of $\pi$ of the shape
$a_{i_j}x_{j+1}a_{i_{j+1}}$, define $\psi(x_{j+1})$
according to the following case distinction.
\begin{description}[leftmargin=0cm]
\item[Case i:] $i_j$ and $i_{j+1}$ have equal parities.
Set $\psi(x_{j+1}) = \ve$.
\item[Case ii:] $i_j$ is even and $i_{j+1}$ is odd.
\begin{description}[leftmargin=0cm]
\item[Case ii.1:] 
$j > 1$ and $i_{j-1}$ is even.
Pick any odd $j' \in \{1,\ldots,k\}$ 
such that 
$a_{j'} \neq a_{j+1}$, and set $\psi(x_{j+1}) = a_{j'}$.
\item[Case ii.2:] $j > 1$ and $i_{j-1}$ is odd, or $j = 1$.
Pick any even $j' \in \{1,\ldots,k\}$ 
and pick any odd $j'' \in \{1,\ldots,k\}$ such that 
$a_{j''} \neq a_{i_{j+1}}$, 
and set $\psi(x_{j+1}) = a_{j'}a_{j''}$.  
\end{description}
\item[Case iii:] $i_j$ is odd and $i_{j+1}$ is even.
Pick any odd $j' \in \{1,\ldots,k\}$ such that 
$a_{j'} \neq a_{i_j}$, and set $\psi(x_{j+1}) = a_{j'}$.
\end{description}
Furthermore, pick $j_1,j_2 \in \{1,\ldots,k\}$ such that
$a_{j_1} \notin \{a_{i_1},a_{i_2}\}$ and $a_{j_2} \notin \{a_{i_{n-1}},a_{i_{n-2}}\}$;
set $\psi(x_1) = a_{j_1}$ and $\psi(x_n) = a_{j_2}$.\footnote{Such $j_1$ and $j_2$ must exist since $|\Sigma| \geq 3$.}
For all other variables $x$ occurring in $\pi$, set $\psi(x) = \ve$.
Set $w_2 = \psi(\pi)$.



\item[$w_3$:] Arguing as in the proof of Lemma \ref{lem:tdsimplebregpatbinary},
the consistency of $\tau$ with $(w_1,+)$ and $(w_2,+)$ implies that $\tau$
is of the shape $x_1A_1x_2A_2\ldots A_{k-1}x_k$, where every maximal 
constant block $A_i$ has length at most $2$; furthermore, if $A_i = a_{\ell}a_{\ell'}$,
then $\ell$ and $\ell'$ have opposite parities.

Note that Lemma \ref{lem:simplebrnegative} cannot be directly applied
here since the consistency of $\tau$ with $(w_1,+)$ and $(w_2,+)$
does not imply that $\tau$ is simple block-regular.
We will, however, give a different construction of $w_3$ by analysing
a decomposition of $w_2$ containing subwords $\beta_1,\beta_2,\ldots,\beta_{n-2}$
such that any maximal constant block of $\tau$ is a subword
of some $\beta_j$ (details are to follow).

For each $j \in \{1,\ldots,n-2\}$, define $\beta_j := a_{i_j} \psi(x_{j+1}) a_{i_{j+1}}$.
The positions of $\beta_1,\ldots,\beta_{n-2}$ are illustrated below.

\begin{equation}\label{eqn:w2decomposebeta}
w_2 = \psi(x_1)\overbrace{a_{i_1}\psi(x_2)a_{i_2}}^{\beta_1} \ldots
\overbrace{a_{i_j}\psi(x_{j+1})a_{i_{j+1}}}^{\beta_j} \ldots
\overbrace{a_{i_{n-2}}\psi(x_{n-1})a_{i_{n-1}}}^{\beta_{n-2}}\psi(x_n).
\end{equation} 

Corresponding to each $\beta_j$, where $j \in \{1,\ldots,n-2\}$, we define
a word $\alpha_j$ based on the following case distinction. 

\begin{description}[leftmargin=0cm]
\item[Case i:] $\beta_j = a_{i_j}a_{i_{j+1}}$, where $i_j$ and $i_{j+1}$
have equal parities. 
\begin{description}[leftmargin=0cm]
\item[Case i.1:] $i_j$ and $i_{j+1}$ are even.
\begin{description}[leftmargin=0cm]
\item[Case i.1.1:] $j-1 \geq 1$ and $i_{j-1}$ is odd, $j+2 \leq n-1$
and $i_{j+2}$ is odd.
Then $\psi(x_j) = a_{j'}$ for some odd $j'$ such that $a_{j'} \neq a_{i_{j-1}}$
and $\psi(x_{j+2}) = a_{j''}$ for some odd $j''$ such that $a_{j''} \neq a_{i_{j+2}}$.  
Set
$$
\alpha_j = \left\{\begin{array}{ll}
a_{i_{j+1}}a_{j''}a_{j'}a_{i_j} & \mbox{if $a_{i_j} \neq a_{i_{j+1}}$;} \\
a_{j'}a_{i_j}a_{j''} & \mbox{if $a_{i_j} = a_{i_{j+1}}$.}\end{array}\right. 
$$ 
\item[Case i.1.2:] $j-1 \geq 1$ and $i_{j-1}$ is odd; either 
$j+2 \leq n-1$ and $i_{j+2}$ is even, or $j+2 > n-1$.
Then $\psi(x_j) = a_{j'}$ for some odd $j'$ such that $a_{j'} \neq a_{i_{j-1}}$.  
If $j+2 \leq n-1$ and $i_{j+2}$ is even, define $\alpha_j$ as in Case i.1.1 but 
with all occurrences of $a_{j''}$ deleted.
If $j+2 > n-1$, define $\alpha_j$ as in Case i.1.1 but with all occurrences of
$a_{j''}$ replaced with $\psi(x_n)$ and $\psi(x_n)$ appended to $\alpha_j$.
\item[Case i.1.3:] $j+2 \leq n-1$ and $i_{j+2}$ is odd; either
$j-1 \geq 1$ and $i_{j-1}$ is even, or $j-1 < 1$.  Then
$\psi(x_{j+2}) = a_{j''}$ for some odd $j''$ such that $a_{j''} \neq a_{i_{j+2}}$.
If $j-1 \geq 1$ and $i_{j-1}$ is even, define $\alpha_j$ as in Case i.1.1 but 
with all occurrences of $a_{j'}$ deleted.  If $j-1 < 1$,
define $\alpha_j$ as in Case i.1.1 but with all occurrences of $a_{j'}$
replaced with $\psi(x_1)$ and $\psi(x_1)$ prepended to $\alpha_j$.
\item[Case i.1.4:] $j-1 \geq 1$ and $i_{j-1}$ is even, or $j-1 < 1$;
$j+2 \leq n-1$ and $i_{j+2}$ is even, or $j+2 > n-1$.
If $j-1 \geq 1, j+2 \leq n-1$ and both $i_{j-1},i_{j+2}$ are even, set
$$
\alpha_j = \left\{\begin{array}{ll}
a_{i_{j+1}}a_{i_j} & \mbox{if $a_{i_j} \neq a_{i_{j+1}}$;} \\
a_{i_j} & \mbox{if $a_{i_j} = a_{i_{j+1}}$.}\end{array}\right.
$$
If $j-1 < 1$, set 
$$
\alpha_j = \left\{\begin{array}{ll}
\psi(x_1)a_{i_{j+1}}\psi(x_1)a_{i_j} & \mbox{if $a_{i_j} \neq a_{i_{j+1}}$;} \\
\psi(x_1)a_{i_j} & \mbox{if $a_{i_j} = a_{i_{j+1}}$.}\end{array}\right.
$$
If $j+2 > n-1$, set
$$
\alpha_j = \left\{\begin{array}{ll}
a_{i_{j+1}}\psi(x_n)a_{i_j}\psi(x_n) & \mbox{if $a_{i_j} \neq a_{i_{j+1}}$;} \\
a_{i_j}\psi(x_n) & \mbox{if $a_{i_j} = a_{i_{j+1}}$.}\end{array}\right.
$$
\end{description}
\item[Case i.2:] $i_j$ and $i_{j+1}$ are odd.
If $j-1 \geq 1$ and $j+2 \leq n-1$, set 
$$
\alpha_j = \left\{\begin{array}{ll}
a_{i_{j+1}}a_{i_j} & \mbox{if $a_{i_j} \neq a_{i_{j+1}}$;} \\
a_{i_j} & \mbox{if $a_{i_j} = a_{i_{j+1}}$.}\end{array}\right.
$$
If $j-1 < 1$, set 
$$
\alpha_j = \left\{\begin{array}{ll}
\psi(x_1)a_{i_{j+1}}\psi(x_1)a_{i_j} & \mbox{if $a_{i_j} \neq a_{i_{j+1}}$;} \\
\psi(x_1)a_{i_j} & \mbox{if $a_{i_j} = a_{i_{j+1}}$.}\end{array}\right.
$$
If $j+2 > n-1$, set 
$$
\alpha_j = \left\{\begin{array}{ll}
a_{i_{j+1}}\psi(x_n)a_{i_j}\psi(x_n) & \mbox{if $a_{i_j} \neq a_{i_{j+1}}$;} \\
a_{i_j}\psi(x_n) & \mbox{if $a_{i_j} = a_{i_{j+1}}$.}\end{array}\right.
$$
\end{description}
\item[Case ii:] $i_j$ is odd and $i_{j+1}$ is even.
\begin{description}[leftmargin=0cm]
\item[Case ii.1:] $j+2 \leq n-1$ and $i_{j+2}$ is odd;
$j-1 \geq 1$ and $i_{j-1}$ is even.
Suppose 
$\beta_j = a_{i_j} a_{j_1} a_{i_{j+1}}$ and $\beta_{j+1} = a_{i_{j+1}}a_{j_2} 
a_{j_3} a_{i_{j+2}}$ for some even $j_2$ and odd $j_1$ and $j_3$,
where $a_{j_1} \neq a_{i_j}$ and $a_{j_3} \neq a_{i_{j+2}}$.
Set 
$$
\alpha_j = \left\{\begin{array}{ll} 
a_{j_1} a_{i_{j+1}} a_{j_2} \psi(x_j) a_{j_2} a_{i_j} a_{j_1} & \mbox{if $a_{j_3} = a_{i_j}$;} \\
a_{j_1} a_{i_{j+1}} a_{j_2} \psi(x_j) a_{j_2} a_{j_3} a_{i_j} a_{j_1} & \mbox{if $a_{j_3} \neq a_{i_j}$.}\end{array}\right.
$$

\item[Case ii.2:] $j+2 \leq n-1$ and $i_{j+2}$ is odd; either
$j-1 \geq 1$ and $i_{j-1}$ is odd, or $j-1 < 1$.
If $j-1 \geq 1$ and $i_{j-1}$ is odd, define $\alpha_j$ as in Case ii.1 
(note that $\psi(x_j) = \ve$ in this case).
If $j-1 < 1$, define $\alpha_j$ as in Case ii.1 but with $\psi(x_1)$
prepended to $\alpha_j$.

\item[Case ii.3:] $j-1 \geq 1$ and $i_{j-1}$ is even; either
$j+2 \leq n-1$ and $i_{j+2}$ is even, or $j+2 > n-1$.
Suppose $\beta_j = a_{i_j}a_{j_1}a_{i_{j+1}}$
for some odd $j_1$ such that $a_{j_1} \neq a_{i_j}$.
If $j+2 \leq n-1$ and $i_{j+2}$ is even, set $\alpha_j = 
a_{j_1}a_{i_{j+1}}\psi(x_j)a_{i_j}a_{j_1}$.  
If $j+2 > n-1$, set $\alpha_j = a_{j_1}a_{i_{j+1}}\psi(x_n)
\psi(x_j)\psi(x_n)a_{i_j}a_{j_1}\psi(x_n)$.

\item[Case ii.4:] $j-1 \geq 1$ and $i_{j-1}$ is odd, or $j-1 < 1$;
$j+2 \leq n-1$ and $i_{j+2}$ is even, or $j+2 > n-1$.
Suppose $\beta_j = a_{i_j}a_{j'}a_{i_{j+1}}$, where $j'$ is
odd and $a_{j'} \neq a_{i_j}$.
If $j-1 \geq 1$, $i_{j-1}$ is odd, $j+2 \leq n-1$
and $i_{j+2}$ is even, set $\alpha_j = a_{j'}a_{i_{j+1}}a_{i_j}a_{j'}$.
If $j-1 < 1$, set $\alpha_j = \psi(x_1)a_{j'}a_{i_{j+1}}\psi(x_1)a_{i_j}a_{j'}$.
If $j+2 > n-1$, set $\alpha_j = a_{j'}a_{i_{j+1}}\psi(x_n)a_{i_j}a_{j'}\psi(x_n)$. 
\end{description}
\item[Case iii:] $i_j$ is even and $i_{j+1}$ is odd.
\begin{description}[leftmargin=0cm]
\item[Case iii.1:] $\beta_j = a_{i_j}a_{j_1}a_{j_2}a_{i_{j+1}}$
for some even $j_1$ and odd $j_2$ such that $a_{j_2} \neq a_{i_{j+1}}$.
Set 
$$
\alpha_j = \left\{\begin{array}{ll}
a_{j_2}a_{i_{j+1}}\psi(x_{j+2})\psi(x_j)a_{i_j}a_{j_1}a_{j_2} & \mbox{if $2 \leq j \leq n-3$;} \\
\psi(x_1)a_{j_2}a_{i_{j+1}}\psi(x_{j+2})\psi(x_1)a_{i_j}a_{j_1}a_{j_2} & \mbox{if $j-1 < 1$;} \\
a_{j_2}a_{i_{j+1}}\psi(x_n)\psi(x_j)a_{i_j}a_{j_1}a_{j_2}\psi(x_n) & \mbox{if $j+2 > n-1$.}\end{array}\right. 
$$

\item[Case iii.2:] $\beta_j = a_{i_j}a_{j_2}a_{i_{j+1}}$ for some
odd $j_2$ such that $a_{j_2} \neq a_{i_{j+1}}$ (note that if $j-1 \geq 1$,
then $i_{j-1}$ is even and so $\psi(x_j) = \ve$).
Define $\alpha_j$ as in Case iii.1, but with all occurrences of $a_{j_1}$ deleted.

\end{description}
\end{description}

\medskip
Set $w_3 := \alpha_1\alpha_2\ldots\alpha_{n-2}$.

\end{description}
By construction, $w_1 \in L(\pi)$ and $w_2 \in L(\pi)$.
Furthermore, induction on $j = 1,\ldots,n-2$ shows that
the longest prefix of $x_1a_{i_1}x_2a_{i_2}x_2\ldots a_{i_{n-1}}x_n$
matching $\alpha_1\ldots\alpha_j$ is $x_1a_{i_1}x_2\ldots a_{i_j}x_{j+1}$.
Hence $w_3 \notin L(\pi)$.  The lemma will follow from the next two claims.


\begin{subclaim}\label{clm:taumorphismw2}
Suppose $h,g:(X \cup \Sigma)^* \mapsto \Sigma^*$ are constant-preserving morphisms witnessing
$w_1 \in L(\tau)$ and $w_2 \in L(\tau)$ respectively, and 
suppose $\pi(\ve) = a_{i_1}a_{i_2}\ldots$ $a_{i_{n-1}} \sqsubseteq \tau(\ve)$.
Let $\langle p_1,p_2,\ldots,$ $p_{n-1}\rangle$ be a sequence of positions of $\tau$
such that $\tau[p_j] = a_{i_j}$ for all $j \in \{1,\ldots,n-1\}$.
For each $j \in \{1,\ldots,n-1\}$, let $q_j$ be the position of $w_1$ occupied by
the specific occurrence of $a_{i_j}$ indicated with braces in Equation (\ref{eqn:w1decomposeqi}).
\begin{equation}\label{eqn:w1decomposeqi}
w_1 := \varphi(x_1)\overbrace{a_{i_1}}^{q_1} \varphi(x_2) \overbrace{a_{i_2}}^{q_2} \ldots 
\varphi(x_{i_j}) \overbrace{a_{i_j}}^{q_j} \varphi(x_{i_{j+1}})  \ldots 
\varphi(x_{n-1}) \overbrace{a_{i_{n-1}}}^{q_{n-1}} \varphi(x_n).
\end{equation}     
Similarly, let $R_j$ be the sequence of positions of $w_2$ indicated with braces in 
Equation (\ref{eqn:w1decomposeri}). 
\begin{equation}\label{eqn:w1decomposeri}
w_2 := \rlap{$\underbrace{\phantom{\psi(x_1)a_{i_1}\psi(x_2)}}_{R_1}$} \psi(x_1)a_{i_1} \overbrace{\psi(x_2) a_{i_2} 
\psi(x_3)}^{R_2} \ldots \overbrace{\psi(x_j) a_{i_j} \psi(x_{j+1})}^{R_j} \ldots \rlap{$\overbrace{\phantom{\psi(x_{n-2}) a_{i_{n-2}}
\psi(x_{n-1})}}^{R_{n-2}}$}\psi(x_{n-2}) a_{i_{n-2}} \underbrace{\psi(x_{n-1})a_{i_{n-1}}\psi(x_n)}_{R_{n-1}}.
\end{equation}     
Let $I^{const}_{h,\tau}$ (resp.~$I^{const}_{g,\tau}$) be the mapping of sequences 
of positions of constants in $\tau$ to sequences of positions of $w_1$ 
(resp.~$w_2$) induced by $h$ (resp.~$g$).  
Then for all $j \in \{1,\ldots,n-1\}$, $I^{const}_{h,\tau}(\langle p_j \rangle) 
= \langle q_j \rangle$ and $I^{const}_{g,\tau}(\langle p_j\rangle)$
is a subsequence of $R_j$.

In particular, if $a_{i_1}a_{i_2}\ldots a_{i_{n-1}} \sqsubseteq \tau(\ve)$, then
$L(\tau) = L(\pi)$.
\end{subclaim}

\begin{subclaim}\label{clm:w3match}
Let $\eta$ be any regular pattern such that $\{w_1,w_2\} \subset L(\eta)$ and 
$a_{i_1}a_{i_2}\ldots$ $a_{i_{n-1}} \not\sqsubseteq \eta(\ve)$.
Then $w_3 \in L(\eta)$.
\end{subclaim} 

\noindent\emph{Proof of Claim \ref{clm:taumorphismw2}.} Let $P_1,P_2,\ldots,P_{n-1}$ denote
the sequences of positions of $w_1$ indicated by braces in Equation (\ref{eqn:morphismscopew1}).
\begin{equation}\label{eqn:morphismscopew1}
w_1 := \rlap{$\underbrace{\phantom{\varphi(x_1)a_{i_1}\varphi(x_2)}}_{P_1}$} \varphi(x_1)a_{i_1} \overbrace{\varphi(x_2) a_{i_2} \varphi(x_3)}^{P_2} \ldots 
\overbrace{\varphi(x_j) a_{i_j} \varphi(x_{j+1})}^{P_j} \ldots \rlap{$\overbrace{\phantom{\varphi(x_{n-2}) a_{i_{n-2}}
\varphi(x_{n-1})}}^{P_{n-2}}$}\varphi(x_{n-2}) a_{i_{n-2}} \underbrace{\varphi(x_{n-1})a_{i_{n-1}}\varphi(x_n)}_{P_{n-1}}.
\end{equation}
It suffices to show that whenever $j \in \{1,\ldots,n-1\}$, $I^{const}_{h,\tau}(\langle p_j\rangle)$ 
is a subsequence of $P_j$; the claim that $I^{const}_{h,\tau}(\langle p_j \rangle) = \langle q_j\rangle$ 
will then follow from the fact that $\varphi(x_j) \notin N(x_j,\pi)$ for all $j \in \{1,\ldots,n-1\}$.  
So assume, by way of contradiction, that there were a least $\ell \in \{1,\ldots,n-1\}$ such that 
$I^{const}_{h,\tau}(\langle p_{\ell} \rangle)$ is not a subsequence
of $P_{\ell}$.  First, suppose that $I^{const}_{h,\tau}(\langle p_{\ell} \rangle)$
were a subsequence of some $P_{\ell'}$ with $\ell' < \ell$.  Then,
since $\varphi(x_{\ell}) \notin \{a_{i_{\ell}},a_{i_{\ell-1}}\}$ and
$\varphi(x_{\ell-1}) \neq a_{i_{\ell-1}}$, $I^{const}_{h,\tau}(\langle p_{\ell-1} \rangle)$
is not a subsequence of $P_{\ell-1}$.  Iterating the preceding argument
then gives that for all $j \leq \ell$, $I^{const}_{h,\tau}(\langle p_j \rangle)$
is not a subsequence of $P_j$, a contradiction. 
A similar argument holds if $I^{const}_{h,\tau}(\langle p_{\ell} \rangle)$ were
a subsequence of some $P_{\ell''}$ with $\ell'' > \ell$.

The proof that $I^{const}_{g,\tau}(\langle p_j\rangle)$ is a subsequence of $R_j$ 
is similar (making crucial use of the definition of $\psi$).  
This establishes the first part of the claim.

Now we establish the second part of the claim.  Note that from the first 
part of the claim, if $i_j$ is odd, then $I^{const}_{g,\tau}(\langle p_j\rangle)$
cannot be a subsequence of the sequence of positions of $w_2$
corresponding to $\psi(x_j)$ (resp.~$\psi(x_{j+1})$).
If $i_j$ is even, then $I^{const}_{g,\tau}(\langle p_j\rangle)$ cannot 
be a subsequence of the sequence of positions of $w_2$ corresponding
to $\psi(x_j)$.  Furthermore, suppose $I^{const}_{g,\tau}(\langle p_j\rangle)$
were a subsequence of the sequence of positions of $w_2$ corresponding
to $\psi(x_{j+1})$; then if $j+1 \leq n-1$, $i_{j+1}$ must be odd and
therefore $I^{const}_{g,\tau}(\langle p_{j+1}\rangle)$ equals $\langle q' \rangle$, 
where $q'$ is the position of $w_2$ occupied by $a_{i_{j+1}}$ in
$R_{j+1}$.  

From the fact that $\{w_1,w_2\} \subset L(\tau)$, we know that
$\tau$ must start as well as end with variables.
For any $\alpha \in (X \cup \Sigma)^*$, let $o(\alpha)$ denote
the number of substrings of $\alpha$ of the shape $b_1xb_2$,
where $x \in X \cup \{\ve\}$, $b_1,b_2 \in \Sigma$ and
$b_1,b_2$ have opposite parities.  Note that $o(w_2) = o(\pi)$. 
Since $I^{const}_{h,\tau}(\langle p_j\rangle) = \langle q_j \rangle$ 
whenever $j \in \{1,\ldots,n-1\}$, it follows that if 
$a_{i_1}\ldots a_{i_{n-1}} \sqsubset \tau(\ve)$, then there is some
position $p'$ of $\tau$ such that for some $j \in \{1,\ldots,n-2\}$,
$p_j < p' < p_{j+1}$ and $\tau[p'] = \varphi(x_{i_{j+1}}) \in \Sigma$.
By the definition of $\varphi$, if $\varphi(x_{i_{j+1}}) = a_{j'}$, then
$j'$ has parity opposite to that of $i_j$ as well as $i_{j+1}$.  Thus   
$o(\tau) > o(\pi)$.  But $w_2 \in L(\tau)$ implies $o(\tau) \leq o(\pi)$,
and therefore $\tau(\ve) = a_{i_1}\ldots a_{i_{n-1}}$.
The fact that $w_1 \in L(\tau)$ (resp.~$w_2 \in L(\tau)$) implies 
that a variable occurs in $\tau$ between every pair $a_{i_j},a_{i_{j+1}}$ such
that $i_j$ and $i_{j+1}$ have equal (resp.~opposite) parities.   
Thus $L(\tau) = L(\pi)$.\qed~(Claim \ref{clm:taumorphismw2})

\medskip
\noindent\emph{Proof of Claim \ref{clm:w3match}.}
Our strategy to show $w_3 \in L(\eta)$ is as follows.
First, fix some constant-preserving morphism $g:(X \cup \Sigma)^* \mapsto \Sigma^*$
such that $g(\eta) = w_2$.  Then $g$ induces a mapping
$\mathcal{I}_{g,\eta}$ of closed intervals of $\{1,\ldots,|\eta|\}$ to
closed intervals of $\{1,\ldots,w_2\}$ such that for all
$[p_1,p_2] \subseteq \{1,\ldots,|\eta|\}$, $g(\eta[p_1]\ldots
\eta[p_2]) = w_2[\mathcal{I}_{g,\eta}([p_1,p_2])]$.  One may take the ``inverse''
$\overline{\cI}_{g,\eta}$ of $\mathcal{I}_{g,\eta}$, where, for all $[q_1,q_2] \subseteq \{1,\ldots,|w_2|\}$,
$\overline{\cI}_{g,\eta}([q_1,q_2]) = [s_1,s_2]$ for some $s_1,s_2 \in \{1,\ldots,|\eta|\}$
such that $[q_1,q_2]$ is a subinterval of $\mathcal{I}_{g,\eta}([s_1,s_2])$
and for all proper subintervals $R$ of $[s_1,s_2]$, $[q_1,q_2]$ is not a subinterval of
$\mathcal{I}_{g,\eta}(R)$. 
Let $r_1,\ldots,r_{n-1}$ be the positions of $a_{i_1},\ldots,a_{i_{n-1}}$
respectively in $w_2$ marked with braces in Equation (\ref{eqn:w2decomposeqi}).
\begin{equation}\label{eqn:w2decomposeqi}
w_2 := \psi(x_1) \overbrace{a_{i_1}}^{r_1} \psi(x_2) \overbrace{a_{i_2}}^{r_2} \psi(x_3) 
\ldots \psi(x_j) \overbrace{a_{i_j}}^{r_j} \psi(x_{j+1}) \ldots \psi(x_{n-2}) 
\overbrace{a_{i_{n-2}}}^{r_{n-2}} \psi(x_{n-1}) \overbrace{a_{i_{n-1}}}^{r_{n-1}}\psi(x_n).
\end{equation}     
By our assumption on $\eta$, there is a least $\ell \in \{1,\ldots,n-1\}$
such that $\eta[\overline{\cI}_{g,\eta}([r_{\ell},r_{\ell}])]$ is a variable and
if there is a least $r' > r_{\ell}$ such that $\eta[\overline{\cI}_{g,\eta}([r',r'])]$ is a constant,
then $\eta[\overline{\cI}_{g,\eta}([r',r'])] \neq a_{i_{\ell}}$.
As argued at the beginning of the construction of $w_3$,
$\eta$ starts and ends with variables, and every maximal constant 
block $A$ of $\eta$ has length at most $2$; furthermore, if the 
length of $A$ is exactly $2$, then $A = a_{j_1}a_{j_2}$
for some $j_1,j_2 \in \{1,\ldots,k\}$ such that $j_1$ and $j_2$ have
opposite parities.  

We define a set $\cC$ 
consisting of all possible intervals of positions of $w_2$ of length at most $2$
such that for every maximal constant block of $\eta$, say $\eta[J]$ for some
closed interval $J \subseteq \{1,\ldots,|\eta|\}$, there is an $I \in \cC$
for which $\cI_{g,\eta}(J) \subseteq I$.

First, suppose $i_{\ell}$ is even and the first letter of $\psi(x_{\ell+1})$ 
equals $a_{i_{\ell}}$.  Then $\cC$ consists of all intervals of 
positions of $w_2$ of the form
\begin{enumerate}[label=\roman*]
\item $[q,q+1]$, where $q < r_{\ell}-1$ and $w_2[q]w_2[q+1] = a_{j_1}a_{j_2}$ 
for some $j_1,j_2 \in \{1,\ldots,k\}$ with opposite parities, or 
\item $[q,q+1]$, where $q > r_{\ell}+1$ 
and $w_2[q]w_2[q+1] = a_{j_1}a_{j_2}$ for some $j_1,j_2 \in \{1,\ldots,k\}$ 
with opposite parities, or 
\item $[q,q]$, where $q < r_{\ell}-1$ and if $q \geq 2$, then $w_2[q-1]w_2[q]w_2[q+1] = 
b a_{j_3}a_{j_4}$ for some $b \in \Sigma$ 
and $j_3,j_4 \in \{1,\ldots,k\}$ such that $j_3$ and $j_4$ have equal parities,
and 
$b = a_{j_5}$ for some $j_5 \in \{1,\ldots,k\}$ such that 
$j_5$ and $j_3$ have equal parities; if $q < 2$, then the same holds with 
$w_2[q-1]$ and $b$ replaced with $\ve$, or
\item $[q,q]$, where $q = r_{\ell}-1$ and if $q \geq 2$, then $w_2[q-1]w_2[q] = ba_{j_6}$ 
for some $j_6 \in \{1,\ldots,k\}$ and $b \in \Sigma$ such that 
if $b = a_{j_7}$ for some $j_7 \in \{1,\ldots,k\}$,
then $j_7$ and $j_6$ have equal parities; if $q < 2$, then the same holds with
$w_2[q-1]$ and $b$ replaced with $\ve$, or 
\item $[q,q]$ for some $q > r_{\ell}+2$ such that if $q+1 \leq |w_2|$, then $w_2[q-1]w_2[q]w_2[q+1] = 
a_{j_8}a_{j_9}b$ for some $j_8,j_9 \in \{1,\ldots,k\}$
with equal parities and $b \in \Sigma$ such that if $b = a_{j_{10}}$ for some
$j_{10} \in \{1,\ldots,k\}$, then $j_{10}$ and $j_9$ have equal parities;
if $q+1 > |w_2|$, then the same holds with $w_2[q+1]$ and $b$ replaced with $\ve$,
or 
\item $[q,q]$, where $q = r_{\ell}+2$ and if $q+1 \leq |w_2$, then $w_2[q]w_2[q+1] = a_{j_{11}}b$
for some $j_{11} \in \{1,\ldots,k\}$ and $b \in \Sigma$
such that if $b = a_{j_{12}}$ for some $j_{12} \in \{1,\ldots,k\}$, then
$j_{12}$ and $j_{11}$ have equal parities; if $q+1 > |w_2|$, then
the same holds with $w_2[q+1]$ and $b$ replaced with $\ve$.    
\end{enumerate}  
Second, suppose either $i_{\ell}$ is odd or the first letter of 
$\psi(x_{\ell+1})$ is not equal to $a_{i_{\ell}}$.
Then we define $\cC$ exactly as above but with three differences:
first, $q > \ell+1$ is replaced with $q > \ell$ in (ii); second, $q > \ell+2$ is 
replaced with $q > \ell+1$ in (v); third, $q = \ell+2$ is replaced
with $q = \ell+1$ in (vi).
We next define a one-one mapping $F$ from $\cC$ to the set of all intervals
of positions of $w_3$ satisfying the following conditions for all $[q,q],[q,q+1] \in \cC$:
\begin{itemize}
\item $F([q,q+1]) = [q',q'+1]$ for some $q' \in \{1,\ldots,|w_3|-1\}$ with 
$w_2[q]w_2[q+1] = w_3[q']w_3[q'+1]$.
\item $F([q,q]) = [q',q']$ for some $q' \in \{1,\ldots,|w_3|\}$ with
$w_2[q] = w_3[q']$.
\item Suppose $q_1$ and $q_2$ are the left endpoints of $I_1$ and
$I_2$ respectively, where $I_1,I_2 \in \cC$, $I_1 \neq I_2$ and $q_1 < q_2$ (note that
no two distinct members of $\cC$ intersect).  Let $q'_1$ and $q'_2$
be the left endpoints of $F(I_1)$ and $F(I_2)$ respectively.  
Then $q'_1 < q'_2$ and $F([I_1]) \cap F([I_2]) = \emptyset$.  
\end{itemize} 
Note that the existence of an $F$ satisfying the above three conditions implies
that for any sequence $\langle I_1,I_2,\ldots,I_m \rangle$ of intervals of positions
of $w_2$ such that every $I_i$ corresponds to a maximal constant block
of $\eta$ and for all $i,j \in \{1,\ldots,m\}$ with $i < j$,
$I_i \cap I_j = \emptyset$, and the left endpoint of $I_i$ is smaller than that
of $I_j$, there is a corresponding sequence $\langle I'_1,I'_2,\ldots,I'_m\rangle$
of intervals of positions of $w_3$ such that for all $i,j \in \{1,\ldots,m\}$
with $i < j$, $w_2(I_i) = w_3(I'_i)$, $I'_i \cap I'_j = \emptyset$,
and the left endpoint of $I_{i'}$ is smaller than that of $I_{j'}$. 
Thus, since $\eta$ starts as well as ends with variables, the existence of
such an $F$ will suffice to show that $w_3 \in L(\eta)$.  We consider
a case distinction based on the earlier definition of $\cC$. 
Let $Q_1,\ldots,Q_{n-2}$ be the closed intervals of positions of $w_3$
corresponding to the occurrences of $\alpha_1,\ldots,\alpha_{n-2}$ respectively
as shown in Equation (\ref{eqn:w3decomposealpha}).
\begin{equation}\label{eqn:w3decomposealpha}
w_3 := \overbrace{\alpha_1}^{Q_1} \ldots \overbrace{\alpha_j}^{Q_j} \ldots \overbrace{\alpha_{n-2}}^{Q_{n-2}}.
\end{equation} 
Consider any $I \in \cC$.
\begin{description}[leftmargin=0cm]
\item[Case 1:] $I = [r_j,r_j+1]$ for some $j < \ell$, where, if $w_2[r_j,r_j+1] = a_{j'}a_{j''}$
for some $j',j'' \in \{1,\ldots,k\}$, then $j'$ and $j''$ have opposite parities.
Note that if $i_j$ were odd, then by Cases i and iii in the construction of $w_2$,
$w_2[r_j+1] = a_{j'}$ would imply that $j'$ is odd, which is impossible
by Conditions i and ii in the definition of $\cC$.  Hence $i_j$ is even.  
Furthermore, an inspection of Cases i and ii in the construction
of $w_2$ shows that $r_j+1 \neq r_{j+1}$, and therefore $r_j+1$ is the position of 
the first letter of $\psi(x_{j+1})$ in $w_2$; moreover, $i_{j+1}$ is odd.
Suppose $\beta_j = a_{i_j}a_{j_1}a_{i_{j+1}}$
for some odd $j_1$ such that $a_{j_1} \neq a_{i_{j+1}}$ (the positions
of $\beta_1,\ldots,\beta_{n-2}$ are illustrated in Equation (\ref{eqn:w2decomposebeta})).
From Case iii.2 in the construction of $w_3$, one sees that
$\alpha_j = \ga a_{i_j}a_{j_1}$ for some $\ga \in \Sigma^*$;
fix $\ga$.
Set $F(I) = \left[\sum_{1\leq l < j}|\alpha_l|+|\ga|+1,
\sum_{1\leq l < j}|\alpha_l|+|\ga|+2\right]$. 

\item[Case 2:] $I = [r_j-1,r_j]$ for some $j < \ell$.
First, suppose $j-1 \geq 1$.  Then an argument similar to that in Case 1.1 shows that
$i_j$ must be even and $i_{j-1}$ must be odd.
From Cases i.1.1 and iii in the construction of $w_3$,
one sees that 
$\alpha_j = \ga_1 w_2[r_j-1] a_{i_{j}} \ga_2$
for some $\ga_1,\ga_2 \in \Sigma^*$; fix such $\ga_1$ and $\ga_2$.
Set $F(I) = \left[\sum_{1 \leq l < j}|\alpha_l|+|\ga_1|+1,\sum_{1 \leq l < j}|\alpha_l|+|\ga_1|+2\right]$.        

Second, suppose $j-1 < 1$.  From Cases i.1.3, i.1.4, i.2, ii.4 and iii
in the construction of $w_3$, we deduce that
there are $\ga_1,\ga_2 \in \Sigma^*$ such that
$\alpha_1 = \ga_1 \psi(x_1) a_{i_1} \ga_2$; fix such $\ga_1$ and $\ga_2$.
Set $F(I) = \left[|\ga_1|+1,|\ga_1|+2\right]$. 

\item[Case 3:] $I = [r_j+1,r_j+2]$ for some $j < \ell$
such that $\psi(x_{j+1}) = w_2[r_j+1]w_2[r_j+2]$.
Based on the case distinction in the construction of $w_2$,
one sees that $i_j$ must be even and if $j+2 \leq n-1$,
then $a_{i_{j+1}}$ must be odd.
From Case iii.1, we deduce that $\alpha_j = \ga \psi(x_{j+1})$
for some $\ga \in \Sigma^*$.  Set
$F(I) = \left[\sum_{1 \leq l < j}|\alpha_l|+|\ga|+1,\sum_{1 \leq l < j}
|\alpha_l|+|\ga|+2\right]$.  

\item[Case 4:] $I = [r_j-1,r_j]$ for some $j > \ell$.
Arguing as in the earlier cases, $i_j$ must be even
and $i_{j-1}$ must be odd.  By examining Case ii
in the construction of $w_3$, one sees that
$\alpha_{j-1} = w_2[r_j-1]a_{i_j}\ga$ for some
$\ga \in \Sigma^*$.  Set $F(I) = \left[\sum_{1 \leq l < j-1}
|\alpha_l|+1,\sum_{1 \leq l < j-1}|\alpha_l|+2\right]$.

\item[Case 5:] $I = [r_j,r_j+1]$ for some $j > \ell$.
First, suppose $j+1 \leq n-1$.  Arguing as before, 
$i_j$ and $i_{j-1}$ must be even while $i_{j+1}$ must be odd.
It follows from Case i.1 in the construction of $w_3$
that $\alpha_{j-1} = \ga_1 a_{i_j}w_2[r_j+1] \ga_2$ for some
$\ga_1,\ga_2 \in \Sigma^*$; fix such $\ga_1$ and $\ga_2$.
Set $F(I) = \left[\sum_{1 \leq l < j-1}|\alpha_l|+|\ga_1|+1,
\sum_{1 \leq l < j-1}|\alpha_l|+|\ga_1|+2 \right]$.

Second, suppose $j+1 > n-1$, i.e. $j = n-1$.  It follows from Cases
i.1.2 and i.1.4 in the construction of $w_3$ that for some $\ga_1,\ga_2 \in \Sigma^*$, 
$\alpha_{n-2} = \ga_1 a_{i_{n-1}} \psi(x_n) \ga_2$; fix
such $\ga_1$ and $\ga_2$.  Set $F(I) = \left[\sum_{1 \leq l 
< j-1}|\alpha_l|+|\ga_1|+1,\right.$ $\left.\sum_{1 \leq l < j-1}\right.$ $\left.|\alpha_l|+|\ga_1|+2 
\right]$.   

\item[Case 6:] $I = [r_j+1,r_j+2]$ for some $j > \ell$ such
that $\psi(x_{j+1}) = w_2[r_j+1]w_2[r_j+2]$.
Based on the case distinction in the construction of $w_2$,
we deduce that $i_{j-1}$ and $i_{j+1}$ are odd while 
$i_j$ is even.  It follows from Case ii in the construction of
$w_3$ that $\alpha_{j-1} = \ga_1 a_{i_j} \ga_2 w_2[r_j+1]
w_2[r_j+2] \ga_3$ for some $\ga_1,\ga_2,\ga_3 \in \Sigma^*$;
fix such $\ga_1,\ga_2$ and $\ga_3$.  Set
$F(I) = \left[\sum_{1 \leq l < j-1}|\alpha_l|+|\ga_1|+|\ga_2|+2,
\sum_{1 \leq l < j-1}\right.$ $\left.|\alpha_l|+|\ga_1|+|\ga_2|+3 \right]$.  

\item[Case 7:] $I = [r_j,r_j]$ for some $j < \ell$.
\begin{description} 
\item[Case 7.1:] $i_j$ is even.
First, suppose $j-1 \geq 1$.
Then both $i_{j-1}$ and $i_{j+1}$ must be even.
From Case i.1 in the construction of $w_3$, we deduce
that there exist $\ga_1,\ga_2 \in \Sigma^*$ such that 
$\alpha_j = \ga_1 a_{i_j} \ga_2$ and $|\ga_2| \leq 1$;
fix such $\ga_1$ and $\ga_2$.
Set $F(I) = \left[\sum_{1 \leq l < j}|\alpha_l|+|\ga_1|+1,
\sum_{1 \leq l < j}|\alpha_l|+|\ga_1|+1\right]$. 

Second, suppose $j-1 < 1$.  It follows from Cases i.1
and iii in the construction of $w_3$ that there exist
$\ga_1,\ga_2 \in \Sigma^*$ such that $\alpha_1 = \ga_1 a_{i_1} \ga_2$
and $|\ga_2| \leq 2$; fix such $\ga_1$ and $\ga_2$.
Set $F(I) = \left[|\ga_1|+1,|\ga_1|+1\right]$. 

\item[Case 7.2:] $i_j$ is odd. 
It follows from Cases i.2 and ii in the construction of $w_3$ that
there exist $\ga_1,\ga_2 \in \Sigma^*$ such that
$\alpha_j = \ga_1 a_{i_j} \ga_2$ and $|\ga_2| \leq 1$;
fix such $\ga_1$ and $\ga_2$.  Set $F(I) = \left[\sum_{1 \leq l < j}
|\alpha_l|+|\ga_1|+1,\right.$ $\left.\sum_{1 \leq l < j}|\alpha_l|+|\ga_1|+1 \right]$. 
\end{description}

\item[Case 8:] $I = [r_j,r_j]$ for some $j > \ell$.
From the case distinction in the construction of $w_3$, we deduce
that there exist $\ga_1,\ga_2 \in \Sigma^*$ with $|\ga_1| \leq 1$
such that $\alpha_{j-1} = \ga_1 a_{i_j} \ga_2$; fix such $\ga_1$
and $\ga_2$.  Set $F(I) = \left[\sum_{1 \leq l < j-1}|\alpha_l|+|\ga_1|+1,
\sum_{1 \leq l < j-1}|\alpha_l|\right.$ $\left.+|\ga_1|+1\right]$.

\item[Case 9:] $I = [1,1]$.
Observe from 
the construction
of $w_3$ that $\alpha_1$ starts with $\psi(x_1)$.  Set $F(I) = [1,1]$. 

\item[Case 10:] $I = [|w_2|,|w_2|]$. 
Observe from 
the construction
of $w_3$ that $\alpha_{n-2}$ ends with $\psi(x_n)$.
Set $F(I) = [|w_3|,|w_3|]$.

\end{description}

This completes the definition of $F$. 
By Claim \ref{clm:w3match}, since $\{w_1,w_2\} \subset L(\tau)$ and
$w_3 \notin L(\tau)$, one has that $a_{i_1}a_{i_2}\ldots a_{i_{n-1}} \sqsubseteq \tau(\ve)$.
Thus by Claim \ref{clm:taumorphismw2}, $L(\tau) = L(\pi)$.
Therefore $T = \{(w_1,+),(w_2,+),(w_3,-)\}$ is indeed a teaching set
for $\pi$ w.r.t.\ $R\Pi^z$.~\qed


\subsection{Example for Lemma \ref{lem:tdsimplebregpatatleast3}}

We give an example to illustrate the construction of the teaching set
in the proof of Lemma \ref{lem:tdsimplebregpatatleast3}.

\begin{exmps}\label{exmp:simpleblockregulartdternary}
Suppose $\Sigma = \{0,1,2\}$.  Following the notation of Lemma \ref{lem:tdsimplebregpatatleast3},
set $a_1 = 0, a_2 = 1$ and $a_3 = 2$.  Let $\pi = x_1 0 x_2 1 x_3 2 x_4 1 x_5 1 x_6$.
According to the construction in the proof of Lemma \ref{lem:tdsimplebregpatatleast3},
$\pi$ has the teaching set $\{(w_1,+),(w_2,+),(w_3,$ $-)\}$ w.r.t.\ $R\Pi^3$, where 
$w_1,w_2$ and $w_3$ are defined as follows ($\varphi,\psi$ and $\alpha_i$ are defined
as in the proof of Lemma \ref{lem:tdsimplebregpatatleast3}):
\begin{itemize}
\item $w_1 = 0121\underbrace{0}_{\varphi(x_5)}1$.
\item $w_2 = \underbrace{1}_{\psi(x_1)} 0 \underbrace{2}_{\psi(x_2)} 1 \underbrace{10}_{\psi(x_3)} 
2 \underbrace{0}_{\psi(x_4)} 11 \underbrace{0}_{\psi(x_6)} $.
\item $w_3 = \underbrace{211102}_{\alpha_1} \underbrace{0202110}_{\alpha_2} \underbrace{011020}_{\alpha_3}
\underbrace{010}_{\alpha_4}$.
\end{itemize}   
\end{exmps}

\subsection{Proof of Theorem \ref{thm:tdsimplebregclassofpatterns}}

\proofthmtdsimplebregclassofpatterns

\subsection{Example for Theorem \ref{thm:tdsimplebregclassofpatterns}}

\begin{exmps}\label{exmp:linearubsimplebregtight}
We exhibit a family of simple block-regular patterns for which the
lower bound given in Theorem \ref{thm:tdsimplebregclassofpatterns}(ii) 
is tight (up to numerical constant factors).

Suppose $z = |\Sigma| \geq 2$ and $0,1 \in \Sigma$.
For all $n \in \natnum$, let $\pi_n$ be the simple block-regular
pattern $x_1 0 x_2 0 \ldots 0 x_{n+1}$; in particular, $\pi_n(\ve) = 0^n$.
We construct a teaching set $T$ for $\pi_n$ w.r.t.\ $\Pi^z$ 
as follows.  Let $\tau$ denote any pattern that is consistent with $T$.  
First, put $(0^n,+)$ into $T$; this example ensures that $\tau(\ve)
\sqsubseteq 0^n$.  Next, for each $k \in \{0,\ldots,n-1\}$, put $(0^k,-)$ 
into $T$.  The examples put into $T$ so far ensure that $\tau(\ve) = 0^n$.
Now for all $i \in \{1,\ldots,n+1\}$, put $\left(\pi[x_i \ra 1, x_j \ra \ve,
j \in \{1,\ldots,n+1\} \sm \{i\}],+\right)$ into $T$.  The last set of examples
will ensure that every maximal variable block of $\tau$ contains at
least one free variable.  Thus $L(\tau) = L(\pi)$, and this proves
that $\pi_n$ has a teaching set w.r.t.\ $\Pi^z$ of size $O(n)$.
\end{exmps}

\subsection{Proof of Lemma \ref{lem:mquasiregnotsubsetbound}}

\prooflemmquasiregnotsubsetbound

\subsection{Proof of Theorem \ref{thm:cfqrplfinitetd}}

\proofthmcfqrplfinitetd



\subsection{Proof of Theorem \ref{thm:pbtdmquasiregularcf}}

We first observe a basic fact about graph colourings.
We recall that for any finite, simple graph $G = (V,E)$, the \emph{distance} between any two vertices $u$ and $v$,
denoted $d_G(u,v)$, is the length of a shortest path in $G$ from $u$ to $v$ (or vice-versa;
if no such path exists, then $d_G(u,v) = \infty$),
and for any $\ell \geq 1$, the \emph{$\ell$-distance chromatic number of $G$}, denoted $\chi_{\ell}(G)$, 
is the smallest $k$ for which there exists a $k$-colouring of $G$ such that for any pair of vertices $s,t$ of 
$G$ with $d_G(s,t) \leq \ell$, $s$ and $t$ receive distinct colours; such a colouring is called
an \emph{$\ell$-distance colouring of $G$} \cite{Jensen94,Kramer08}.

\begin{lems}\label{lem:graphcolor2distance}
Let $G = (V,E)$ be any finite, simple graph with vertex set $V$, edge set $E$
and maximum degree $\Delta(G)$.  Then $\chi_2(G) \leq \Delta(G)^2+1$; equality
occurs if $G$ is the $5$-cycle. 
\end{lems}

\proof
We note that $\chi_2(G)$ is equal to $\chi_1(G^2)$, the (ordinary) chromatic number of
the square of $G$; $G^2$ is the graph whose vertex set is equal to that
of $G$ and for all distinct vertices $v_1,v_2$ of $G$, $(v_1,v_2)$ is an edge of $G^2$ iff
$d_G(v_1,v_2) \leq 2$.  The maximum degree of any vertex of $G^2$ is at most 
$\Delta(G)+\Delta(G)\cdot(\Delta(G)-1) = \Delta(G)^2$, and so by Brook's theorem \cite[Theorem 11]{Jensen94},
$\chi_1(G^2) \leq \Delta(G)^2+1$; equality occurs if $G$ is the $5$-cycle.~\qed~(Lemma 
\ref{lem:graphcolor2distance})

\medskip
\proofthmpbtdmquasiregularcf

\begin{rems}
The notion of the \emph{adjacency graph} of a (constant-free) pattern was 
introduced in the study of pattern avoidance \cite[Chapter 3]{Lothaire02}. 
We do not know whether the lower bound on $|\Sigma|$ in Theorem \ref{thm:pbtdmquasiregularcf}
is tight.  The minimum number of colours needed to satisfy Conditions
1 and 2 in the proof of Theorem \ref{thm:pbtdmquasiregularcf} might be smaller
than $4m^2+1$; if so, this would give a reduction in the minimum alphabet
size needed for the theorem to hold.  In fact, the upper bound on $k$
in the proof of Theorem \ref{thm:pbtdmquasiregularcf} would still hold if the second condition on $c$
is weakened as follows:
if there are distinct $j_1,j_2 \in \{1,\ldots,n\}$ such that
(i) $(x_i^L,x_{j_1}^R) \in E(\AG(\pi))$ and $(x_i^L,x_{j_2}^R) \in E(\AG(\pi))$,
(resp.~$(x_{j_1}^L,x_i^R) \in E(\AG(\pi))$ and $(x_{j_2}^L,x_i^R) \in E(\AG(\pi))$),
then $x_i^L$ (resp.~$x_i^R$) is adjacent to at least two vertices that are assigned different
colours.  
\end{rems}

\subsection{Proof of Proposition \ref{prop:pbtdqrunary}}

\proofproppbtdqrunary

\subsection{Proof of Lemma \ref{lem:noncrossstructure}}

\prooflemnoncrossstructure

\subsection{Proof of Theorem \ref{thm:tdnoncrossbounded}}\label{appen:proofthmtdnoncrossbounded}

\proofthmtdnoncrossbounded

\subsection{Example \ref{exmp:noncrosslowerbound}}\label{appen:noncrosslowerbound}

Suppose $\{0,1\} \subseteq \Sigma$.  Let $\pi = x_1^4 x_2^8 x_3^9$.
There are $3$ maximal proper prime power factors of $4,8$ and $9$, namely,
$2,4$ and $3$, and so by the proof of Theorem \ref{thm:tdnoncrossbounded},
the \TD\ 
of $\pi$ w.r.t.\ $\NC\Pi^{|\Sigma|}_{\infty,9}$ is at most
$2+3 = 5$.  However, one can build a teaching set $T$ of size $4$ for $\pi$ as follows.
As in the proof of Theorem \ref{thm:tdnoncrossbounded}(ii), put $(v_1,+)$ and $(v_2,-)$ into $T$, where
$v_1 := (01)^4 (001)^8 (0001)^9$     
and $v_2 := (01)^{9!} (001)^{9!}(0001)^{9!} (00001)^{9!}$.  
Arguing as in the proof of Theorem \ref{thm:tdnoncrossbounded}(ii), any pattern $\tau \in 
\NC\Pi^{|\Sigma|}_{\infty,9}$ that is consistent with both $(v_1,+)$ and $(v_2,-)$ must be of the
shape $x_1^{k_1}x_2^{k_2}x_3^{k_3}$, where $k_1 \mid 4, k_2 \mid 8$ and $k_3 \mid 9$.
Thus at this stage, it suffices to distinguish $\pi$ from the three patterns $x_1^2 x_2^8 x_3^9$,
$x_1^4 x_2^4 x_3^9$ and $x_1^4 x_2^8 x_3^3$.  Put $(v_3,-)$ into $T$, where
$$
v_3 := 0^8 (10^6)^8 0^3 = 0^2 (0^6 1)^8 0^9.
$$
To see that $v_3 \notin L(\pi)$, assume, by way of contradiction, that some morphism $\psi: X^* \mapsto \Sigma^*$
satisfies $\psi(\pi) = v_3$.  Since $v_3$ is not a $4$-th, $8$-th or $9$-th power, at least two variables of $\pi$ are not
mapped to the empty word by $\psi$.  

First, suppose $\psi(x_1) \neq \ve$.  Then $\psi(x_1^4)$ must be equal to
either $0^8$ or $0^4$.  If $\psi(x_1^4) = 0^8$, then, since $(10^6)^8 0^3$ is not a $9$-th power, 
$\psi(x_2^8) = (10^6)^8$ and therefore $\psi(x_3^9) = 0^3$, which is impossible.
The argument for the case $\psi(x_1^4) = 0^4$ is similar.
Second, suppose $\psi(x_1) = \ve$.  Then $\psi(x_2^8) \neq \ve$ and $\psi(x_3^9) \neq \ve$.
Hence $\psi(x_2^8) = 0^8$ and so $\psi(x_3^9) = (10^6)^8 0^3$, which is impossible.

On the other hand, $v_3 \in L(x_1^2 x_2^8 x_3^9) \cap L(x_1^4 x_2^8 x_3^3)$.
Thus it only remains to distinguish $\pi$ from $x_1^4 x_2^4 x_3^9$, and this may be done
with a single negative example, say $v_4 := (01)^4 (001)^4 (0001)^9$.  

\subsection{Remark on Theorem \ref{thm:tdnoncrossbounded}}

Establishing the exact \TD\ 
of any given pattern in $\NC\Pi^z_{\infty,m}$ (for any fixed finite $z \geq 2$
and $m \geq 2$) seems to be quite difficult in general.  We highlight a potential difficulty faced when one tries to 
apply a natural method to determine a lower bound on the \TD\ 
of such a pattern.  
Suppose $\pi := x_1^{n_1} \ldots x_k^{n_k}$, where $n_1,\ldots,n_k \geq 2$ and $k \geq 1$.  For each maximal 
proper prime power factor $q^r$ of $n_i$, let $\pi_{i,q}$ be the pattern derived from $\pi$ by replacing $x_i^{n_i}$ 
with $x_i^{q}$, and let $P$ be the finite class of all patterns so obtained.  For the sake of convenience, assume
the variables of patterns in $P$ are renamed so that for all $P_i,P_j \in P$ with $i \neq j$, $\Var(P_i) \cap \Var(P_j) = 
\emptyset$ and $\Var(P_i) \cap \{x_1,\ldots,x_k\} = \emptyset$. 
For each partition $\mathcal{P}$ of $P$ and every member $\{P_1,\ldots,P_d\}$ of $\mathcal{P}$, 
let $y_1,\ldots,y_{\ell'}$ be all the variables occurring in $P_1,\ldots,P_d$.  Then $\pi$ is distinguishable from 
$\{P_1,\ldots,P_d\}$ with a single negative example iff the sentence 
\begin{equation}\label{eqn:decidewordeqn}
\begin{aligned}
& (\exists y_1,\ldots,y_{\ell'})(\forall x_1,\ldots,x_k)
\left[\left(\bigwedge^d_{j=2} (P_1(y_1,\ldots,y_{\ell'}) = P_j(y_1,\ldots,y_{\ell'}))\right) \wedge P_1(y_1,\right. \\
& \left.\ldots,y_{\ell'}) \neq \pi(x_1,\ldots,x_k)\right]
\end{aligned}
\end{equation}
holds.  As implied by the work of Karhum\"{a}ki et al.\ \cite{Karhumaki00}, there is a 
word equation $E$ with variables in $\Var(P_1) \cup 
\Var(\pi) \cup \{z_1,\ldots,z_{\ell''}\}$ (for some additional variables $z_1,\ldots,z_{\ell''}$) such that the
inequation $P_1(y_1,\ldots,y_{\ell'}) \neq \pi(x_1,\ldots,x_k)$ is equivalent to $(\exists z_1,\ldots,z_{\ell''})E$.  Consequently,
(\ref{eqn:decidewordeqn}) is equivalent to a sentence whose prenex normal form has quantifier prefix
$\exists\forall\exists$ (call this an $\exists\forall\exists$-sentence; a $\forall\exists$-sentence
is defined analogously) over a conjunction of word equations.  If a decidability procedure exists for all such $\exists\forall\exists$-sentences, then one could decide whether or not $\pi$ is distinguishable from 
$\{P_1,\ldots,P_d\}$ with a single example.  More generally, one 
could find a largest number $f \leq |P|$ such that for all partitions $\mathcal{P}$ of $P$ of size $f' 
< f$ (that is, $\mathcal{P}$ has exactly $f'$ members), there is a member $\{P_1,\ldots,P_d\}$ of $\mathcal{P}$
from which $\pi$ is not distinguishable with a single example.  Then $f$ would be a lower bound on the teaching
dimension of $\pi$ w.r.t.\ $\NC\Pi^z_{\infty,m}$.  However, this method does not seem feasible because 
the set of all $\forall\exists$-sentences over positive word equations (combinations of word equations
using $\wedge$ or $\vee$) is already undecidable \cite{GaneshMSR12}.    

\subsection{Proof of Theorem \ref{thm:pbtdgeneralinfinite}}

\proofthmpbtdgeneralinfinite

\subsection{Proof of Lemma \ref{lem:succinctinfinitealphabet}}

\prooflemsuccinctinfinitealphabet

\subsection{Proof of Lemma \ref{lem:subsetrestrictedvarwitness}}

\prooflemsubsetrestrictedvarwitness

\subsection{Proof of Theorem \ref{thm:subclassfinitetd}}

\proofthmsubclassfinitetd

\medskip
\noindent\emph{Assertion (ii).}  The first part of this assertion follows quite directly from the 
proof of a result in \cite{bayeh18}.  As the latter reference is currently under review, we reproduce the 
proof here (with a few minor modifications). 

In this proof, $\Pi^z_k$ denotes the class of all $k$-variable patterns.
Let $\pi$ be a given $(k-1)$-variable pattern in which every variable occurs at most $m$ times, and suppose 
for the sake of a contradiction that $\pi$ is not simple block-regular but it has a finite teaching set $T$ 
w.r.t.\ $\Pi^z_k$.
Let 
\begin{equation}\label{eqn:decomposepi}
\pi = Y_1 \underbrace{c_1}_{I_1} Y_2 \ldots Y_i \underbrace{c_i}_{I_i} Y_{i+1} \ldots \underbrace{c_{n-1}}_{I_{n-1}} Y_n,
\end{equation} 
where $Y_1,Y_n \in X^*$, $Y_2,\ldots,Y_{n-1} \in X^+$, $c_1,\ldots,c_{n-1} \in \Sigma^+$
and $I_1,\ldots,I_{n-1}$ are the closed intervals of positions of $\pi$ corresponding,
respectively, to the particular occurrences of the constant blocks $c_1,\ldots,c_{n-1}$
as marked in Equation (\ref{eqn:decomposepi}).  
Fix some $s > \max(\{|\alpha|:\alpha \in T^+\cup T^- \cup \{\pi\}\})$ and pick a 
variable $y \in X \sm \Var(\pi)$.  We consider three cases.
\begin{description}[leftmargin=0cm]
\item[Case 1:] There is a least $i \in \{1,\ldots,n\}$ such that $Y_i \neq \ve$ and every 
variable in $Y_i$ occurs at least twice in $\pi$.  
We will assume that $2 \leq i \leq n-1$, as the cases $i=1$ and $i = n$ can be handled very similarly.
Fix some distinct $a,b \in \Sigma$ such that both $a$ and $b$ differ from the last
symbol of $c_{i-1}$ as well as the first symbol of $c_i$.\footnote{Such choices of $a$ and $b$
are possible because $|\Sigma| \geq 4$.}
Suppose $Y_i$ starts at the $p^{th}$ position of $\pi$.
We consider two subcases.
\begin{description}[leftmargin=0cm]
\item[Case 1.1:] For every variable $x$ occurring in $Y_i$, $x$ occurs
in some $Y_{i'}$ with $i' \neq i$.
Let $\pi'$ be the pattern derived from $\pi$ by inserting $y^s$ between
the $p^{th}$ and the $(p+1)^{st}$ positions of $\pi$; $\pi' \in 1\Pi^z_m$
because no variable of $\pi$ occurs more than $m$ times.
Note that $L(\pi) \subseteq L(\pi')$ by construction, and so $\pi'$ is
consistent with $\{(v,+): v \in T^+\}$.  Moreover, since $|w| > \max(\{|\alpha|:
\alpha \in T^-\})$ for all $w \in L(\pi') \sm L(\pi)$, $\pi'$ is also
consistent with $\{(v,-): v \in T^-\}$.  
Hence $\pi'$ is consistent with $T$.  Furthermore, let $w = \pi'[y \ra a]$.  
Decompose $w$ as 
\begin{equation}\label{eqn:wdecomposecase1}
w = \underbrace{c_1}_{J_1} \ldots \underbrace{c_{i-1}}_{J_{i-1}} a^s \underbrace{c_i}_{J_i} \ldots \underbrace{c_{n-1}}_{J_{n-1}},
\end{equation}
where $J_1,\ldots,J_i,\ldots,J_{n-1}$ are the closed intervals of positions of $w$ corresponding,
respectively, to the particular occurrences of the constant blocks $c_1,\ldots,c_i,\ldots,$ $c_{n-1}$ as marked in 
Equation (\ref{eqn:wdecomposecase1}).  Assume, by way of contradiction, that there
exists a substitution $h:X \mapsto \Sigma^*$ such that $h(\pi) = w$. By the choice of $a$,
$\cI_{h,\pi}(I_{i-1})$ cannot be an interval starting or ending between $J_{i-1}$ and $J_i$.
Furthermore, $\cI_{h,\pi}(I_{i-1})$ cannot intersect any of the intervals $J_1,\ldots,J_{i-2}$
because otherwise $\sum_{j=1}^{i-2}\left|\cI_{h,\pi}(I_j)\right|$ would be smaller than
$\sum_{j=1}^{i-2}\left|J_j\right|$, which is impossible.  Similarly, $\cI_{h,\pi}(I_{i-1})$
cannot intersect any of the intervals $J_i,\ldots,J_{n-1}$.  Hence $\cI_{h,\pi}(I_{i-1}) = J_{i-1}$.
An analogous argument shows that $\cI_{h,\pi}(I_i) = J_i$.  It follows that for all
$j \in \{1,\ldots,n-1\}$, $\cI_{h,\pi}(I_j) = J_j$.  Thus there is a subsequence $Y^{\prime}_i$ of $Y_i$ such that $Y^{\prime}_i \neq \varepsilon$ and $h(Y^{\prime}_i)=a^s$. However, based on Equation (\ref{eqn:wdecomposecase1}) and the fact that every variable
of $Y_i$ occurs in some $Y_{i'}$ with $i' \neq i$, it can be concluded that $h(Y^{\prime}_i)=\varepsilon$, which contradicts $h(Y^{\prime}_i)=a^s$.
Thus $w \in L(\pi') \sm L(\pi)$, and so $T$ cannot be a teaching
set for $\pi$ w.r.t.\ $\Pi^z_k$.   
\item[Case 1.2:] $Y_i$ contains at least one variable that
does not occur in any $Y_j$ with $j \neq i$.  Let $x_{j_1},\ldots,x_{j_{\ell}}$
be all the variables of $Y_i$ that do not occur outside $Y_i$,
and let $p_1,p_2,\ldots,p_{\ell'}$ be all the positions of $\pi$
that are occupied by some $x_{j_q}$ with $q \in \{1,\ldots,\ell\}$,
where $p_1 < p_2 < \ldots < p_{\ell'}$.
Let $\pi'$ be the pattern derived from $\pi$ by simultaneously inserting 
$y^{2s-j+1}$ between the $(p_j-1)^{st}$
and the $p_j^{th}$ positions of $\pi$ for all $j \in \{1,\ldots,\ell'\}$.
For example, if $\pi = x_1x_2 a x_2 x_3 x_3 b x_4$ and $i = 2$, then 
$\pi' = x_1 x_2 a x_2 y^{2s}x_3 y^{2s-1} x_3 b x_4$.
Note that $\pi' \in 1\Pi^z_m$. 
By construction, $L(\pi) \subseteq L(\pi')$ and $\pi'$
is consistent with $T$.  Now set $\beta = \pi'[y \ra a, x_{j_q} \ra b,
1 \leq q \leq \ell]$.  We argue that $\beta \notin L(\pi)$.
One has that
\begin{equation}\label{eqn:wsubyandxij}
\beta := c_1 \ldots c_{i-1} \underbrace{a^{2s} b a^{2s-1} b \ldots b a^{2s-\ell'+1} b}_{\ga} c_i \ldots c_{n-1}.
\end{equation} 
\end{description}
By arguing as in Case 1.1, the choice of $a,b$ implies that if 
$Y'_i$ is the restriction of $Y_i$ to $\{x_{j_1},\ldots,x_{j_{\ell}}\}$, 
then there is a substitution $h:X \mapsto \Sigma^*$ such that $\ga = h(Y'_i)$, where $\ga$
is as defined in Equation (\ref{eqn:wsubyandxij}).  Note that $|Y'_i| = \ell'$.

That $\ga \neq h\left(Y'_i\right)$ will follow from Lemma \ref{lem:numberofcutsandpositionsrelation}
and the following claim.  


\begin{subclaim}\label{clm:numberofcutsofga}
If $\ga = h(Y'_i)$, then $\ga$ has at least $\left|Y'_i\right|$ cuts relative to $(h,Y'_i)$.
\end{subclaim}

\noindent\emph{Proof of Claim \ref{clm:numberofcutsofga}.}
We first decompose $\ga$ as follows:
\begin{equation}\label{eqn:gadecompose}
\ga := \underbrace{a^{2s}}_{I_1} \underbrace{b a^{2s-1} b}_{I_2} \ldots
\underbrace{b a^{2s-j+1} b}_{I_j} \ldots \underbrace{b a^{2s-\ell'+1} b}_{I_{\ell'}}.  
\end{equation}
Claim \ref{clm:numberofcutsofga} will follow from the fact that for all 
$j \in \{1,\ldots,\ell'\}$, $I_j$ contains at least one cut-point.
First, observe that since $a^{2s}$ occurs exactly once as a substring
of $\ga$ and every variable of $Y'_i$ occurs at least twice in $Y'_i$, 
there cannot exist any $q \in \{1,\ldots,\ell\}$ such that
$a^{2s}$ is a substring of $h(x_{j_q})$.  Thus $I_1$ must contain
at least one cut-point of $\ga$.  Second, for all $j \in \{2,\ldots,\ell'\}$,
$b a^{2s-j+1} b$ occurs exactly once as a substring of $\ga$.
Arguing as before, we conclude that $I_j$ contains at least one cut-point of 
$\ga$.~\qed~(Claim \ref{clm:numberofcutsofga}) 

It follows from Lemma \ref{lem:numberofcutsandpositionsrelation} and 
Claim \ref{clm:numberofcutsofga} that $\ga \neq h\left(Y'_i\right)$
and therefore $\beta \notin L(\pi')$, as desired.

\item[Case 2:] $\pi$ contains a substring of the shape $ab$, where $a,b \in \Sigma$
($a$ and $b$ are not necessarily distinct). 
Since $|\Sigma| \geq 4$, one can
fix some $c \in \Sigma$ with $c\notin\{a,b\}$.
Let $j_3$ be a position of $\pi$ such that 
$\pi[j_3]\pi[j_3+1] = ab$.  If $L(\pi)$ had a finite
teaching set $T$ w.r.t.\ $\Pi^z$, then one can argue
as in Case 1 that there is a positive $s$ so
large that if $\pi'$ is obtained from $\pi$ by inserting
$y^s$ between the $j_3^{th}$ and $(j_3+1)^{st}$ positions
of $\pi$, 
then $\pi'$ would be consistent with $T$.
On the other hand, let $\gamma$ be the string derived
from $\pi'$ by substituting $c$ for $y$ and $\ve$ for every other variable;
note that the number of times the substring $ab$
occurs in $\gamma$ is strictly less than the number
of times that $ab$ occurs in $\pi$, which implies
$\gamma \notin L(\pi)$ and so $L(\pi') \neq
L(\pi)$.  Therefore $\TD(\pi,\Pi^z)=\infty$.

\item[Case 3:] $\pi$ does not start or end with variables.  
Suppose $\pi$ starts with the constant symbol $a$.
The proof that $L(\pi)$ has no finite teaching set w.r.t.\
$\Pi^z$ is very similar to that in Case 2;
the only difference here is that one chooses some
$b \in \Sigma\sm\{a\}$ and considers $\pi' = y^s\pi$
for some variable $y \notin \Var(\pi)$ and a sufficiently
large $s$.  In this case, $b^s\pi(\ve) \in L(\pi')\sm L(\pi)$, and therefore
$L(\pi') \neq L(\pi)$.  An analogous argument holds if $\pi$ ends with a 
constant symbol. 
\end{description} 

Next, we prove the second part of the assertion.  Suppose $\pi$ contains a variable $x$ that
occurs $\ell$ times for some $\ell > m$.  We build a teaching set $T$ for $\pi$ w.r.t.\ $1\Pi^{\infty}_m$.  
First, put the sample $\{(\pi(\ve),+)\} \cup \{(w',-): w' \sqsubset \pi(\ve)\}$ into $T$; this
sample uniquely identifies the constant part of $\pi$ (i.e.\ $\pi(\ve)$).
Second, pick some $a \in \Sigma \sm \Const(\pi)$ and put $(\pi[x \ra a],+)$ into $T$;
this additional example reduces the version space to all patterns in $1\Pi^{\infty}_m
\cap \Pi^{\infty}_{\infty,\ell}$.  Since, by Assertion (i), every pattern in $\Pi^{\infty}_{\infty,\ell}$    
has a finite \TD, this implies that $\TD(\pi,1\Pi^{\infty}_m) \leq \TD(\pi,\Pi^{\infty}_{\infty,\ell}) < \infty$,
as required.  Furthermore, if $\pi$ is simple block-regular, then it follows from \cite[Proposition 4]{bayeh17}
that $\TD(\pi,1\Pi^{\infty}_m) \leq \TD(\pi,\Pi^{\infty}) < \infty$.~\qed 

\subsection{Proof of Theorem \ref{thm:sizetwoboundedvarfreq}}

\proofthmsizetwoboundedvarfreq

\subsection{Remark \ref{rem:noncrosslowerbound}}\label{appen:binarynoncrosslowerbound}

We prove that $T := \{(\ve,+),(0^2 1^2 0^2,+),(0,-),(01^20,$ $-),(0^3,-),((01)^2(0^21)^2$ 
$(0^31)^2$ $(0^41)^2,-)\}$ 
is a teaching set for $\pi := x_1^2 x_2^2 x_3^2$ w.r.t.\ $\Pi^2_{\infty,3}$.  
Let $\tau$ be any pattern 
in $\Pi^2_{\infty,3}$ that is consistent with $T$.  Since $\ve \in L(\tau)$, $\tau$ does not contain
any constant symbols.  The negative examples $(0,-)$ and
$(0^3,-)$ ensure that every variable of $\tau$ occurs exactly twice.
The consistency of $\tau$ with $(01^20,-)$ then implies that $\tau$ is a non-cross
pattern, i.e., $\tau$ is equivalent to a pattern of the shape
$x_1^2 x_2^2 \ldots x_k^2$ for some $k$.  Since $0^2 1^2 0^2 \in L(\tau)$, $k \geq 3$.
Finally, $(01)^2(0^21)^2(0^31)^2(0^41)^2 \notin L(\tau)$ implies that $k \leq 3$.
Hence $\tau$ is equivalent to $x_1^2 x_2^2 x_3^2$. 

\end{document}